\documentclass[review,12pt]{elsarticle}

\usepackage{graphicx}
\usepackage{caption}
\usepackage{subcaption}
\usepackage{amsbsy}
\usepackage{xcolor}
\usepackage{amsmath}
\numberwithin{equation}{section}
\usepackage{hyperref}

\newcommand{\difft}[1]{\mathop{}\!\frac{\mathrm{d}#1}{\mathrm{d}t}}
\newcommand{\diffu}[1]{\mathop{}\!\frac{\mathrm{d}#1}{\mathrm{d}u}}
\newcommand{\diffint}[1]{\mathop{}\!\mathrm{d}#1}
\newcommand{\partdev}[2]{\mathop{}\!\frac{\partial #1}{\partial #2}}
\newcommand{\TextInMath}[1]{\mathop{}\!\mathrm{#1}}

\begin{document}

\begin{frontmatter}

	\title{Interplay between exogenous triggers and endogenous behavioral changes in contagion processes on social networks}

        \author[1]{Clara Eminente\corref{aaa}}
        \ead{clara.eminente@gmail.com}
        \author[2]{Oriol Artime}
        \author[1,3]{Manlio De Domenico}
        
        \cortext[aaa]{Corresponding Author}
        \address[1]{Department of Physics and Astronomy `Galileo Galilei', University
of Padua, Padova, Veneto, Italy}
        \address[2]{CHuB Lab, Fondazione Bruno Kessler, Via Sommarive 18, Povo, TN 38123, Italy}
        \address[3]{Padua Center for Network Medicine, University of Padua, Padova, Veneto, Italy}

	\begin{abstract}
		In recent years, statistical physics' methodologies have proven extremely successful in offering insights into the mechanisms that govern social interactions. However, the question of whether these models are able to capture trends observed in real-world datasets is hardly addressed in the current literature. With this work we aim at bridging the gap between theoretical modeling and validation with data. In particular, we propose a model for opinion dynamics on a social network in the presence of external triggers, framing the interpretation of the model in the context of misbehavior spreading. We divide our population in aware, unaware and zealot/educated agents. Individuals change their status according to two competing dynamics, referred to as behavioral dynamics and broadcasting. The former accounts for information spreading through contact among individuals whereas broadcasting plays the role of an external agent, modeling the effect of mainstream media outlets. Through both simulations and analytical computations we find that the stationary distribution of the fraction of unaware agents in the system undergoes a phase transition when an all-to-all approximation is considered. Surprisingly, such a phase transition disappears in the presence of a minimum fraction of educated agents. Finally, we validate our model using data collected from the public discussion on Twitter, including millions of posts, about the potential adverse effects of the AstraZeneca vaccine against COVID-19. We show that the intervention of external agents, as accounted for in our model, is able to reproduce some key features that are found in this real-world dataset. 
	\end{abstract}


	\end{frontmatter}
	\href{https://doi.org/10.1016/j.chaos.2022.112759}{https://doi.org/10.1016/j.chaos.2022.112759}

	\section{Introduction}\label{section:intro}
	
	The effects that external, often traumatic events have on collective attention and public opinion are of utmost importance due to their societal, economical and political impact and, accordingly, have been studied from different points of view in various disciplines~\cite{starbird2014, infodemy, vidgen2020}. The role played by mainstream media in disseminating information is particularly crucial during periods of crisis~\cite{barry2009}, as citizens' perception of the news can alter the way they process and share information, ultimately leading to behavioral changes that might be harmful from both the individual and the collective viewpoints. This inevitably entangles the role of mainstream media with another increasingly important phenomenon: the unprecedented speed and reach of content spreading, such as rumors and fake news, on online social networks~\cite{wardle2017}. However, recent studies have shown that often mainstream media fail to bridge the public discourse with the accurate and objective representation of external events~\cite{sanford2021}, calling for a study of the effects of their influence: do mainstream media influence online discussion both in terms of topics (what is discussed online) and sentiment (how is it addressed)? Can the way media report an event be more relevant in shaping public discussion than the occurrence of the event itself? Providing solid and convincing answers to this type of questions would be an important step toward a more efficient, reliable and democratic information ecosystem.
	
	Previous works have already investigated the role of mainstream media from a mechanistic perspective. For example, Quattrociocchi \textit{et al.}~\cite{quattrociocchi2014} highlighted how different communication strategies can determine the reach of consensus in the population; Gonz\'{a}lez-Avella \textit{et al.}~\cite{gonzalez2007information} linked the cultural diversity of a society with the influence strength of the broadcasted messages by the media; Brooks \textit{et al.}~\cite{brooks2020} proposed a model that incorporates mainstream media as part of the social media network, analyzing how to maximize their influence. Moreover, by analyzing real-world data, it has been found that mainstream media agenda and online discussion tend to align, especially in periods of crises~\cite{gozzi2020}. Finally, Pires \textit{et al.}~\cite{pires2021antivax} showed how the sentiment towards a topic (vaccination) can indeed have disruptive effects when coupled to another dynamical process (disease spreading). The strands of research can be roughly divided between those works that propose simple models where isolated socially-inspired mechanisms are tested in order to establish cause-effect relations that go beyond statistical correlations~\cite{artime2019herding} (see, e.g.,~\cite{castellano2009statistical, sen2014sociophysics} for good reviews), and those works that aim at giving insights directly from the data analysis~\cite{gallotti2020assessing, cha2010measuring}, that sometimes are oriented toward accurate future predictions, e.g., via machine learning algorithms. Of course, the frontier is blurry and there is a continuous spectrum of works that lie between these two approaches~\cite{d2022epidemic}, even though they are still scarce in number. 
	
	With the aim of bridging the two aforementioned strands in the context of information spreading and mass media influence, here we study a variation of one of these statistical mechanics flavoured models -the so-called Kirman~\cite{kirman1993ants} or noisy voter model~\cite{carro2016noisy}- and evaluate its performance at reproducing the onset of the online debate in Twitter about the \textit{AstraZeneca} vaccine ban in the late winter of 2021. Hence, from a theoretical and modeling standpoint, we address how exogenous events influence the behavioral dynamics on a social network via a model that, with an appropriate fitting procedure, is able to reproduce some trends observed in a real event. 
	
	The article is organized as follows. In the first section we introduce our proposed model, explaining how we account for both social interactions and the external drive of mainstream media. We provide a summary of results, highlighting different regimes of the dynamics using agent-based simulations and offering a description of the model analytically in its all-to-all approximation. We next move to the comparison of the model to a real-world scenario. To close, we offer the conclusions.
	
	\section{Modeling the role of broadcasters in a social contagion process}\label{section:model}
	The phenomenon we are investigating lies at the intersection between opinion and information (e.g., rumors) spreading~\cite{castellano2009statistical}. We refer to the spreading phenomenon taking place on the social network as \textbf{behavioral dynamics}. This accounts for standard communication between individuals via their social contacts. On top of that, we consider an additional process that, in first approximation, is not bounded by the topology. This, what we call \textbf{broadcasting dynamics}, is related to the role played by the mainstream media outlets, such as newspapers, television and radio, whose direct influence disregards social connections. Last but not least, we want to take into account those individuals with a strong moral, who would never engage in the misbehavior. They can also be interpreted as a fraction of ``uninterested'' individuals, who might, in general, be connected to individuals with certain interests (e.g. the discussion around vaccines) but that decide to not take part in a particular discussion (e.g. hesitancy toward the \textit{AstraZeneca} vaccine).
	
	\subsection{Mathematical modeling of behavioral and broadcasting dynamics}\label{section:mathModel}
	Let us consider a population of $N$ individual, also called agents, each one endowed with a binary variable: $U$ and $A$. In the framework of misconduct dynamics, $U$ stands for \textit{unaware}, whereas $A$ stands for \textit{aware} of the misconduct. The unaware population is split, in turn, into two types of agents $ U(t) = \tilde{U} + U_0(t) $, such that $ 0 \leq \tilde{U} + U_0(t) \leq N$. $\tilde{U}$ represents the number of \textit{zealots}, agents that, by external reasons, will not change their state~\cite{zealots,xie2011social, khalil2018}. On the contrary, $U_0$ is the corresponding population of unaware individuals that can change state, therefore all the temporal evolution of the unaware population is precisely due to $U_0$.
	
	We consider two mechanisms of state change. The first is an SIS-like dynamics~\cite{pastor2015epidemic}, given by the transitions
	\begin{align}{}
		& A + U_0 \quad \overset{\lambda}{\longrightarrow} \quad  2A \label{eq:transitionBehavUA}\\
		& A + U \quad  \overset{\mu}{\longrightarrow} \quad U + U_0 \label{eq:transitionBehavAU}.
	\end{align}
 
	These transitions correspond to the \textbf{behavioral spreading}. Note that, at odds with epidemic models, both transitions occur by contact. In fact, when the number of zealots $\tilde{U} = 0$, the model reduces to well-known models such as the asymmetric (or biased) voter model~\cite{lanchier2007voter, sood2008voter}, the Abrams-Strogatz model of language competition with neutral volatility~\cite{abrams2003modelling, vazquez2010agent} or the asymmetric Hubbell's Community Drift Model~\cite{zhang1997effects}. 
	
    The second mechanism is the \textbf{broadcasting}, given by the transitions rates
	\begin{align}
		& U_0 \quad  \overset{\beta}{\longrightarrow} \quad A \label{eq:transitionBroadcUA}\\
		& A \quad  \overset{\gamma}{\longrightarrow} \quad U_0 \label{eq:transitionBroadcAU}.
	\end{align}

	These state changes occur spontaneously, i.e., without a contact of individuals in the contrary state. We make two further assumptions in the broadcasting dynamics. The first is that the transitions \eqref{eq:transitionBroadcUA} and \eqref{eq:transitionBroadcAU} are not always available but they are activated only when broadcasting is on and the media are informing about the topic of interest. The second is that the broadcasting dynamics can only affect a fraction $B$ of users at each time, which we choose uniformly at random. In other words, $B$ is the fraction of agents reached by the news. These spontaneous changes of state have been considered in other mechanistic models, in contexts such as economy~\cite{kirman1993ants, alfarano2005estimation, carro2015markets}, consensus formation~\cite{carro2016noisy, artime2018aging}, catalytic reactions~\cite{fichthorn1989noise, considine1989comment} or percolation in strongly correlated systems~\cite{lebowitz1986percolation}. Summarizing, the set of variables in our model are the transition rates $\lambda$, $\mu$, $\beta$, $\gamma$, the number of educated individuals $\tilde{U}$, and the fraction of agents susceptible to the broadcasting $B$. Regarding the broadcast switching, for simplicity we assume that it turns on at a certain time $\tau$ and keeps active until the end of the simulation. We do so because we are primarily interested in describing the initial effects that the external trigger has on population of individuals that operates normally, i.e., the transition from a non-perturbed to perturbed communication ecosystem. 
	
	The variables we use to describe the temporal evolution of the state of a single node are $p_i^{U_0} (t)$ and $p_i^A(t)$, the probability to find at time $ t $ the node $i$ in state $U_0$ and $A$, respectively. Note that at any time $p_i^{U_0} (t) + p_i^A (t) + p_i^{\tilde{U}} = 1$. We will be interested in the prevalence
	\begin{equation}
		\rho(t) = \frac{1}{N} \sum_{i=1}^N p_i^A(t),
		\label{eq:prevalenceShort}
	\end{equation}
	which is nothing else than the average fraction of aware agents at a given time, and maps to the fraction of infected agents in epidemics models. The dynamics runs on top of a social network, which is defined by an adjacency matrix $\mathbf{A}$ whose elements $A_{ij}$ are $1$ if individuals $i$ and $j$ are connected, $0$ otherwise. We assume that the connections between nodes are unweighted and undirected.
	
	If only the behavioral dynamics is considered, the temporal evolution of the probabilities to find an agent in either state $A$ or $U$ at time $t$, $p_i^A(t)$ and $p_i^{U_0}(t)$, are given by
	
	\begin{eqnarray}
		 \difft{} p_i^{U_0}(t) = - \left[1 - \prod_{j=1}^N \left(1-\lambda A_{ij}p_j^A \right) \right] p_i^{U_0}(t) + p_i^A(t) \left[1 - \prod_{j=1}^N \left( 1-\mu A_{ij}p_j^U \right) \right] \label{eq:evolutionProbBehavU}\\
		\difft{} p_i^A(t) = + \left[1 - \prod_{j=1}^N \left(1-\lambda A_{ij}p_j^A \right) \right] p_i^{U_0}(t) - p_i^A(t) \left[1 - \prod_{j=1}^N \left( 1-\mu A_{ij}p_j^U \right) \right].
		\label{eq:evolutionProbBehavA}
	\end{eqnarray}

    Similarly, the evolution of $p_i^A(t)$ and $p_i^{U_0}(t)$ if we only consider broadcasting is
	\begin{eqnarray}
		\difft{} p_i^{U_0}(t) & = \left[-\beta p_i^{U_0}(t) + \gamma p_i^A(t) \right] \Theta(t - \tau)p_i^b \label{eq:evolutionProbBroadcU} \\
		\difft{} p_i^A(t) & = \left[ + \beta p_i^{U_0}(t) - \gamma p_i^A(t) \right] \Theta(t - \tau)p_i^b
		\label{eq:evolutionProbBroadcA}
	\end{eqnarray}
	where $\Theta(x)$ is the Heaviside step function and $p_i^b$ is the probability that node $i$ is reached during broadcasting. Note Equations~\eqref{eq:evolutionProbBehavU}--\eqref{eq:evolutionProbBroadcA} sum up $0$, as expected.
	
	In order to compute the prevalence from these equations we can take the derivative of Equation~\eqref{eq:prevalenceShort}. Now, the derivative of the prevalence is expressed as the mean of the time derivative of $p_i^A(t)$, which we can compute using Equations~\eqref{eq:evolutionProbBehavA} and \eqref{eq:evolutionProbBroadcA}. Then, by substituting $p_i^{U_0}(t) = 1 - p_i^{\tilde{U}} - p_i^A(t)$ we obtain:	
	\begin{align}
		\difft{} p_i^A(t) =& - p_i^A(t) \left[ 2 + (\gamma + \beta)\Theta(t - \tau)p_i^b   - \prod_{j=1}^N \left( 1-\lambda A_{ij}p_j^A(t) \right)  + \right. \nonumber \\
		&  \left. - \prod_{j=1}^N \left ( 1-\mu A_{ij}(1 - p_j^A(t)) \right) \right] + \nonumber \\
		& + ( 1 - p_i^{\tilde{U}}) \left[ 1 - \prod_{j=1}^N(1-\lambda A_{ij}p_j^A) + \beta \Theta(t - \tau)p_i^b  \right].
	\label{eq:evolutionProbBehavBroadc}
	\end{align}
	
    Following Equation~\eqref{eq:prevalenceShort} and integrating it is possible to compute the prevalence. In~\ref{section:appendix:preliminaryAnalysis} we offer an overview of some properties of Equation~\eqref{eq:evolutionProbBehavBroadc}. We start by showing the adherence between its numerical solution and the outcome of the dynamics simulated via direct Monte Carlo methods. We then explore the long term role of broadcasting on the prevalence. We find that, depending on the topology of the interacting network, the introduction of the external trigger on top of the behavioral dynamics can have an enhancing (higher prevalence) or diminishing (lower prevalence) effect.
	
	\subsection{Analysis in the complete-graph limit}\label{section:completeGraphModel}
	An analysis of the networked version of our model would be in principle possible, for example, via the standard techniques such as  pair approximations~\cite{pugliese2009heterogeneous, peralta2020binary} or approximate master equations~\cite{gleeson2013binary}. However, analytical insights can already be given in the limit of all-to-all connectivity, where it is assumed \textit{well-mixed population} so the process is not \textit{individual-based} anymore, as every node in the same state is now equivalent. It is important to underline that the complete-graph limit is convenient due to its mathematical tractability, but fails to capture some empirical features of real social networks. We stick for now to this limit, thus disregarding the role of the network parameters but gaining, nonetheless, mathematical understanding, and leave the analysis of uncorrelated networked topologies to~\ref{section:appendix:preliminaryAnalysis}. As we state in the final discussion, the further inclusion of empirical features such as topological correlations is among the most relevant directions this work may take.
	
    Let us start by analysing the stationary state of the model by computing $\rho^*(u)$, the stationary probability density function to find a fraction $u$ of unaware agents once the system has stabilized. The detailed calculations can be found in~\ref{section:appendix:stationarySol}. Here we report the result, which reads	
	\begin{equation}
		\left\{
		\begin{array}{lr}
			\rho^*(u) = D_2(u)^{-1}k \exp\left(2N \int_{\tilde{u}}^{u} \diffint u' D_1(u')D_2(u')^{-1}\right)\\
			\\
			k = \left[ \int_{\tilde{u}}^{1} \diffint u D_2(u)^{-1}k \exp\left( 2N \int_{\tilde{u}}^{u} \diffint u' D_1(u')D_2(u')^{-1} \right) \right]^{-1}, 
		\end{array}
		\right.  
		\label{eq:stationaryPrevalence}
	\end{equation}
	where the auxiliary functions $D_1(u)$ and $D_2(u)$ are defined such that
	\begin{equation}
		\left\{
		\begin{array}{lr}
			D_1(u) \equiv \Omega^+ (u) - \Omega^- (u)\\
			\\
			D_2(u) \equiv \Omega^+ (u) + \Omega^- (u)
		\end{array}
		\right. 
		\label{eq:stationaryPrevalenceAB}
	\end{equation}
	\begin{equation}
		\left\{
		\begin{array}{lr}
			\Omega^- (u) = (u - \tilde{u}) \left(\lambda (1 - u) + B\beta \right) \\
			\\
			\Omega^+ (u) =  (1 - u) \left( \mu u + B\gamma \right).
		\end{array}
		\right.  
		\label{eq:stationaryPrevalenceOmegas}
	\end{equation}
 
	Note the absence in these equations of the onset broadcasting parameters $\tau$. We find that the long-term behavior of the system does not depend on this parameter, which was thus considered to be $\tau=0$ (i.e. broadcasting always available) when computing \eqref{eq:stationaryPrevalence}.
	
	The solution of Equations~\eqref{eq:stationaryPrevalence} turns out to have a fairly cumbersome expression when all the parameters of the process are considered. Therefore, we next proceed step by step, solving several cases of increasing complication and adding one feature at a time, so we can better shed light on the role of every element of the dynamics.
	
	\subsubsection{Symmetric rates and no zealots $(\tilde{u}=0)$}\label{section:completeGraphSymUtilde0}

	We start by considering the simplest case, that is, symmetric rates for both the behavioral and the broadcasting processes, together with the  absence of zealots, i.e. $\tilde{u}=0$. In this case, Equation~\eqref{eq:stationaryPrevalence} can be easily solved, yielding
	\begin{equation}
		\left\{
		\begin{array}{lr}
			\rho^*(u) = D_2(u)^{-1}k D_2(u)^{\frac{N B \beta }{\lambda}} = k \left(  - 2 \lambda u^2 + 2\lambda u + B \beta \right)^{\frac{N B \beta }{\lambda} - 1}\\
			\\
			k = \left[ \int_0^1 \diffint u \left( - 2 \lambda u^2 + 2\lambda u + B \beta \right)^{\frac{N B \beta }{\lambda} - 1}\right]^{-1}.
		\end{array}
		\right.
		\label{eq:statPrevalenceSymUtilde0}
	\end{equation}
	
	The detailed computations can be found in~\ref{section:appendix:StatSolSymATAUtilde0}.
	
\begin{figure}[!ht]
		\centering
		\begin{subfigure}[htb]{0.85\textwidth}
			\centering
			\includegraphics[width=0.8\linewidth]{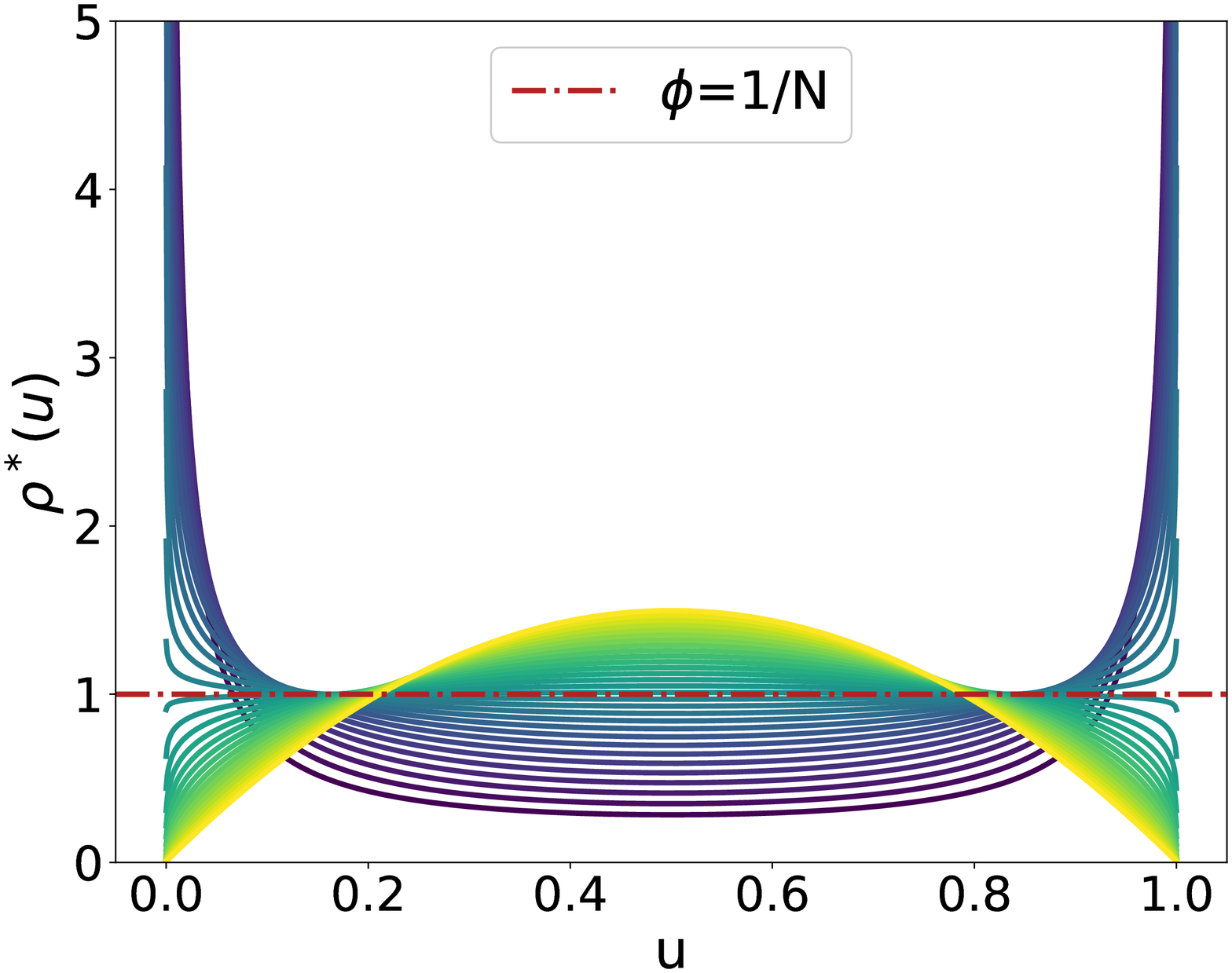}
		\end{subfigure}\hspace{-1cm}
		\begin{subfigure}[htb]{0.1\textwidth}
		\centering
			\includegraphics[width=0.9\linewidth]{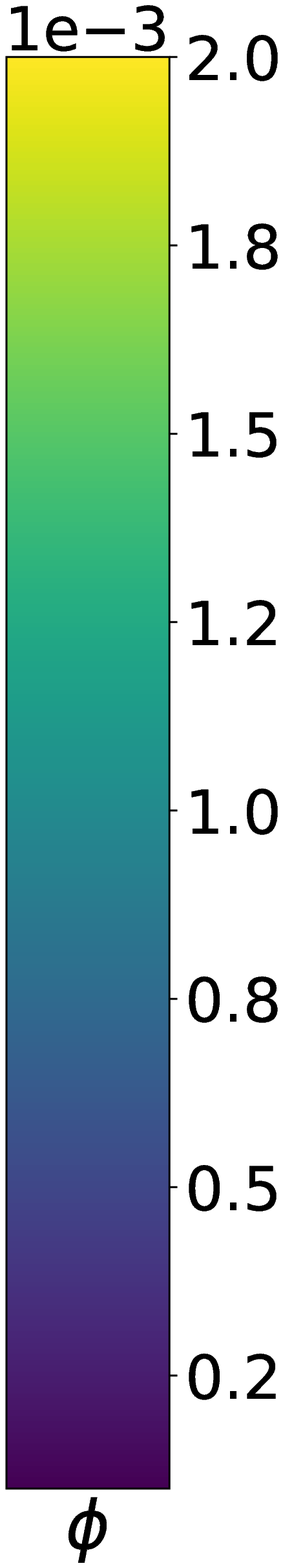}
		\end{subfigure}
	\caption{Stationary probability density function of observing a fraction of unaware agents $u$, $\rho^*(u)$, when varying the comprehensive parameter $\phi=B\beta/\lambda$. The red dashed line corresponds to the case in which the exponent $\alpha = N\phi - 1 = 0$. As explained in the text, the nature and number of extrema depends on $\phi$. We use $N = 1000$.}
	\label{img:Figure1}
	\end{figure}

    Given that the rates are symmetric and given the absence of zealots, the stationary distribution $\rho^*(u)$ is symmetric with respect to $u=\frac{1}{2}$, which corresponds to the case in which half of the population is in the \textit{unaware} state. Moreover we find that the values of the fraction of unaware agents that make $\rho^*(u)$ zero, $u_{\pm}$, fall outside the interval $[0,1]$, making $\rho^*(u) > 0 \quad \forall \quad u \in [0,1]$, as it should be since $\rho^*(u)$ is a probability density function. The state $u = \frac{1}{2}$ is moreover, an extreme point of $\rho^*(u)$ since its derivative vanishes there. Depending on the sign of the exponent $\alpha = NB\beta /\lambda - 1$, it can be a maximum ($\alpha > 0$) or a minimum ($\alpha < 0$). For $\alpha = 0$ $\rho^*(u)$ is constant, marking the transition from a unimodal distribution with a maximum in $u=\frac{1}{2}$ to a bimodal distribution with a minimum in $u=\frac{1}{2}$ and symmetric maxima at the boundaries of the interval, similar to what is observed in the Kirman model~\cite{kirman1993ants}. We define a the ratio $\phi = B\beta /\lambda$, that measures the relative strengths of the broadcasting and the behavioral parameter. We can rewrite $\rho^*(u)$ incorporating all the dependence on the dynamical parameters into the single parameter $\phi$. In Figure~\ref{img:Figure1} we display the transition between the two aforementioned regimes as a function of the $\phi$.
	
	\subsubsection{Symmetric rates and presence of zealots $(\tilde{u}\neq0)$}\label{section:completeGraphSym}
	
	If we drop the assumption of no zealots in the network and assume a finite amount of them it is still possible to compute the stationary probability Equation~\eqref{eq:stationaryPrevalence} in a closed form. Keeping the definition $\phi=B\beta/\lambda$, we obtain
	\begin{equation}
		\left\{
		\begin{array}{lr}
			\rho^*(u) = K_1 \left[ \frac{1 - x(u)}{1 + x(u)}\right] ^{E_1} D_2(u)^{E_2} \\
			\\
			x(u) = \frac{(2-4 u+\tilde{u} )}{{\left[ (\tilde{u} - 2)^2+8 \phi  (1-\tilde{u} )\right]^{1/2}}}\\
			\\
			D_2(u) = -2 u^2 + u (\tilde{u} + 2) + \phi - \tilde{u} - \phi \tilde{u}\\
			\\
			E_1 =  {\frac{N}{2 } \frac{ \left[2 \phi +(2 - \tilde{u} )\right] \tilde{u}}{\left[ (\tilde{u} - 2)^2+8 \phi  (1-\tilde{u} )\right]^{1/2}}}\\
			\\
			E_2 = \frac{N}{2 } (2 \phi + \tilde{u} )- 1.
		\end{array}
		\right.  
		\label{eq:statPrevalenceSym}
	\end{equation}
	The detailed computation can be found in~\ref{section:appendix:StatSolSymATA}.
	
	Compared to the previous case, the introduction of zealots alone makes the stationary probability considerably less trivial to study. Nevertheless, the fact that the whole process is still ascribable to just a few parameters (the comprehensive parameter $\phi$ and the fraction of zealots  $\tilde{u}$) makes it possible for us to highlight their role from a qualitative point of view. 
	
\begin{figure}[!ht]
		\centering
		\begin{subfigure}[h!]{0.85\textwidth}
		   \begin{subfigure}[b]{0.5\textwidth}
    			\centering
    	    	\includegraphics[width=\textwidth]{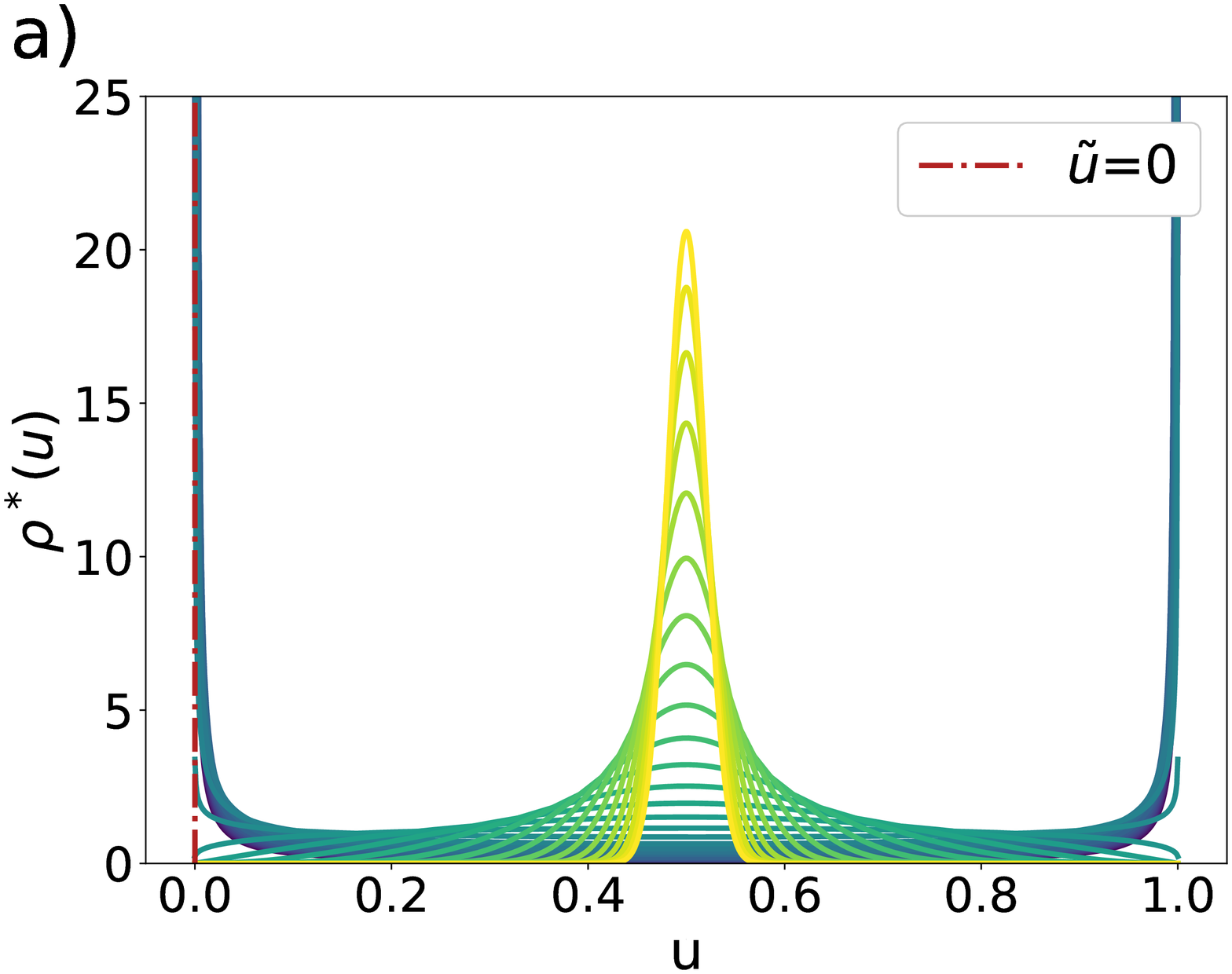}
    		\end{subfigure}
    		\begin{subfigure}[b]{0.5\textwidth}  
    			\centering 
    			\includegraphics[width=\textwidth]{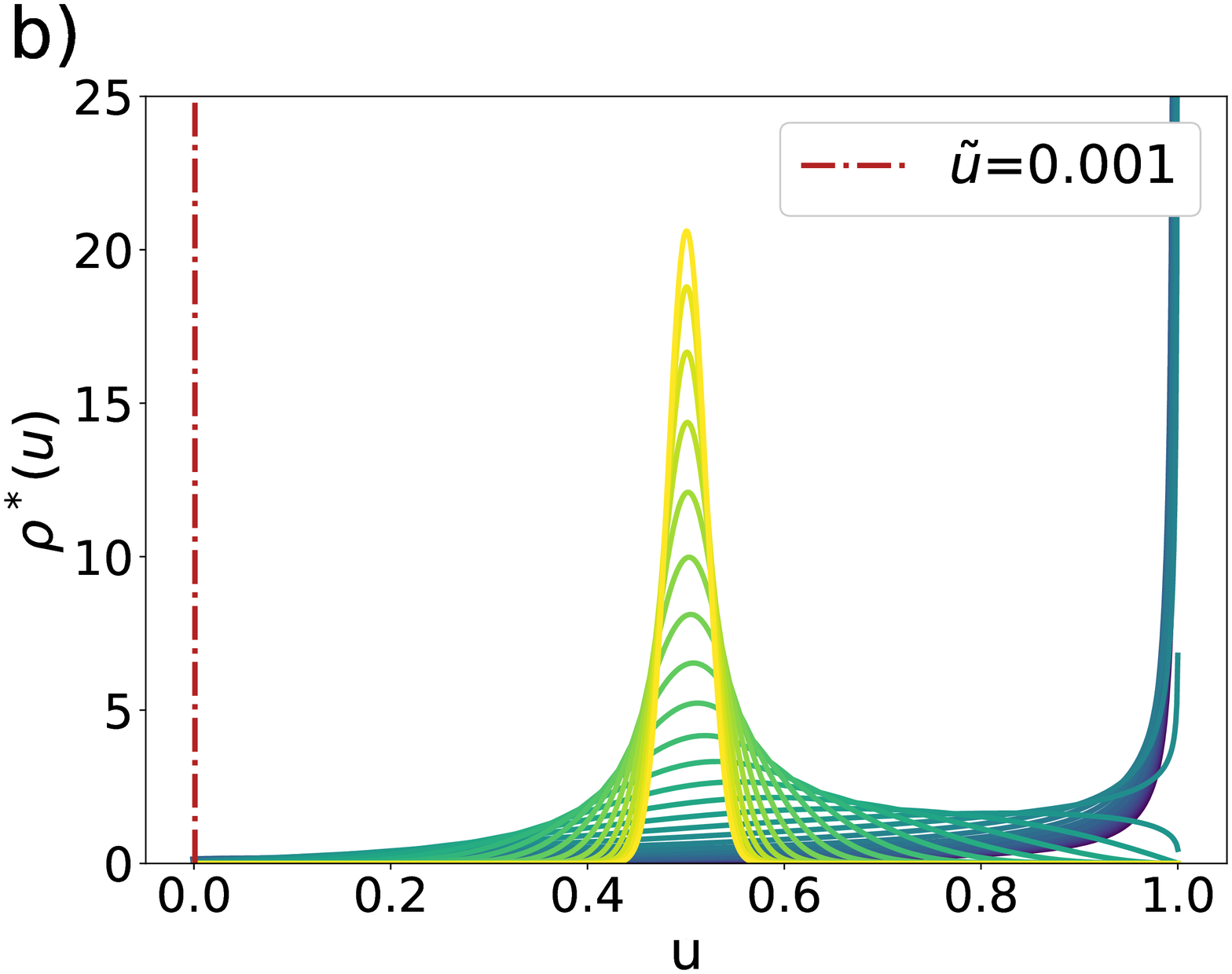}
    		\end{subfigure}
    		\begin{subfigure}[b]{0.5\textwidth}   
    			\centering 
    			\includegraphics[width=\textwidth]{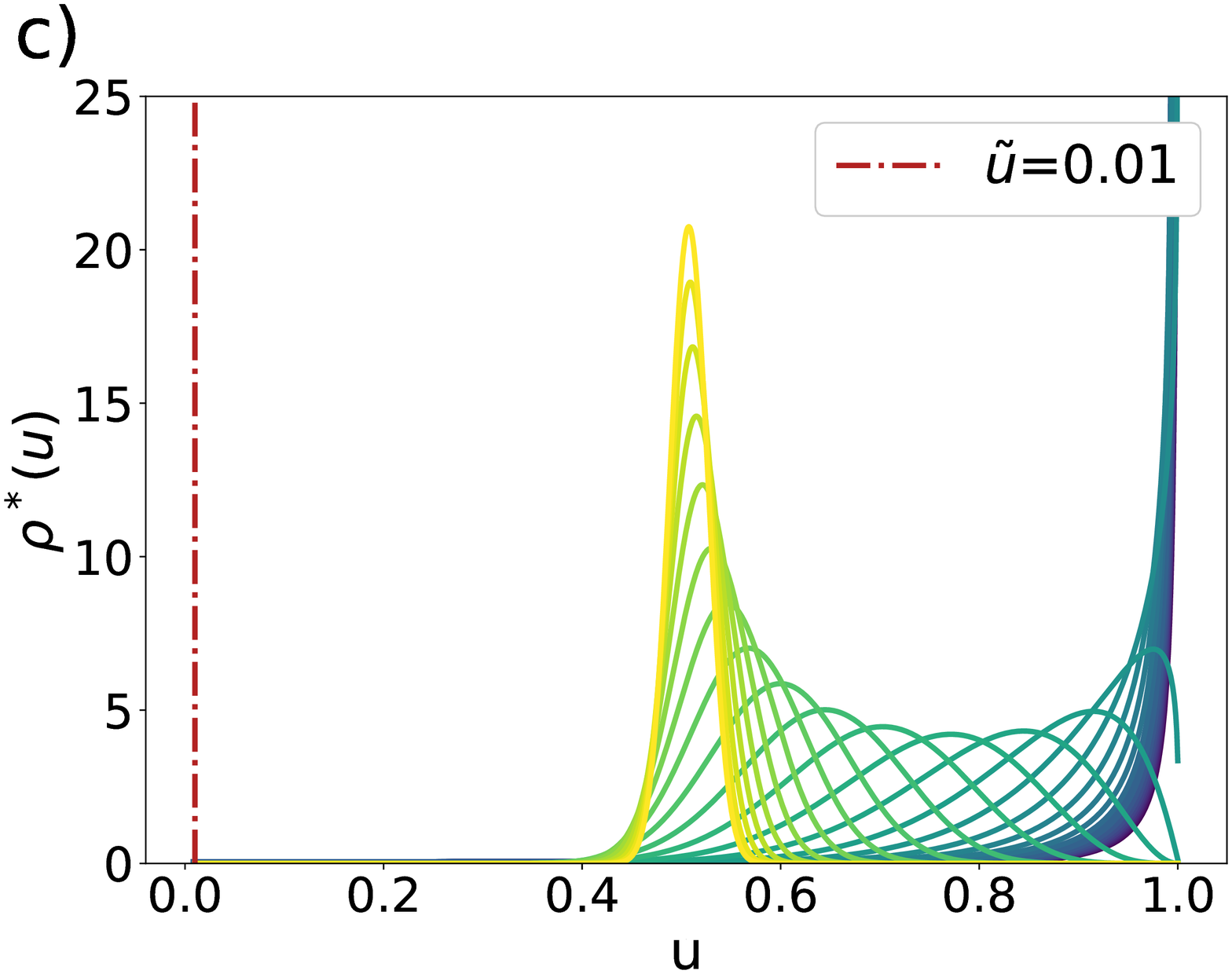}
    		\end{subfigure}
    		\begin{subfigure}[b]{0.5\textwidth}   
    			\centering 
    			\includegraphics[width=\textwidth]{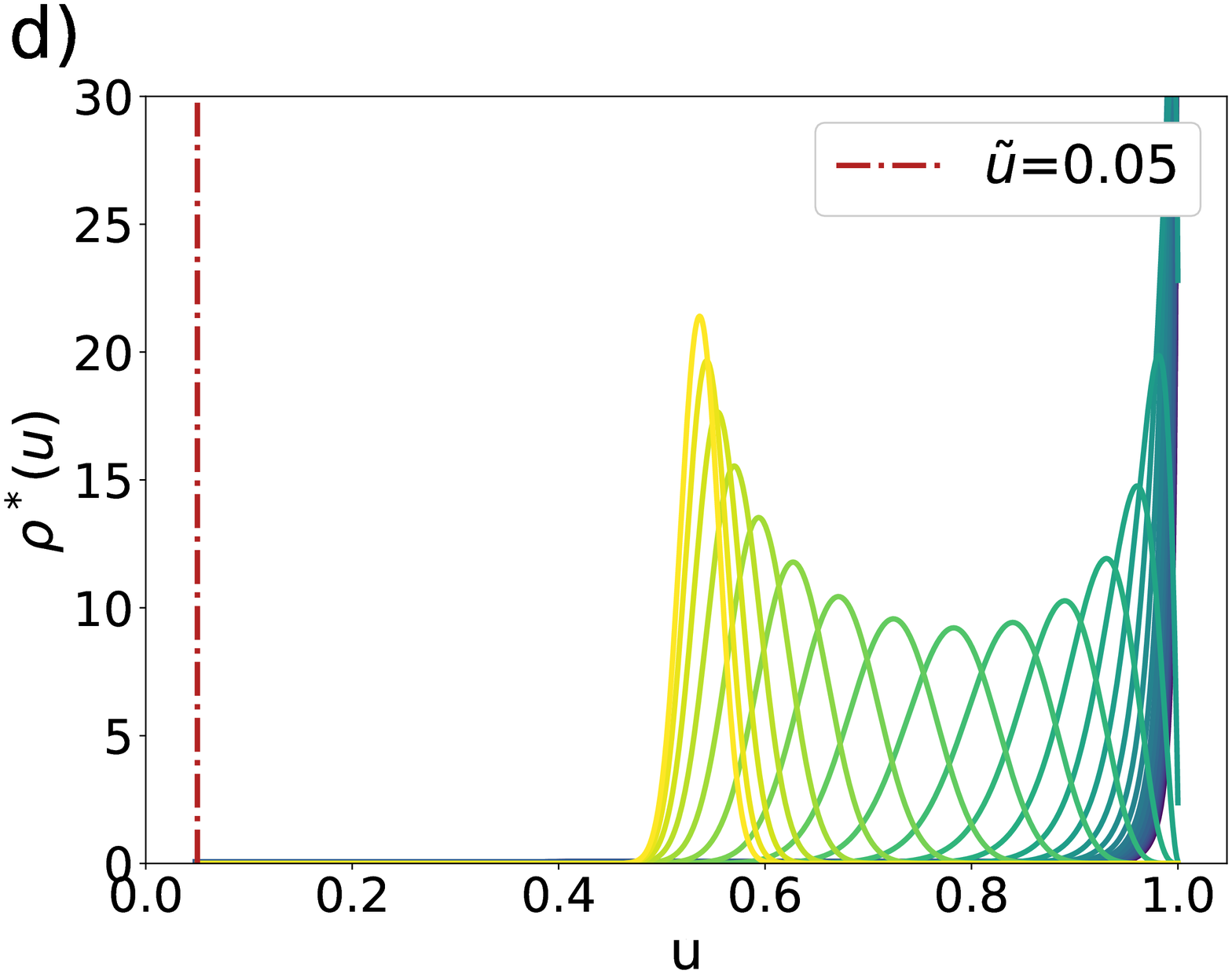}
    		\end{subfigure} 
		\end{subfigure}
		\begin{subfigure}[h!]{0.1\textwidth}
		\centering
			\includegraphics[width=0.8\linewidth]{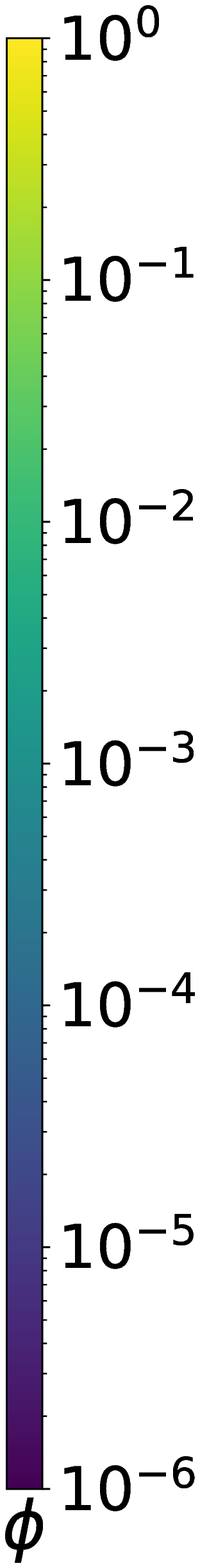}
		\end{subfigure}
		\caption{Stationary probability density function of having a fraction of unaware agents $u$ in the network, $\rho^*(u)$, for fixed fraction of zealots, $\tilde{u}$, varying the comprehensive parameter $\phi$ in the range $[10^{-6}, 1]$. System size is $N=1000$, thus the plots refer to cases with $0, 1, 10, 50$ zealots respectively. The bimodality, characteristic of the scenario with the absence of zealots, panel \textbf{a)}, disappears and symmetry is broken. We observe only one probability maximum that shifts towards the point of perfect coexistence $u=\frac{1}{2}$ without surpassing it as the relative strength between broadcasting and the behavioral dynamics, $\phi$, increases.}
		\label{img:Figure2}
	\end{figure}

\begin{figure}[!ht]
		\centering
		\begin{subfigure}[b]{0.45\textwidth}
			\centering
			\includegraphics[width=\textwidth]{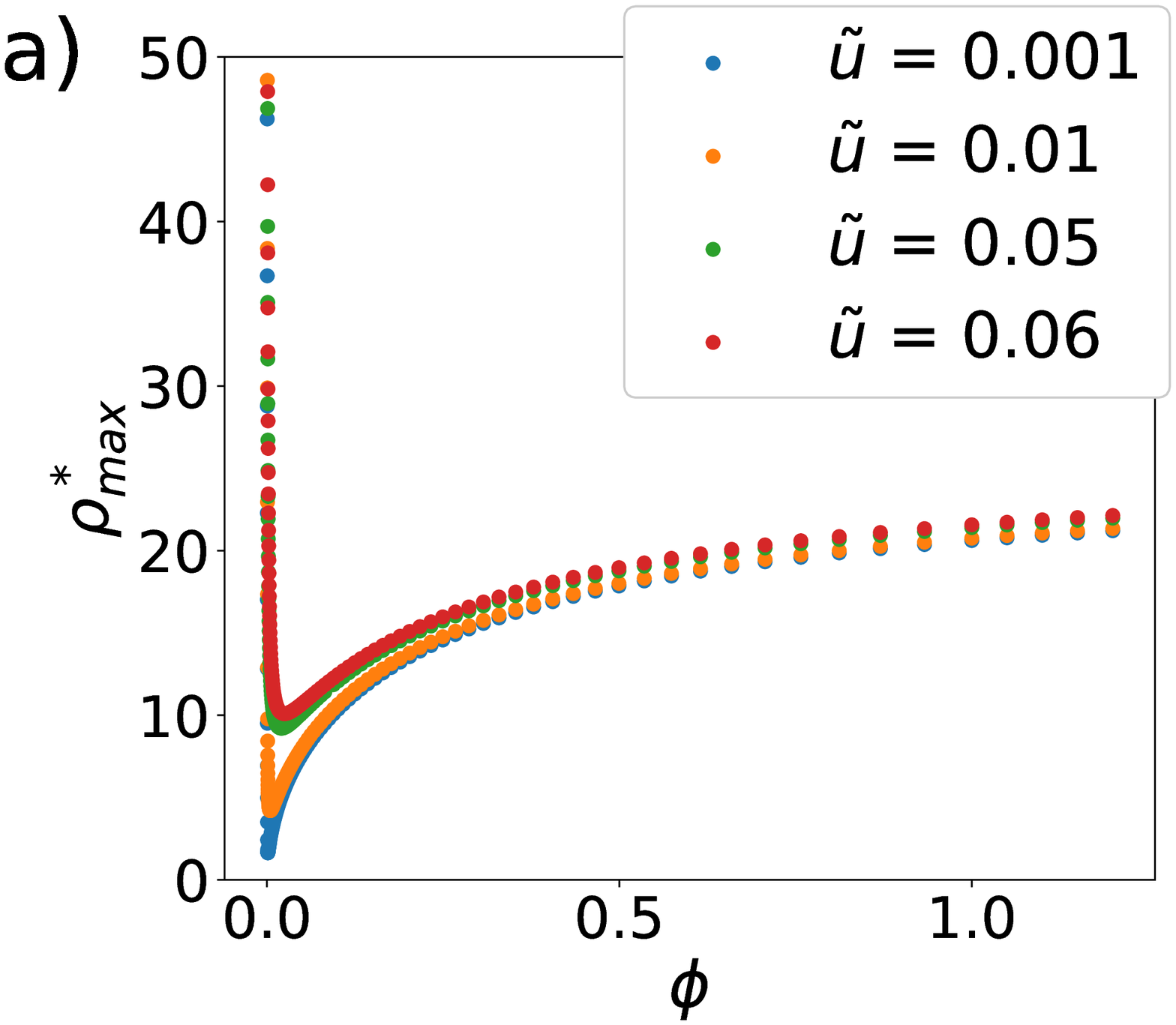}
		\end{subfigure}
		\begin{subfigure}[b]{0.45\textwidth}  
			\centering 
			\includegraphics[width=\textwidth]{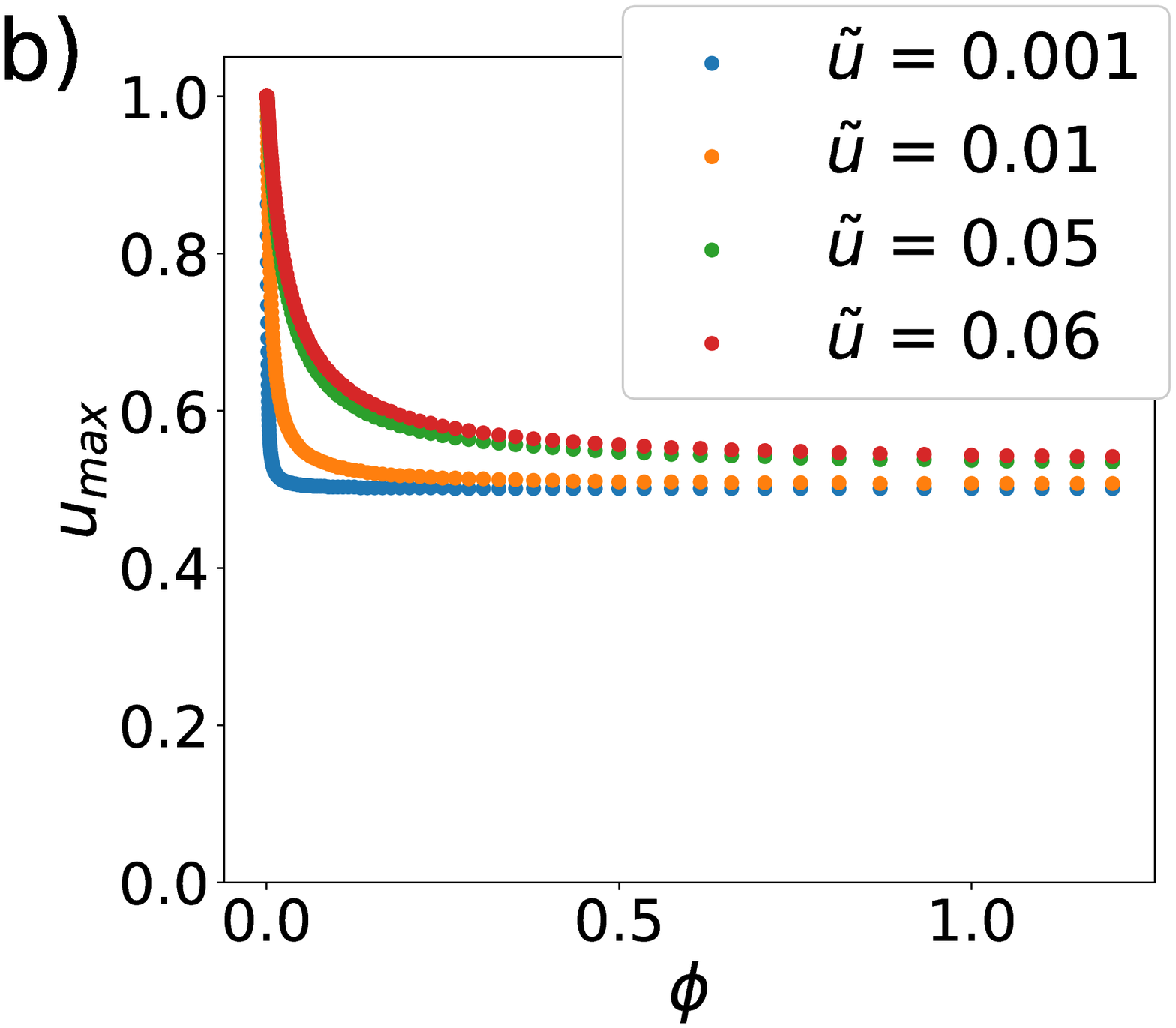}
		\end{subfigure}
		\caption{Trend of the maximum of the stationary probability density function $\rho^*(u)$ and the corresponding fraction of unaware agents $u$, $u_{\TextInMath{max}}$, for different values of the relative strength between broadcasting and behavioral dynamics, $\phi$, and fraction of zealots $\tilde{u}$.}
		\label{img:Figure3}
	\end{figure}

	We show in Figure~\ref{img:Figure2} the stationary solution for different values of $\phi$ and $\tilde{u}$. The first plot shows what happens when no zealot agents are present, a case we already studied in the previous section. When we add a finite amount of zealots ($\tilde{u}\neq0$) the function is asymmetrical and unimodal and for small $\phi$ values has a maximum in $u=1$, where all agents are unaware. As $\phi$ increases the maximum lowers and shifts to the lower values of $u$. This effect can be contained by increasing the fraction of zealots $\tilde{u}$. We notice however that regardless the values of the relative strength of the parameters $\phi$ and $\tilde{u}$ the probability maximum decreases no further than $u=\frac{1}{2}$: increasing $\phi$ will only increase the height of the peak, making the distribution increasingly narrow. Figure~\ref{img:Figure3} shows the trend of the height of the maximum, $\rho^*_{\TextInMath{max}}$, and the corresponding value of $u$, $u_{\TextInMath{max}}$, for different values as a function of $\phi$ at fixed $\tilde{u}$.
	
	It is interesting to notice the effect that the introduction of \textbf{one single zealot agent} has on the stationary behavior. We do expect its presence to break the symmetry between $A$ and $U$ in favour of unaware agents. Nevertheless, the effect is puzzling, as for equal values of $\phi$ the resulting distribution is completely shifted to $u=1$. Regarding the modality, we notice that for any value of the relative strength between broadcasting and the behavioral dynamics, $\phi$, the distribution is always unimodal. A second local maximum in $u=0$ is found only when $\tilde{u} < 1/N$, which is not a physical value.
	
	In summary, in this section we approached the analytical study of our model step by step, introducing one parameter at a time in order to capture their role. Although not complete, this mathematical treatment offered some important insights and showed some non-trivial behaviors, most notably the disappearance of the bimodality in the stationary probability density function to have a fraction $u$ of nodes in the network, $\rho^*(u)$, when zealots are introduced (Figure~\ref{img:Figure2}). Framing this in the context of detrimental behaviors' spreading and adoption, the $u$ agents represent individuals that do not engage in a misconduct (either practically or theoretically) whereas $a$ agents do. Zealots represent inflexible individuals, and if we assume that inflexibility and strong ethics are reached, for example, through education, our result shows how even an incredibly small fraction of well-educated individuals could drastically contain the spreading of the detrimental behavior.
	
	\section{Application to the \textit{AstraZeneca} vaccine debate on Twitter}\label{section:data}

	In the previous section we have shown that our model displays a rich phenomenology and we have identified the role played by some of its parameter. Our goal is now to evaluate whether it is possible to reproduce, within a reasonable error, some trends observed in a timely and relevant empirical scenario. In particular, we look at the discussion in the online social platform \textit{Twitter} about the ban of the \textit{AstraZeneca} vaccine (know in the European Union as the \textit{Vaxzevria} vaccine from the 25ft of March 2021) by the \textit{AIFA}. \textit{AIFA} is the Italian public institution responsible for the approval and the regulation of medicines in the country. Further details about the dataset and the case study can be found in~\ref{section:appendix:dataset}.
	
	The fitting technique we employ is \textit{Particle Swarm Optimization}~\cite{PSO} (see~\ref{section:PSO}), a biological-inspired optimization recipe that proves especially performant to deal with non-convex problems (or problems for which this property is hard to evaluate). Although being a fairly simple algorithm, \textit{Particle Swarm Optimization} has a set of hyperparameters to be tuned and this will be achieved via a random search.
	
    When considering a real-world scenario we need to adapt the model to the particular characteristics of the dataset. In particular, it is necessary to include information not only on the aware and unaware agents at a certain time, but also about how their status translates into active participation to the discussion. As a result, our analysis compares the cumulative number of tweets a certain topic receives (i.e. the cumulative number of tweets including a certain keyword matching the topic) and the same value computed from an extended version of our model.
	
	\subsection{Data-driven model implementation}\label{section:dataDrivenMod}
	Let us re-frame the interpretation of our model in terms of the \textit{AstraZeneca} vaccine debate. In the discussion, \textit{aware} individuals know about the occurrence of some event and are interested in it; conversely, \textit{unaware} agents do not know about it or are not interested in it. Moreover, aware individuals \textbf{may} join the discussion, producing a certain number of tweets. In this context, the role of zealots is that of individuals that under any circumstances will ever join the discussion. The behavioral dynamics corresponds to a phase in which interested parties try to include \textit{unaware/uninterested} individuals in the discussion and vice versa. In this phase \textit{unaware} individuals develop interest in the discussion only if neighbouring \textit{aware} agents ``convince'' them. The broadcasting is triggered by the occurrence of an event. If this event is particularly relevant to the underlying discussion in the social network, we assume that the spreading of its knowledge will overcome the topological constrains of the network, as events perceived as important and valuable for the public discussion will be \textbf{broadcasted} by media outlets.
	
	To be able to compare the outcome of our model with the data, we introduce a new variable, corresponding to the number of produced posts per user. This allows use to ultimately compare the cumulative number of posts produced by \textit{Twitter} users mentioning the \textit{Atrazeneca vaccine} extracted from the dataset with the same quantity computed from our model. 
	
	Incorporating the information of how many posts an user produces is done in two phases. Firstly, we compute the probability to be in the \textit{aware} state for all agents at every point in time by numerically integrating Equation~\eqref{eq:evolutionProbBehavBroadc} for every node $i$ of the network. For a network of $N$ nodes and considering a time interval of $T$ slots, the result is a $N \times T$ matrix whose rows correspond to the evolution over time of the probability of a node to be in the \textit{aware} state. Let us call this matrix $\mathbf{P}$ of entries $P_{i,t}$ for $i=1,..,N$ and $t=0,..,T-1$. The information on how the status of a node translates into activity is obtained sampling the distribution $p_{k,\overline{t}}$ (see~\ref{section:appendix:dataset:pkt}), where $p_{k,\overline{t}}$ stands for the distribution of the number of tweets produced in a timeslot of duration $\overline{t}$ by a user whose degree is $k$ (meaning he/she has $k$ followers). In particular, for every node $i$ in the network the information of its degree is retrieved, and, sampling from $p_{k,\overline{t}}$, we build an array of length $T$ containing the time series of the activity of user $i$. Again, we obtain a $N \times T$ matrix $\mathbf{\Lambda}$ with entries $\Lambda_{i,t}$ for $i=1,..,N$ and $t=0,..,T-1$. It is then possible to compute the number of tweets $N^{\TextInMath{tweets}}_{i,t}$ a user produces given its status at time $t$ by multiplying $\mathbf{P}$ and $\mathbf{\Lambda}$ element wise,
	\begin{equation}
		(\boldsymbol{N}^{\TextInMath{tweets}})_{i,t} = P_{i,t} \Lambda_{i,t}
	\label{eq:nTweetsNodeITimeT}
	\end{equation}
    The overall number of tweets produced at time $t$ is thus obtained summing over the columns of $\boldsymbol{N}^{\TextInMath{tweets}}$, i.e. summing over the activity of all users at fixed time,
	\begin{equation}
		N^{\TextInMath{tweets}}_t = \sum_{i=1}^N(\boldsymbol{N}^{\TextInMath{tweets}})_{i,t},
	\label{eq:nTweetsTimeT}
	\end{equation}
    and their cumulative value up to time $t$ is
	\begin{equation}
		C_t = \sum_{k=0}^t N^{\TextInMath{tweets}}_k.
	\label{eq:cumulativeNTweetsUpToT}
	\end{equation}
	
	We ask next whether there is a configuration of the parameters of our model that is able to produce a series $C_t$ compatible with the one extrapolated from the data. Let us remark now that the parameters to be searched only influence one part of our model, the values of $\mathbf{P}$, through Equation~\eqref{eq:evolutionProbBehavBroadc}. Referring to the terminology of Particle Swarm Optimization (see~\ref{section:PSO}), each candidate set of parameters $\mathbf{g}$ will correspond to a different evaluated cumulative series $C_{\mathbf{g}}(t)$ and the cost function to be minimized during the optimization process is the normalized \textit{mean square error} (MSE) between the data and the generated points. Let $C_{\TextInMath{data}}(t)$ be the value of the cumulative number of tweets produced up to date time $t$ extrapolated from the data, then the cost associated to a set of parameters $\mathbf{g}$ is
	\begin{equation}
		f(C_{\mathbf{g}}, C_{\TextInMath{data}}) = \TextInMath{Var}(C_{\TextInMath{data}})^{-1} T^{-1}\sum_{t=0}^{T-1} \left(C_{\TextInMath{data}}(t) - C_{\mathbf{g}}(t) \right)^2
	\label{eq:PSOCost}
	\end{equation}
    where $\TextInMath{Var}(C_{\TextInMath{data}})$ is the variance in the real data. Dividing by this quantity allows us to quantify the error and its magnitude in comparison to the data.	
	
	\subsection{Results}\label{section:results}	
	Due to the long computational times of the \textit{PSO} searching procedure we did not manage to employ the prevalence model in its full form as presented in Equation~\eqref{eq:evolutionProbBehavBroadc}. The bottleneck of the computations is twofold: we find that, given its dimensions, using the full adjacency matrix $A_{ij}$ considerably slows down computations. Moreover, Equation~\eqref{eq:evolutionProbBehavBroadc} actually refers to a set of coupled differential equations, since we do not use any complete-graph approximation, and this slows down the convergence of the numerical method used to integrate the system of differential equations. Since Equation~\eqref{eq:evolutionProbBehavBroadc} needs to be solved multiple times for a single \textit{PSO} iteration the employment of its full form resulted in unsustainable computational times.
	
	We used the following approach: the prevalence model is considered in the \textit{all-to-all} approximation and with symmetric rates. However, in order to include a more realistic ingredient, to each node we assign a degree $k_i$ (otherwise all the degrees would have value $N-1$) which is not used as an information to compute the dynamical evolution of the prevalence, but is used only to sample the activity (i.e. number of posts produced) by each \textit{aware} node. In this way, the dynamical process according to which agents change their status from \textit{aware} to \textit{unaware} (and back) is still happening in a fully-connected scenario, but the way their status translates into active participation into the discussion (by the creation of posts) takes into account a more realistic scenario, in which the activity of a user is still dependent on the umber of its acquaintances. In practice, we use the degree $k_i$ only to sample the activity of each user $i$ from the distribution of activity $p_{k,\overline{t}}$, where $p_{k,\overline{t}}$ is the distribution of the number of posts produced by a user with degree $k$ in a timeslot of duration $\overline{t}$. Following evidences in the literature~\cite{dedomenico2013, artime2020},  we employ a \textit{scale-free} distribution with exponent $\nu = 2.2$ for building our synthetic degree distribution. For the number of users $N$, we consider the total number of users involved in the discussion surrounding vaccines in the same temporal window considered for our dataset, which is $N=1874$. Further information on the prevalence model and on the covered approximations can be found in~\ref{section:appendix:results}. 
	
\begin{figure}[!ht]
		\centering
		\includegraphics[width=\textwidth]{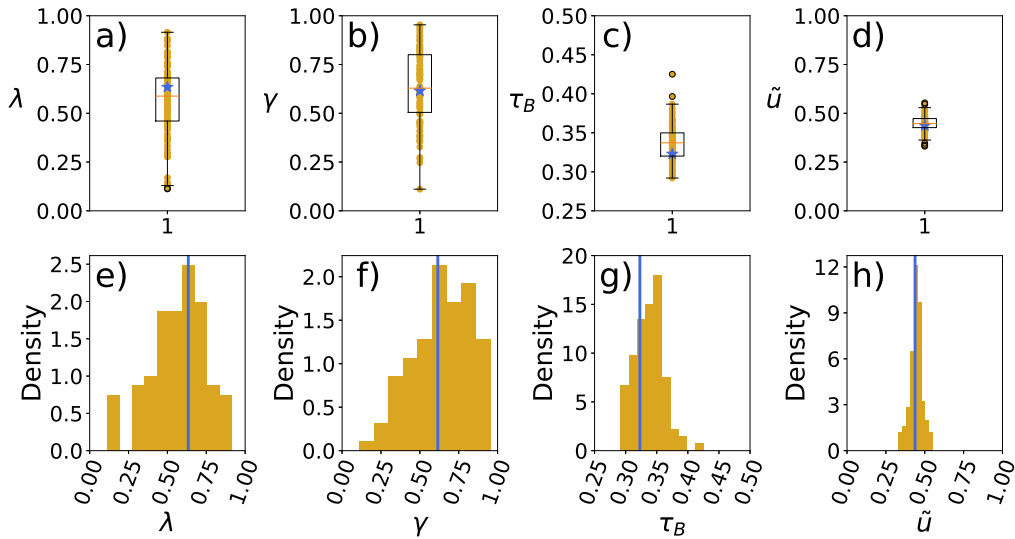}
		\caption{Top row: box plots showing the distribution of the occurrences of the four parameters throughout the repetitions of the \textit{Particle Swarm Optimization} procedure. The four parameters correspond to the behavioral transition rate $\lambda$, the broadcasting transition rate $\gamma$, the broadcasting onset $\tau_B$ and the fraction of unaware nodes $\tilde{u}$. Note that we are considering symmetric rates within the behavioral and broadcasting dynamics. Bottom row: same date shown as histograms. The blue marks represent the values found in the set with lowest cost.}
		\label{img:Figure4}
	\end{figure}

	We search for a good set of hyperparameters following the procedure explained in~\ref{section:PSO}. The results can be found in~\ref{section:appendix:results:sym_ata}. Using this set of hyperparameters, the parameter optimization procedure was repeated $100$ times, and we show in Figure~\ref{img:Figure4} the distributions of the occurrence of each parameter over these realizations. The blue mark correspond to the set of parameters of lowest cost. In addition, we report the lowest cost set of parameters and the medians in Table~\ref{table:chap3:lowest_median}. The parameter $\tau_B$ corresponds to a version of the parameter $\tau$, the onset time of broadcasting, in which $\tau_B=0$ corresponds to $\tau$=\textit{February 9 2021} and $\tau_B=1$ to $\tau$=\textit{April 12 2021}.

	The limits of the box plots in Figure~\ref{img:Figure4} correspond to the lower ($l_i$) and upper ($u_i$) bounds for the search space as explained in~\ref{section:PSO}. Note that we decided to place some boundaries on the value of $\tau$ as it was possible to make some assumptions on it based on graphical considerations. We can nonetheless notice that the bounds turn out to be wider than needed, as the values converge to an even more narrow region. The lowest cost value of each parameter falls between the $Q_1$ and $Q_3$ quantiles and is in general close to the point corresponding to the maximum of its distribution, suggesting that the swarm tends indeed to converge towards the point corresponding to the lowest cost over the repetitions.	
	\begin{table}[!ht]
		\centering
                \begin{tabular}{|l|c|c|c|c|c|}
			\hline
			& $\lambda$ & $\gamma$ & $\tau$             & $\tilde{u}$ & MSE   \\ \hline
			lowest cost & 0.63      & 0.62    & March 1st 2021 & 0.44        & 0.019 \\ \hline
			median      & 0.59      & 0.63    & March 2nd 2021 & 0.45        &       \\ \hline
		\end{tabular}
        \caption{Lowest cost and median parameters. Mind that there is no MSE associated to the median set of parameters as the median is computed for every single parameter and not as a set.}
		\label{table:chap3:lowest_median}
	\end{table}
	
\begin{figure}[!ht]
		\centering
		\includegraphics[width=0.9\textwidth]{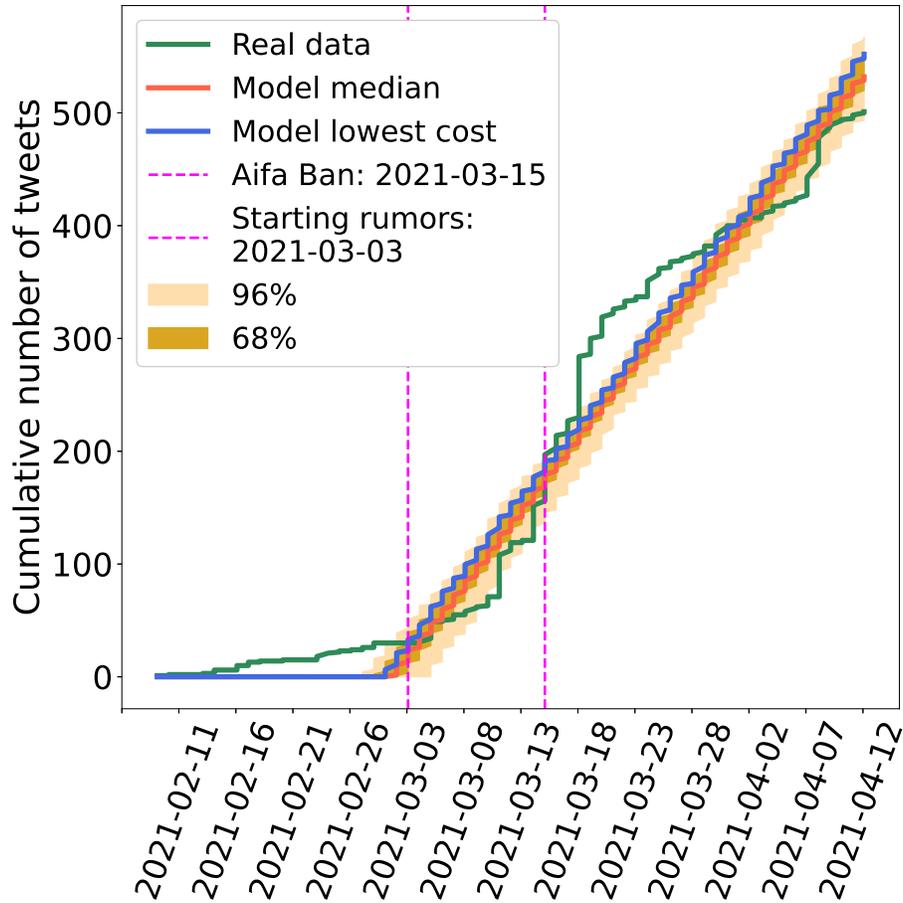}
		\caption{Comparison between the fitted model and the dataset (in green). The blue line corresponds to the model associated to the set of parameters with lowest cost. The red line corresponds to the model obtained using the median parameters. The different shades of yellow aim at reproducing the same information as in the box plot: the cumulative curve was computed for all the $100$ parameters sets, thus the red line correspond to the median values of the curves at each time and the shaded yellow areas to the $68\%$ quantile (dark yellow) and $98\%$ (light yellow) quantile respectively. Mind that the red and yellow lines do not correspond to the model obtained using the median set of parameters and might not correspond to any particular set of parameters in general.}
		\label{img:Figure5}
	\end{figure}

	We next compare the data and the results of the optimization procedure, see Figure~\ref{img:Figure5}. The blue line corresponds to the model with lowest cost (i.e. using the first row of parameters in Table~\ref{table:chap3:lowest_median}). The different shades of yellow reproduce the same information as in the box plot: the cumulative curve was computed for all the $100$ parameters sets, thus the red line corresponds to the median values of the curves at each time and the shaded yellow areas to the $68\%$ quantile (dark yellow) and $98\%$ (light yellow) quantile respectively. Mind that the red and yellow lines do not correspond to the model obtained using the median set of parameters and might not correspond to any particular set of parameters in general.	

	It is possible to notice that the moment in time at which broadcasting activates is able to capture the moment in time at which the interest towards the \textit{AstraZeneca} debate starts to arise. Moreover, our data-driven approach is in general able to reproduce the trend in the data from a quantitative point of view. 
	However, our model is not able to correctly reproduce the initial behavioral dynamics. We believe however that this is mostly due to the symmetry in the rates, as we explain in~\ref{section:appendix:results:asym_ata}.
	
	\section{Conclusions}
	
	In this work we introduced a framework for the study of the influence of external triggers in a spreading dynamics in interconnected social systems. In particular, we aimed to offer a mechanistic framework for the study of the role of mainstream media in shaping public discussion, in particular when it comes to the broadcasting of news that could lead to possibly detrimental behaviors.
	
	To this end, we worked an epidemiological-inspired model that accounts for two types of dynamics that we call \textit{behavioral} and \textit{broadcasting}. The behavioral dynamics aims at describing the interaction between two types of individuals, the \textit{aware} and the \textit{unaware}: in the framework of misbehavior spreading, \textit{aware} agents know about a certain misconduct and promote it to the \textit{unaware}, that, in turn, can persuade the \textit{aware} agents to abandon it. This process is ``bounded'' by the topology of a social network, be it online or offline. The second dynamics, broadcasting, disregards the topological constraints of the network and accounts for the role played by mainstream media: broadcasting is able to shift the status of a node without the necessity of the interaction with an individual of opposite view. This dynamics is not always available, and in the context of news spreading, for example, would correspond to the occurrence of an event of particular interest for the public discussion which gets broadcasted by the media. It is also possible to re-frame the interpretation of the model in a more neutral perspective, with \textit{aware} nodes being active part of a discussion and \textit{unaware} refusing to join it. The last ingredient of the model is the introduction of zealot nodes, that in the context of misconduct spreading represent individuals whose strong moral prevents them from engaging in the misbehavior.
	
	We started by tackling the model analytically in its \textit{all-to-all} approximation: in the further approximation of symmetric rates we observe that the stationary distribution of the fraction of \textit{unaware} nodes displays a transition from unimodal to bimodal depending on a parameter that accounts for the relative strength of broadcasting and behavioral dynamics. We also found that the presence of a single zealot node in the network dramatically changes the distribution, breaking the symmetry between \textit{aware} and \textit{unaware} nodes and determining the disappearance of the bimodality. We then tested the model against a real-world scenario: the discussion surrounding the \textit{AstraZeneca} vaccine ban that took place in many European countries at the beginning of March 2021. We took a data-driven approach to our model so that it is also able to account for the amount of content that an \textit{aware} individual produces. We compared the cumulative number of tweets produced by users in Italy at the turn of the \textit{AstraZeneca} \textit{AIFA} ban with the same quantity computed from the data-driven extension of our model. The comparison was made through the fitting of the parameters of the model and we found that this method succeeds in capturing the shift in popularity observed in the data by means of the activation of the broadcasting dynamics: we found that the model predicts an activation date of the broadcasting dynamics that is compatible with the period in which the discussion surrounding the safety of the \textit{AstraZeneca} vaccine heated up.\\
	
    Further developments of this work could take different directions. Our model is flexible from a topological perspective and we chose to carry out the majority of the analyses presented in this work in a complete-graph limit. By doing so, we found a trade-off between analytical tractability, computational scalability, and, of course, gained insight into the process we wish to study. However, as mentioned in the previous sections, the complete-graph limit is not able to account for many topological features that are typical of real social networks. In \ref{section:appendix:preliminaryAnalysis} and \ref{section:appendix:results:hmf} we take a first step towards including more structured topologies (by considering scale-free and Erdős–Rényi networks) but extending our analyses to account for the presence of clustering, assortativity, and heterogeneity in the communities (just to name a few) will be of utmost importance for future developments of this work.
    
 In order to better account for the nuances of opinion dynamics phenomena, the model could be extended with a bounded-confidence component: the epidemiological-inspired approach we adopted in this work made the model easier to investigate (especially analytically), but we are aware that it lacks the typical complexity of realistic adoption processes. Moreover, additional comparison with real scenarios is needed, even including a sentiment component in the data. Including this element would allow us to interpret the model in the framework of misconduct adoption (pro or against a certain behavior) and to give a meaningful interpretation of the number of aware and unaware individuals, even without the additional information of their activity. Relevant topics could include, for example, the spreading of dangerous trends/challenges on online social media~\cite{sumner2019temporal, mukhra2019blue}.
	
	\section*{Acknowledgements}The authors wish to thank Prof. Samir Suweis, through whom they had the opportunity to establish this fruitful collaboration. The authors also acknowledge Anna Bertani and the whole team of the \textit{COVID-19} \textit{Infodemic Observatory} for the fundamental contribution they offered to this work by providing the data necessary for our analysis and for their availability.
 \section*{Fundings}This research did not receive any specific grant from funding agencies in the public, commercial, or not-for-profit sectors.
	
	\appendix
        \setcounter{figure}{0}
        \setcounter{table}{0}
	\section{Numerical analysis of the model in networked topologies}\label{section:appendix:preliminaryAnalysis}	
	
	In this section we provide some further insights into the behavior of our proposed model. In particular, we show how the numerical solution of Equation~\eqref{eq:prevalenceShort} is well reproduced by Monte Carlo simulations and use this result to investigate the long-term behavior of the model comparing the dynamics at different regimes, with and without broadcasting.\\
	
\begin{figure}[!h]
		\centering
		\hspace{-40pt}\begin{subfigure}{0.9\textwidth}
			\centering
			\begin{subfigure}{0.47\textwidth}
			    \centering
			    \includegraphics[width=0.97\textwidth]{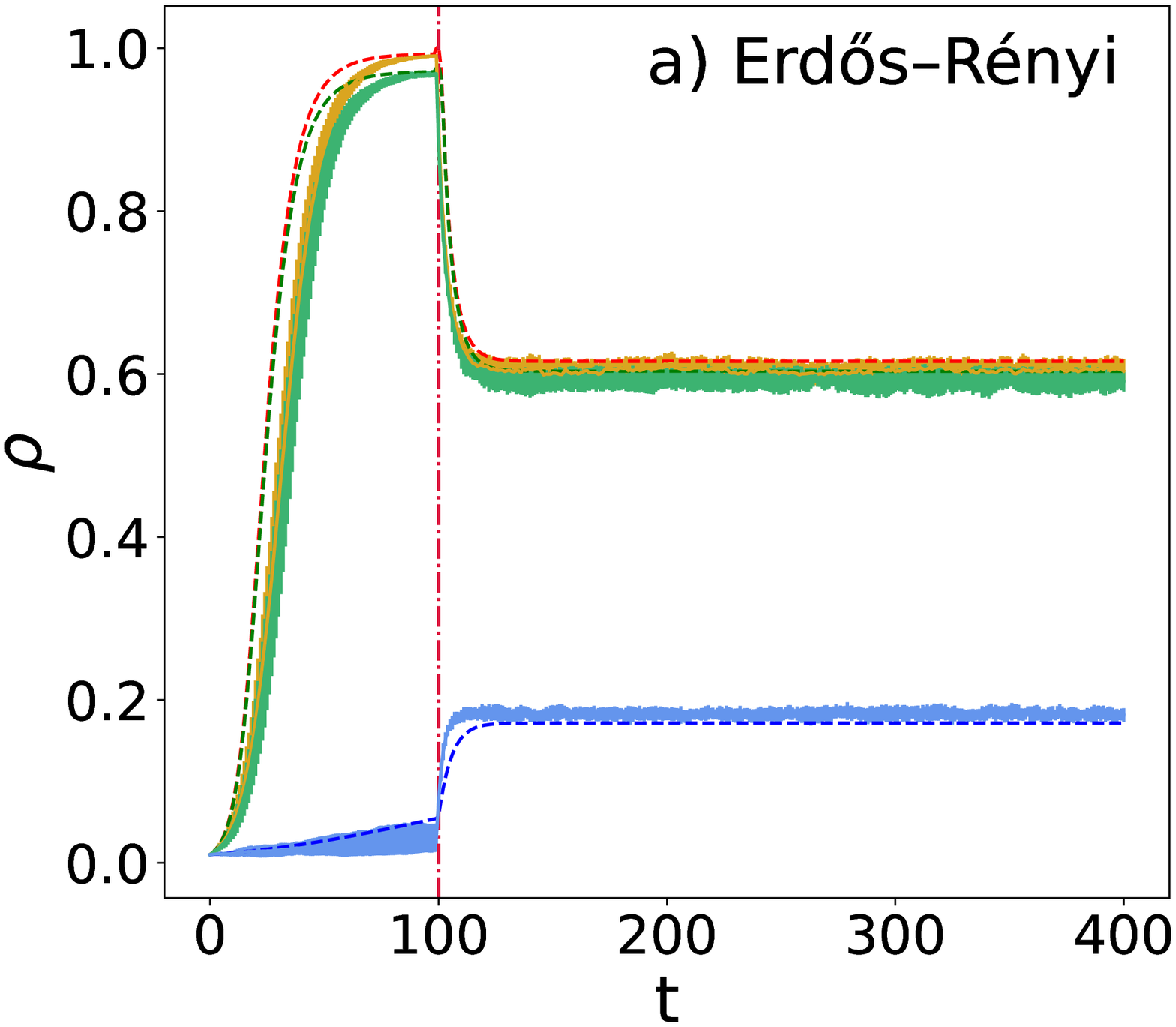}
			\end{subfigure}
			\begin{subfigure}{0.47\textwidth}
			    \centering
			    \includegraphics[width=0.97\textwidth]{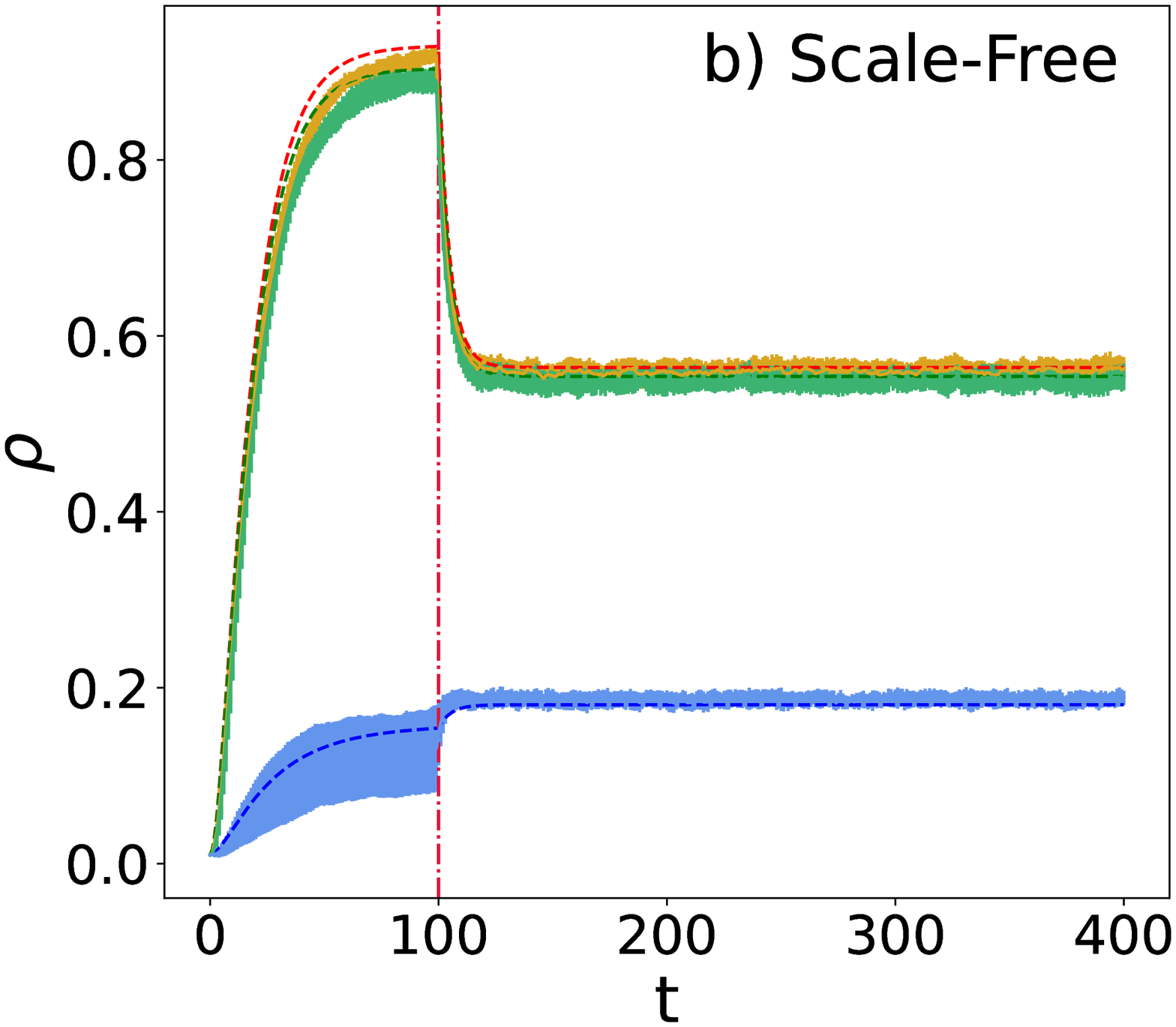}
			\end{subfigure}
		\end{subfigure}
		\begin{subfigure}{0.18\textwidth}
			\centering
		    \includegraphics[width=\textwidth]{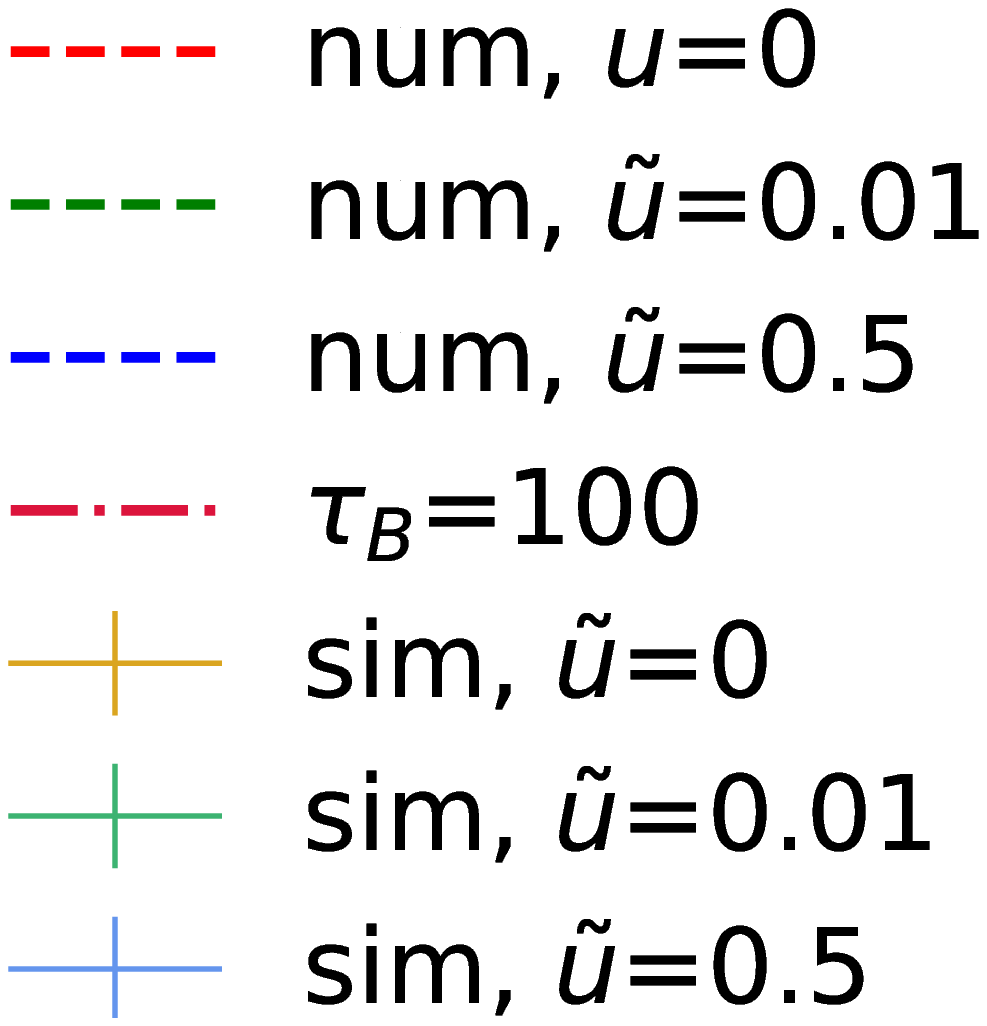}
		\end{subfigure}
		\caption{Simulation (sim., solid lines) and numerical (num., dashed lines) solutions of the prevalence, Equation~\eqref{eq:prevalenceShort} (through Equation~\eqref{eq:evolutionProbBehavBroadc}) for different values of the fraction of zealot nodes $\tilde{u}$ and the following dynamical parameters: $\lambda=0.07$, $\mu=0.03$, $B=0.1$, $\gamma=0.9$, $\beta=0.55$. Topological parameters of the networks are: number of nodes $N = 1800$, \textit{Erdős–Rényi} parameter $p=2.5\times10^{-3}$, and scale-free exponent $\nu=2.2$.}
		\label{img:FigureA1}
	\end{figure}

\begin{figure}[!h]
		\centering
		\hspace{-40pt}\begin{subfigure}{0.9\textwidth}
			\centering
			\begin{subfigure}{0.47\textwidth}
			    \centering
			    \includegraphics[width=0.97\textwidth]{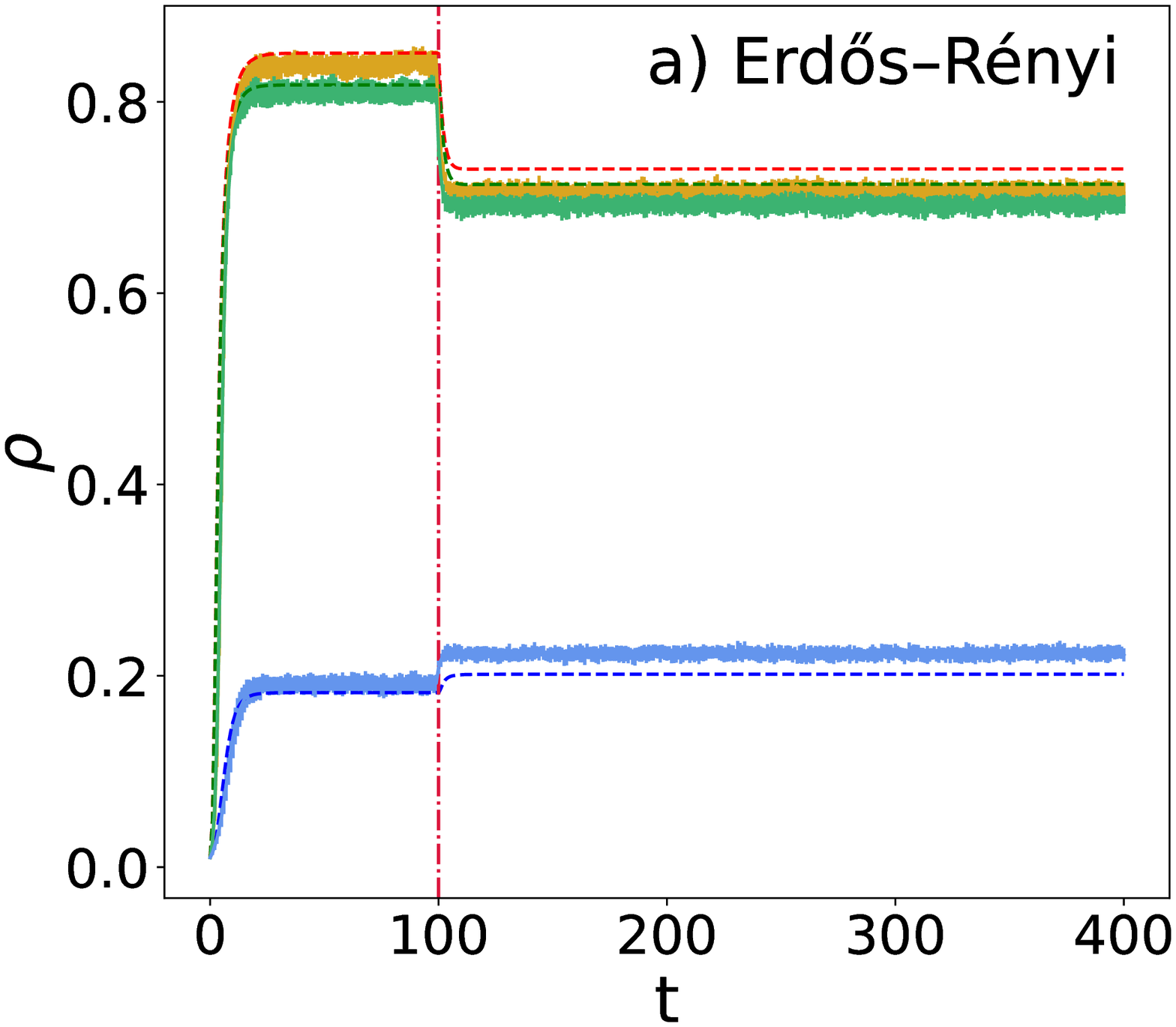}
			\end{subfigure}
			\begin{subfigure}{0.47\textwidth}
			    \centering
			    \includegraphics[width=0.97\textwidth]{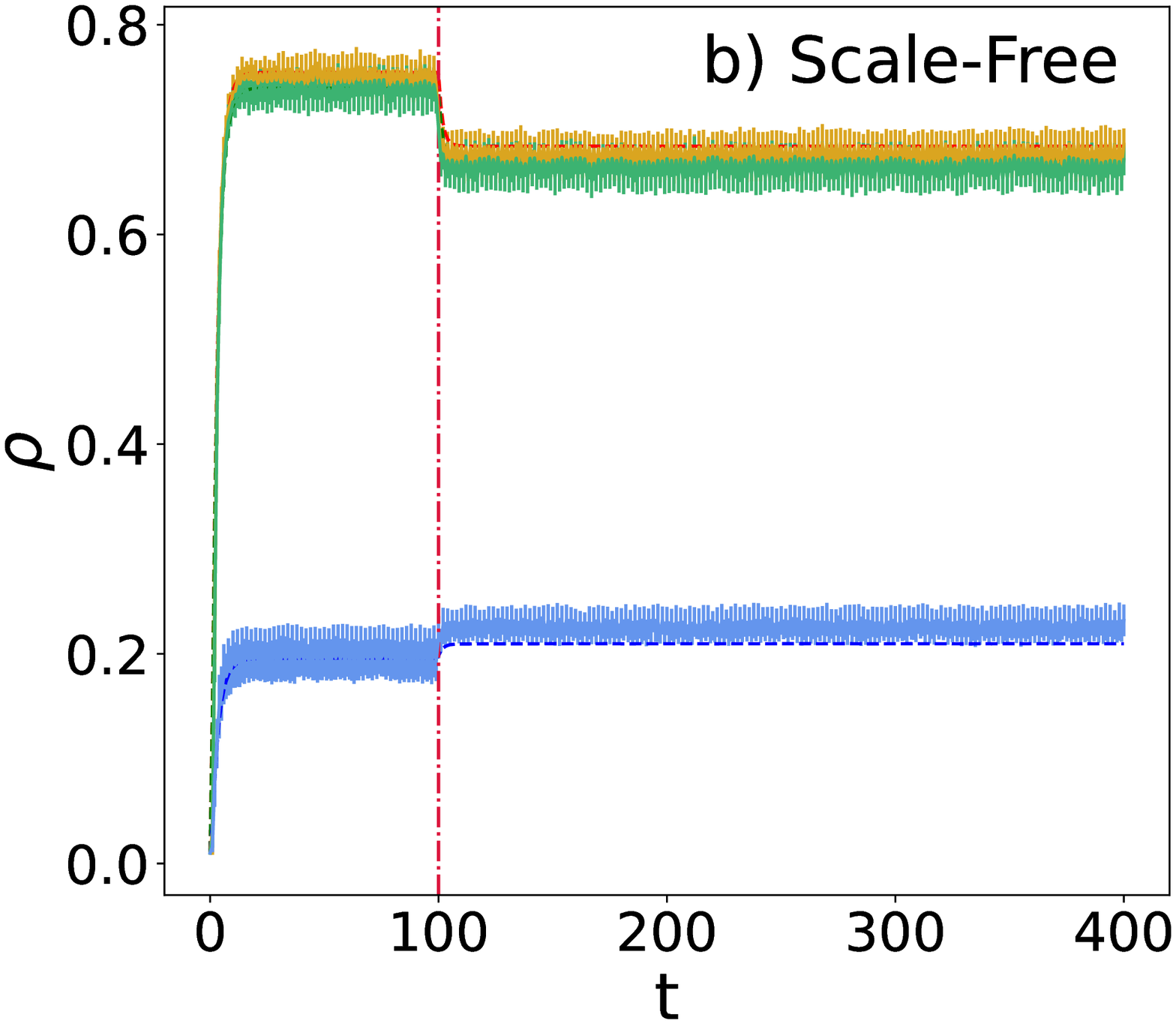}
			\end{subfigure}
		\end{subfigure}
		\begin{subfigure}{0.18\textwidth}
			\centering
		    \includegraphics[width=\textwidth]{FigureA1_legend.eps}
		\end{subfigure}
		\caption{Simulation (sim., solid lines) and numerical (num., dashed lines) solutions of the prevalence, Equation~\eqref{eq:prevalenceShort} (through Equation~\eqref{eq:evolutionProbBehavBroadc}) for different values of the fraction of zealot nodes $\tilde{u}$ and the following dynamical parameters: $\lambda=0.5$, $\mu=0.22$, $B=0.1$, $\gamma=0.8$, $\beta=0.75$.  Topological parameters of the networks are: number of nodes $N = 1800$, \textit{Erdős–Rényi} parameter $p=2.5\times10^{-3}$, and scale-free exponent $\nu=2.2$.}
		\label{img:FigureA2}
	\end{figure}

	Figures~\ref{img:FigureA1} and~\ref{img:FigureA2} show the results of both simulations and the numerical solution of Equation~\eqref{eq:prevalenceShort} (through Equation~\eqref{eq:evolutionProbBehavBroadc}) for two different topology families and different combinations of parameters.
	We checked the adherence between the results obtained via simulation and numerical solution, finding the same behavior.
	
	Simulation were run using a direct Monte Carlo method with random initialization of a fraction $\rho(0)=\rho_0$ of nodes in the \textit{aware} state and assigning a fraction $\tilde{u}$ of zealots, chosen uniformly at random. Numerical integration was performed with uniform initial condition, i.e., $\rho_i(0) = \rho_0$ $\forall$ $i$. To model the underling interaction network we considered an \textit{Erdős–Rényi} network and a \textit{scale-free} network, two paradigmatic examples of homogeneous and heterogeneous connectivity respectively. A \textit{scale-free} (SF) network is a network whose degree distribution $p(k)$ follows a power-law:
	\begin{equation}
	    p(k) \propto k^{-\nu}.
 	\label{eq:probDistSF}
	\end{equation}
	During simulations we built connected network using a configurational model, generating a power-law sequence for the degree $k$. \textit{Erdős–Rényi} (ER) networks~\cite{ER_model} are also referred to as binomial, as their degree distributions follow a binomial distribution:
    \begin{equation}
        p(k) = {n-1 \choose k} p^k (1 - p)^{n-1-k}.
    \label{eq:probDistER}
    \end{equation}
    We built \textit{Erdős–Rényi} networks by choosing the number of nodes in the graph $n$ and assigning connections among them with probability $p$.
	
	Once we established the adherence between the numerical solution of Equation~\eqref{eq:prevalenceShort} and the prevalence computed through simulations we used the numerical solution, which turns out to be faster, to explore possible behaviors of the model. This is not meant to be an exhaustive analysis of the phenomena; we were mostly interested in highlighting possible ranges in which the model behaves differently according to the topology of the network. In particular, we are interested in understanding whether there are differences in how broadcasting affects the dynamics, and a more systematical analysis would be needed in order to fully unravel the role of topology.
	
	We selected a wide range of behavioral parameters $\lambda$ and $\mu$ and numerically computed the prevalence, focusing on the long-time behavior of the model. To model the network of social interactions we considered the two topologies previously introduced plus a third one, a \textit{Barabási–Albert} (BA) network. \textit{Barabási–Albert} networks have power-law degree distributions but are obtained following a different building process, called preferential attachment~\cite{BA_model}. In this part of the analysis we did not include the $\tilde{U}$ zealot nodes to be able to better focus on the role of broadcasting alone.
	
	For each set of parameters we computed $\rho^{\infty}_{\TextInMath{behav}}$, the stationary value of the prevalence when only the behavioral dynamics is considered.
	We focused on sets of parameters that do not determine an ``extreme'' behavior of $\rho^{\infty}_{\TextInMath{behav}}$, meaning that it is not always either $1$ or $0$. We then recomputed $\rho^{\infty}$ for different sets of the broadcasting parameters.
	
\begin{figure}[!ht]
		\centering
		\begin{subfigure}[b]{0.3\textwidth}
			\centering
			\includegraphics[width=1\textwidth]{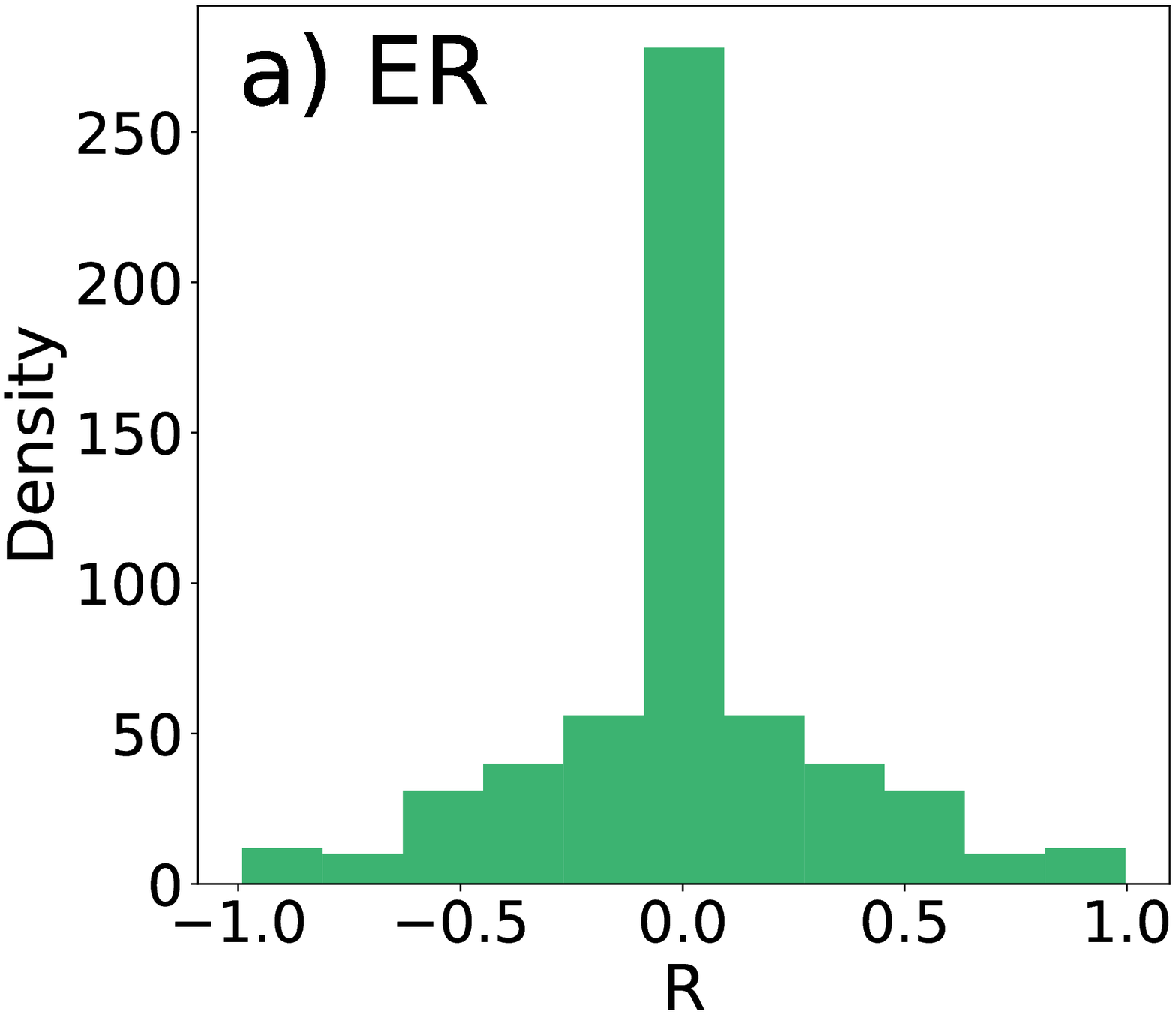}
		\end{subfigure}
		\begin{subfigure}[b]{0.3\textwidth}  
			\centering 
			\includegraphics[width=1\textwidth]{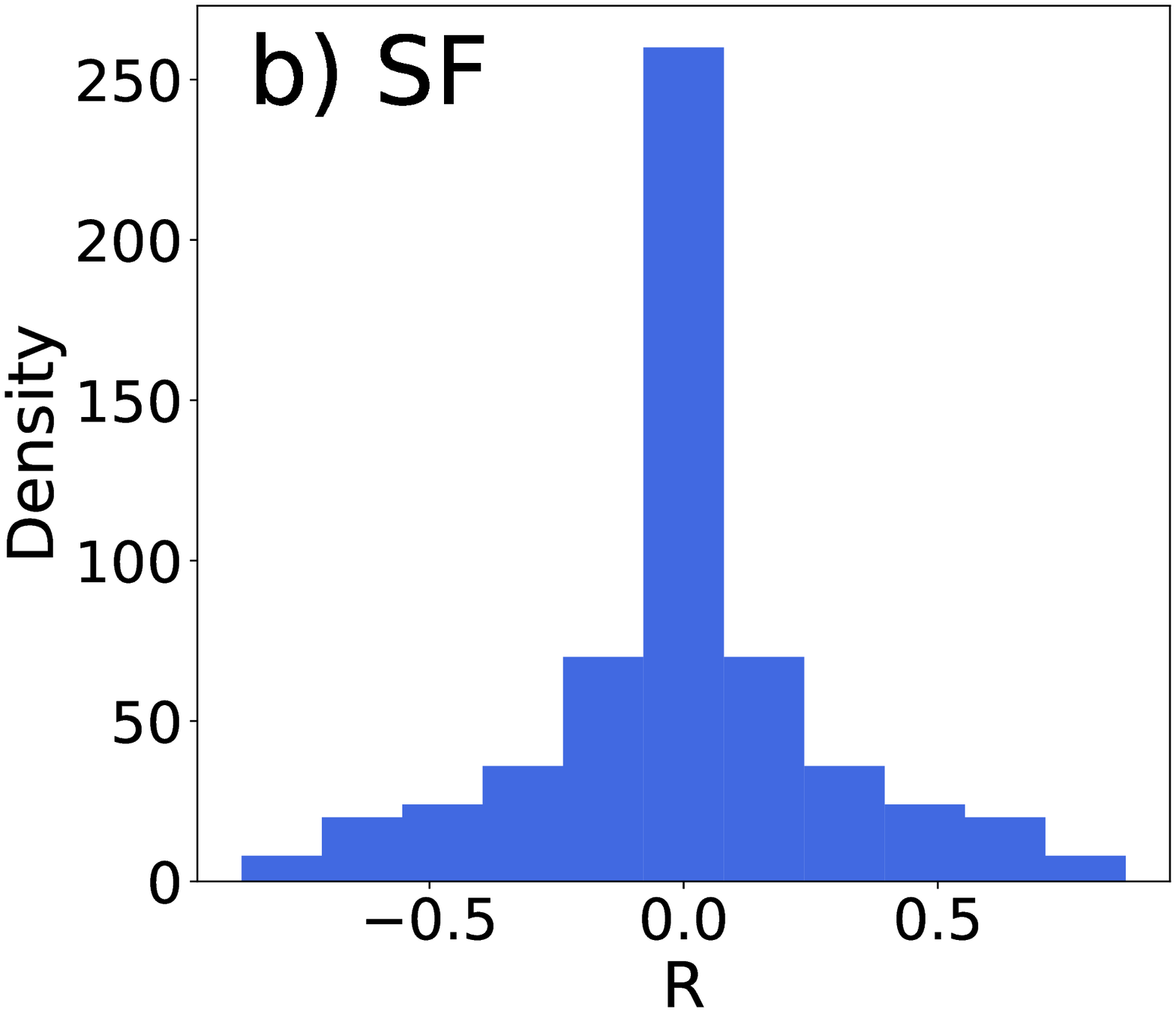}
		\end{subfigure}
		\begin{subfigure}[b]{0.3\textwidth}  
			\centering 
			\includegraphics[width=1\textwidth]{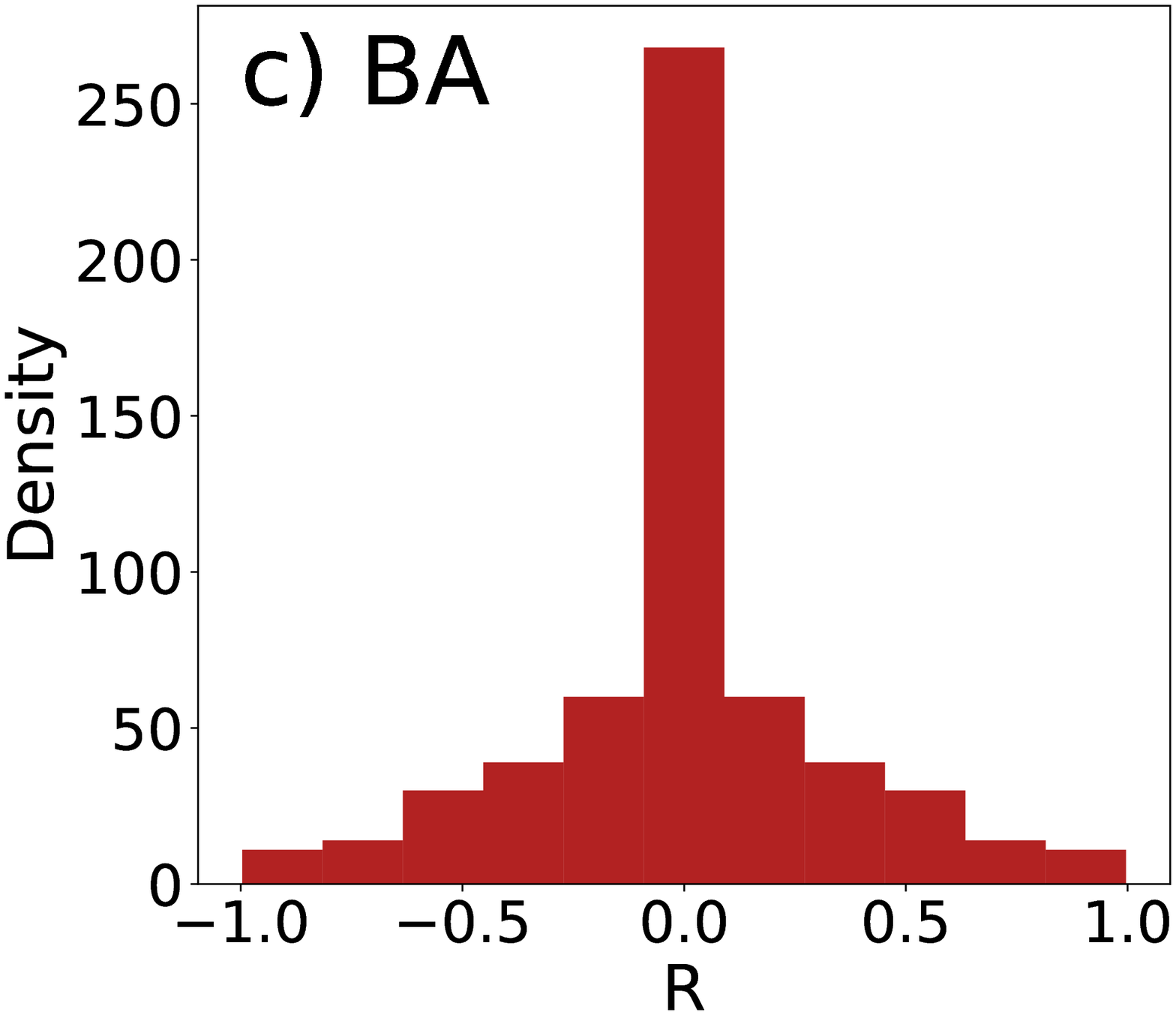}
		\end{subfigure}
		\caption{Distribution of $R$ for different network topologies. The parameter measures the difference between the stationary persistence for the full process and the persistence without the broadcasting dynamics. The three network topologies considered are: an \textit{Erdős–Rényi} network (ER), a generic \textit{scale free} network (SF) and a \textit{Barabasi-Albert} network (BA)}
		\label{img:FigureA3}
	\end{figure}

	We define $R=\rho^{\infty}_{\TextInMath{total}} - \rho^{\infty}_{\TextInMath{behav}}$, i.e. the difference between the persistence of the whole process and that of the broadcasting process alone. Figure~\ref{img:FigureA3} shows the distribution of the values of $R$ for different networks. We can notice how, in general, broadcasting can have both an enhancing effect ($R > 0$) then a mitigating one ($R < 0 $) for all the three topologies. Moreover, we notice that for the \textit{ER} and the \textit{BA} networks broadcasting can determine the expiration of a process even when the broadcasting dynamics alone ended up in $\rho^{\infty}_{\TextInMath{behav}}=1$, as the presence of $R=-1$ suggests. At the same time, also the opposite can happen and an otherwise expiring process can be completely turned around by the introduction of broadcasting ($R=1$). However, on a more detailed analysis, we find that these cases are entirely due to the broadcasting dynamics alone and the behavioral dynamics does not appear to play any particular role.
	
	\begin{figure}[!h]
		\centering
		\includegraphics[width=1\linewidth]{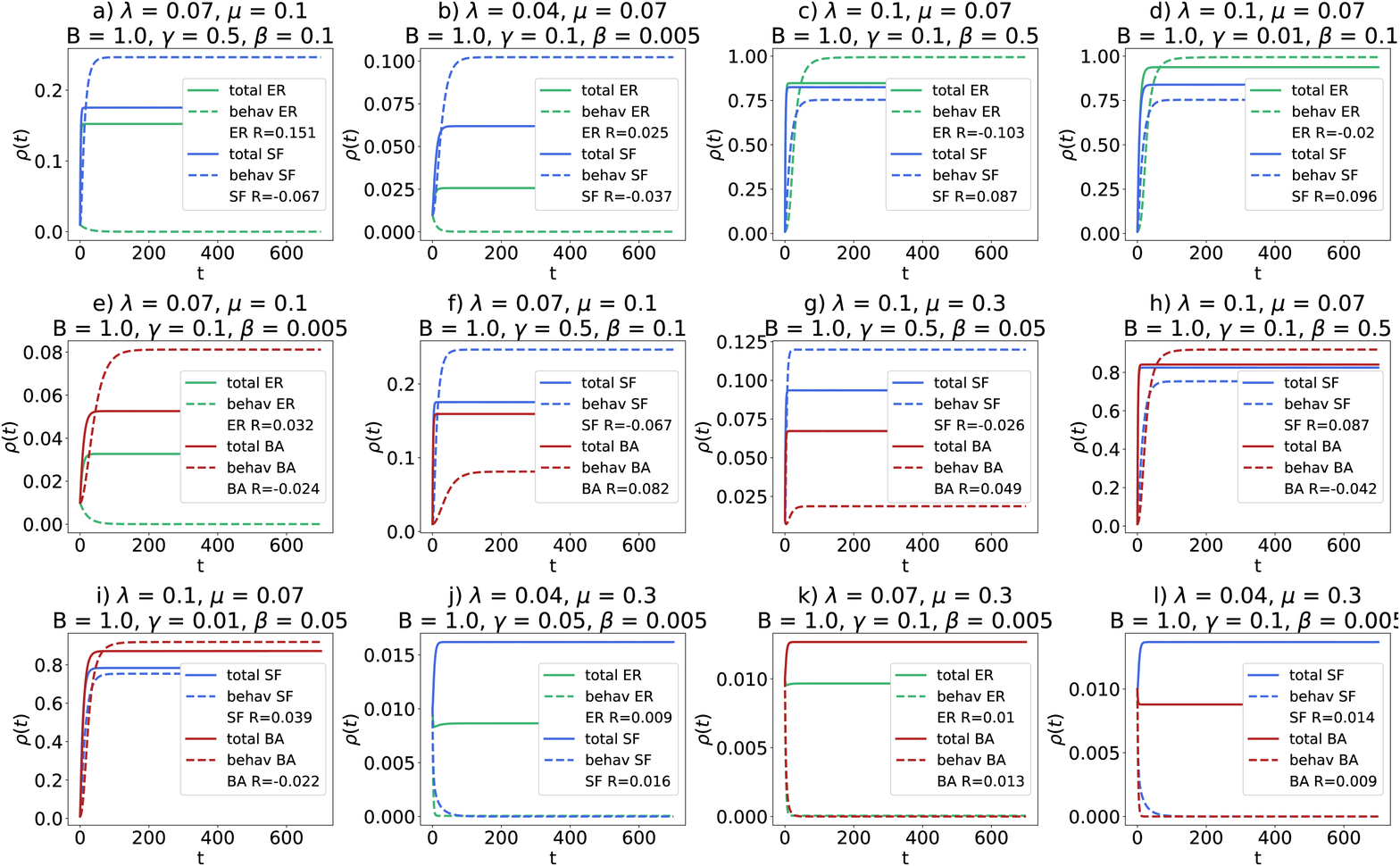}
		\caption{Comparison between the stationary prevalence for couples of networks with different topologies when using the same set of dynamical parameters. For the selected sets of parameters the introduction of broadcasting determines different outcomes for the two networks. Topological parameters of the networks: number of nodes $N = 1200$, \textit{Erdős–Rényi} parameter $p=4\times10^{-3}$, scale-free exponent $\nu=2.2$, \textit{Barabasi-Albert} parameter $m=2$. }
		\label{img:FigureA4}
	\end{figure}
	
	We now proceed to the comparison between topologies, which was carried out by analysing the networks in couples looking for values of $R$ with opposite sign. Figure~\ref{img:FigureA4} shows a selection of such scenarios. Counting row by row from left to right, in the first nine plots the introduction of broadcasting always results in the behaviors of the two networks becoming more similar, i.e. $|\rho^{\infty}_{\TextInMath{total, net_1}}-\rho^{\infty}_{\TextInMath{total, net_2}}| < |\rho^{\infty}_{\TextInMath{behav, net_1}}-\rho^{\infty}_{\TextInMath{behav, net_2}}|$. This trend was found in all cases in which the two networks have values of $R$ of the opposite sign. Among these nine cases we can recognize three different patterns: one of the networks has $\rho^{\infty}_{\TextInMath{behav}} \sim 1$ and the other does not, one of the networks has $\rho^{\infty}_{\TextInMath{behav}} \sim 0$ and the other does not or in nor of the cases the behavioral dynamics ends up in the one of the extremes. 
	
	The last three plots correspond to sets of parameters for which the behavioral dynamics alone ends up in $0$ for both networks but the introduction of broadcasting is such that for one network $\rho^{\infty}_{\TextInMath{total}}$ surpasses the initialization value of the persistence $\rho_0 = 0.01$ and for the other we still obtain $\rho^{\infty}_{\TextInMath{total}} < \rho_0$.

        \setcounter{figure}{0}
        \setcounter{table}{0}
	\section{Stationary distribution in the complete-graph approximation}\label{section:appendix:stationarySol}	

	In this section we compute and then further elaborate on the results obtained in Section~\ref{section:completeGraphModel} in the main text. We first show the steps that led to Equation~\eqref{eq:stationaryPrevalence}, following standard methods~\cite{van1992stochastic}. We show how the same distribution can be reproduced by simulating the dynamics using Monte Carlo simulation. We then expand on~\ref{section:completeGraphModel}, computing the explicit form in the case of symmetric rates, with and without zealot nodes.
	
	Let us slim out notation and denote as $A = A(t)$, $U = U(t)$ the number of $A$-type and $U$-type nodes in the system at time $t$ respectively.
	Looking at the transitions \eqref{eq:transitionBehavUA} -- \eqref{eq:transitionBroadcAU} in Section~\ref{section:model} in the main text we start by computing the rates for the transitions $U_0 \longrightarrow A$ and $A \longrightarrow U_0$ for the complete process (behavioral and broadcasting).
	\begin{align}{}
		\omega^-(U) & = \omega^-(U_0 \longrightarrow A) = \lambda \frac{A}{N} + B\beta = \lambda \frac{N - U}{N} + B\beta \label{eq:transitionRateUA}\\
		\omega^+(U) & = \omega^+(A \longrightarrow U_0) = \mu \frac{U}{N} + B\gamma.
		\label{eq:transitionRateAU}
	\end{align}
	The total rates, i.e. the rates considering the individuals involved, are
	\begin{align}{}
		\Omega^-(U) & = \Omega^-(U  \longrightarrow U-1)= \omega^-(U) U_0
		\label{eq:transitionRateOmegaMInus}\\
		\Omega^+(U) & = \Omega^-(U \longrightarrow  U+1) = \omega^+(U) A .
		\label{eq:transitionRateOmegaPlus}
	\end{align}
	The master equation of this process for the probability of having $U$ unaware nodes at time $t$, $\rho(U,t)$, is, thus,
	\begin{align}
		\rho(U,t + \diffint t) =&  \diffint t\Omega^-(U + 1)\rho(U + 1,t) + \diffint t\Omega^+(U - 1) \rho(U - 1,t) + \nonumber\\
		& + \left[ 1 - \diffint t\Omega^+(U) - \diffint t\Omega^-(U) \right]\rho(U,t)\label{eq:discreteMasterEq}
	\end{align}
	By redefining $u = \frac{U}{N}$, $a = \frac{A}{N}$ and $\tilde{u}=\frac{\tilde{U}}{N}$ so that $1 = a(t) + u(t) = a(t) + \tilde{u} + u_0(t)$ we rewrite the rates as $\Omega^{\pm}(U) = N \Omega^{\pm}(u)$ as
	\begin{align}{}
		\Omega^- (u) & = (u - \tilde{u}) \left(\lambda (1 - u) + B\beta \right)  \label{eq:transitionRateOmegaMInus_frac}\\
		\Omega^+ (u) & =  (1 - u) \left( \mu u + B\gamma \right).
		\label{eq:transitionRateOmegaPlus_frac}
	\end{align}
	By defining $\rho( u,t)$ as the probability to have a fraction of nodes of type $u$ at time $t$ (not to be confused with the prevalence) and performing the required change of variables, Equation~\eqref{eq:discreteMasterEq} becomes
	\begin{align}
		\rho( u,t + \diffint t) \diffint u & =  N\Omega^-(u + \diffint u, t)\rho( u + \diffint u,t)\diffint t + \nonumber \\
  & + N\Omega^+(u - \diffint u) \rho( u - \diffint u,t)\diffint u \diffint t + \nonumber\\
		& - \left[ 1 - N\Omega^+(u)\diffint t - N\Omega^-(u)\diffint t \right]\rho(, u,t)\diffint u	
 \end{align}
	which leads to the Fokker-Planck equation 
	\begin{align}
		\partdev{\rho(u,t)}{t} & = \Omega^-(u + \diffint u, t)\rho(u + \diffint u,t) + \nonumber \\ 
        & + \Omega^+(u - \diffint u) \rho(u - \diffint u,t) -\left( \Omega^+(u) + \Omega^-(u) \right)\rho(u,t)\nonumber\\
		 & = - \partdev{}{u} \left[ \left( \Omega^+ - \Omega^- \right)\rho(u,t) - \frac{1}{2N} \partdev{}{u} \left( \left( \Omega^+ + \Omega^-\right) \rho(u,t) \right)\right]
	\end{align}
	or in a more compact form:
	\begin{equation}
		\left\{
		\begin{array}{lr}
			\partdev{\rho(u,t)}{t} = - \partdev{}{u} J(u,t)\\
			\\
			J(u,t) = \left( \Omega^+ - \Omega^- \right)\rho(u,t) - \frac{1}{2N} \partdev{}{u} \left( \left( \Omega^+ + \Omega^-\right) \rho(u,t) \right).\\
		\end{array}
		\right.  
		\label{eq:FokkerPlank}
	\end{equation}
	The initial (I.C.) and boundary (B.C.) conditions for the differential equation are
	\begin{equation}
		\left\{
		\begin{array}{lr}
			\TextInMath{I.C.} : \quad \rho(u,t=0) = \delta(u-u_{\TextInMath{initial}})\\
			\\
			\TextInMath{B.C.} : \quad J(u = \tilde{u}, t) = J(u = 1, t) = 0 \quad \forall \quad t,\\
		\end{array}
		\right.
	\label{eq:FokkerPlankConditions}
	\end{equation}
	meaning that initially we have a fraction of $u_{\TextInMath{initial}}$ unaware agents, and that for any time our population of unaware agents cannot be negative or become larger than the system size.
	
	At stationarity we have that $\partdev{\rho(u,t)}{t} =0$, $\rho(u,t) = \rho^*(u)$, $J(u,t) = J^*(u)$, and Equations~\eqref{eq:FokkerPlank} become $ 0 = \diffu{} J^*(u)$, which leads to $J^*(u) = \TextInMath{constant}$. Given the boundary conditions, valid at any time $t$, we also obtain that $ J^*(u) = 0$. By defining $D_1(u) = \Omega^+ - \Omega^-$ and $D_2(u) = \Omega^+ + \Omega^-$ the equation to solve is
	\begin{equation}
		0 = D_1(u)\rho^*(u) - \frac{1}{2N} \diffu{}\left( D_2(u) \rho^*(u) \right).
	\end{equation}
	Integrating we obtain Equation~\eqref{eq:stationaryPrevalence} as in the main text
	\begin{equation}
		\left\{
		\begin{array}{lr}
			\rho^*(u) = D_2(u)^{-1}k \exp\left(2N \int_{\tilde{u}}^{u} \diffint u' D_1(u')D_2(u')^{-1}\right)\\
			\\
			k = \left[ \int_{\tilde{u}}^{1} \diffint u D_2(u)^{-1}k \exp\left( 2N \int_{\tilde{u}}^{u} \diffint u'D_1(u')D_2(u')^{-1}\right) \right]^{-1},
		\end{array}
		\right.  
		\label{eq:stationaryPrevalenceAppendix}
	\end{equation}
	where
	\begin{equation}
		\left\{
		\begin{array}{lr}
			D_1(u) = \Omega^+ (u) - \Omega^- (u)\\
			\\
			D_2(u) = \Omega^+ (u) + \Omega^- (u)
		\end{array}
		\right. 
		\label{eq:stationaryPrevalenceABAppendix}
	\end{equation}
	\begin{equation}
		\left\{
		\begin{array}{lr}
			\Omega^- (u) = (u - \tilde{u}) \left(\lambda (1 - u) + B\beta \right) \\
			\\
			\Omega^+ (u) =  (1 - u) \left( \mu u + B\gamma \right).
		\end{array}
		\right.  
		\label{eq:stationaryPrevalenceOmegasAppendix}
	\end{equation}
	
	\subsection{Adherence between simulation results and analytical stationary distribution}\label{section:appendix:symVSAnalytical}
	
\begin{figure}[!ht]
	    \centering
		\begin{subfigure}[b]{0.32\textwidth}
			\centering
			\includegraphics[width=\textwidth]{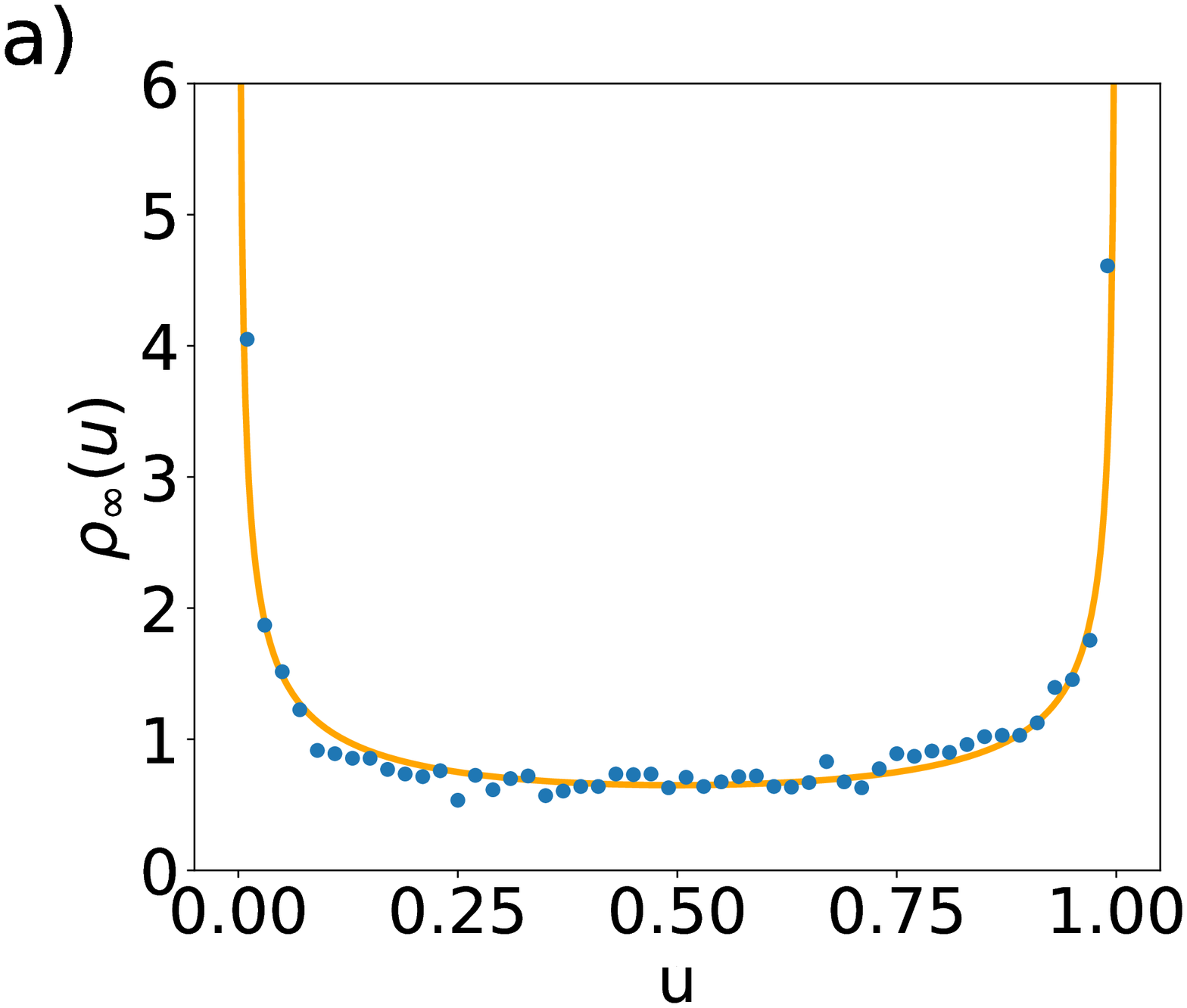}
		\end{subfigure}
		\begin{subfigure}[b]{0.32\textwidth}  
			\centering 
			\includegraphics[width=\textwidth]{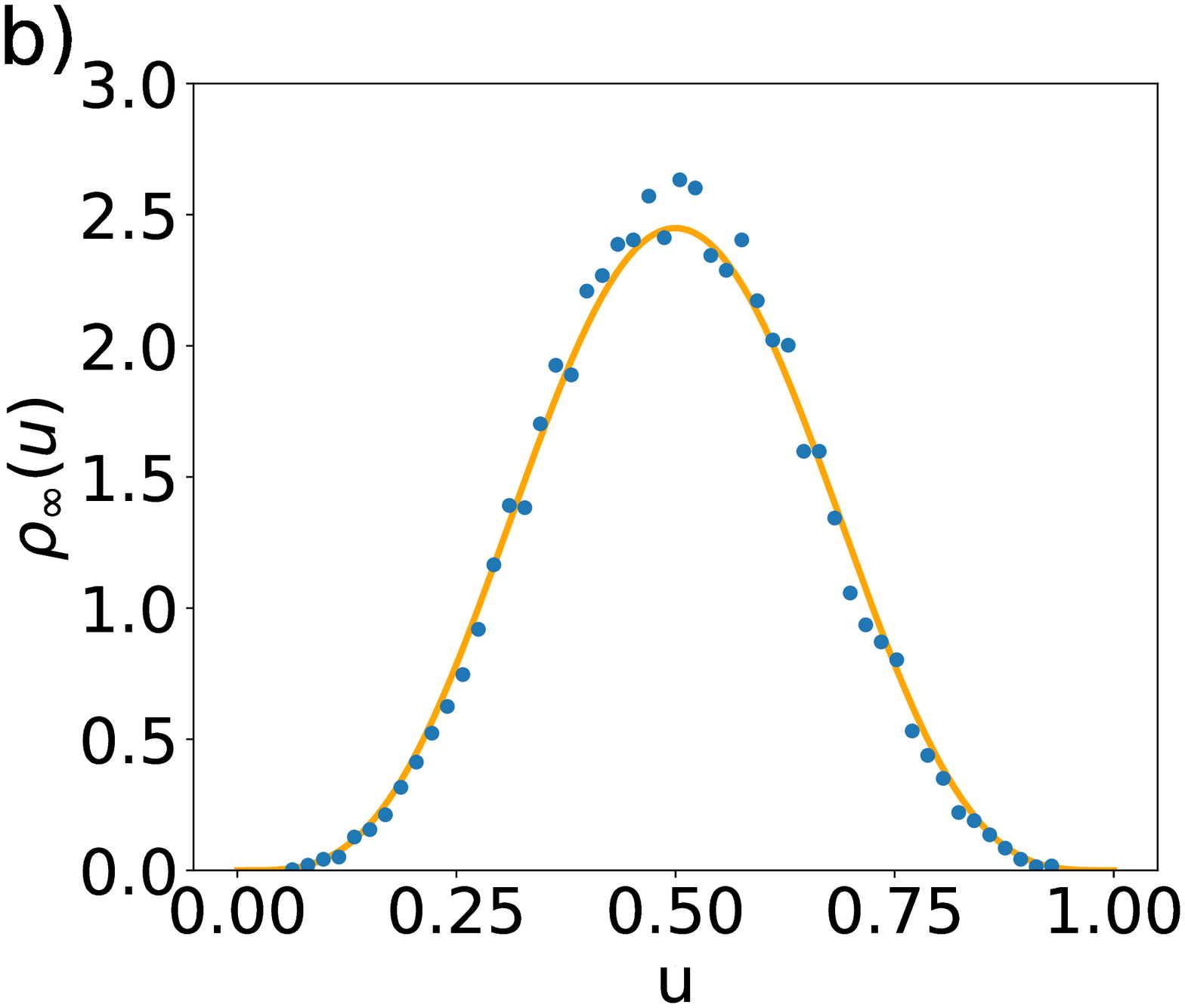}
		\end{subfigure}
		\begin{subfigure}[b]{0.32\textwidth}  
			\centering 
			\includegraphics[width=\textwidth]{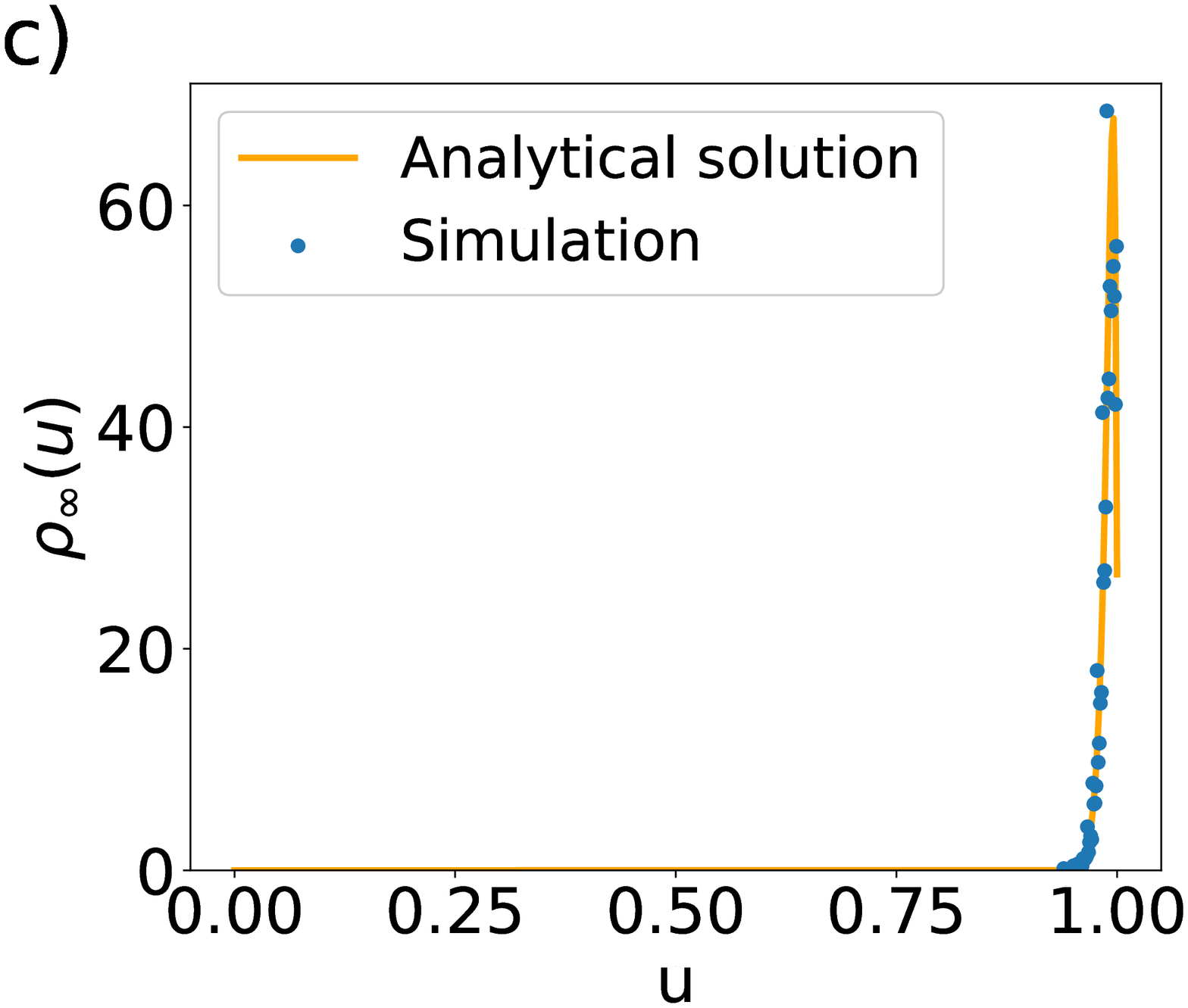}
		\end{subfigure}
		\caption{Comparison between the stationary probability density function $\rho^*(u)$ sampled from 1000 realizations of the simulation's results and the one computed analytically for different values of the dynamical parameters. The parameters set are for Figure~\ref{img:FigureB1}a $\lambda=\mu=0.1, \beta=\gamma=5\cdot10^{-5}, B=1, \tilde{u}=0, \alpha=NB\beta / \lambda-1=-0.5$; for Figure~\ref{img:FigureB1}b $\lambda = \mu=0.1, \beta = \gamma=5\cdot10^{-4}, B=1, \tilde{u}=0, \alpha=NB\beta / \lambda-1=4$; for Figure~\ref{img:FigureB1}c: $\lambda=0.1, \mu=0.12, \beta = 2\cdot10^{-4}, \gamma=10^{-4}, B=1, \tilde{u}=0$.} 
		\label{img:FigureB1}
	\end{figure}

\begin{figure}[!ht]
		\centering
		\begin{subfigure}[b]{0.32\textwidth}
			\centering
			\includegraphics[width=\textwidth]{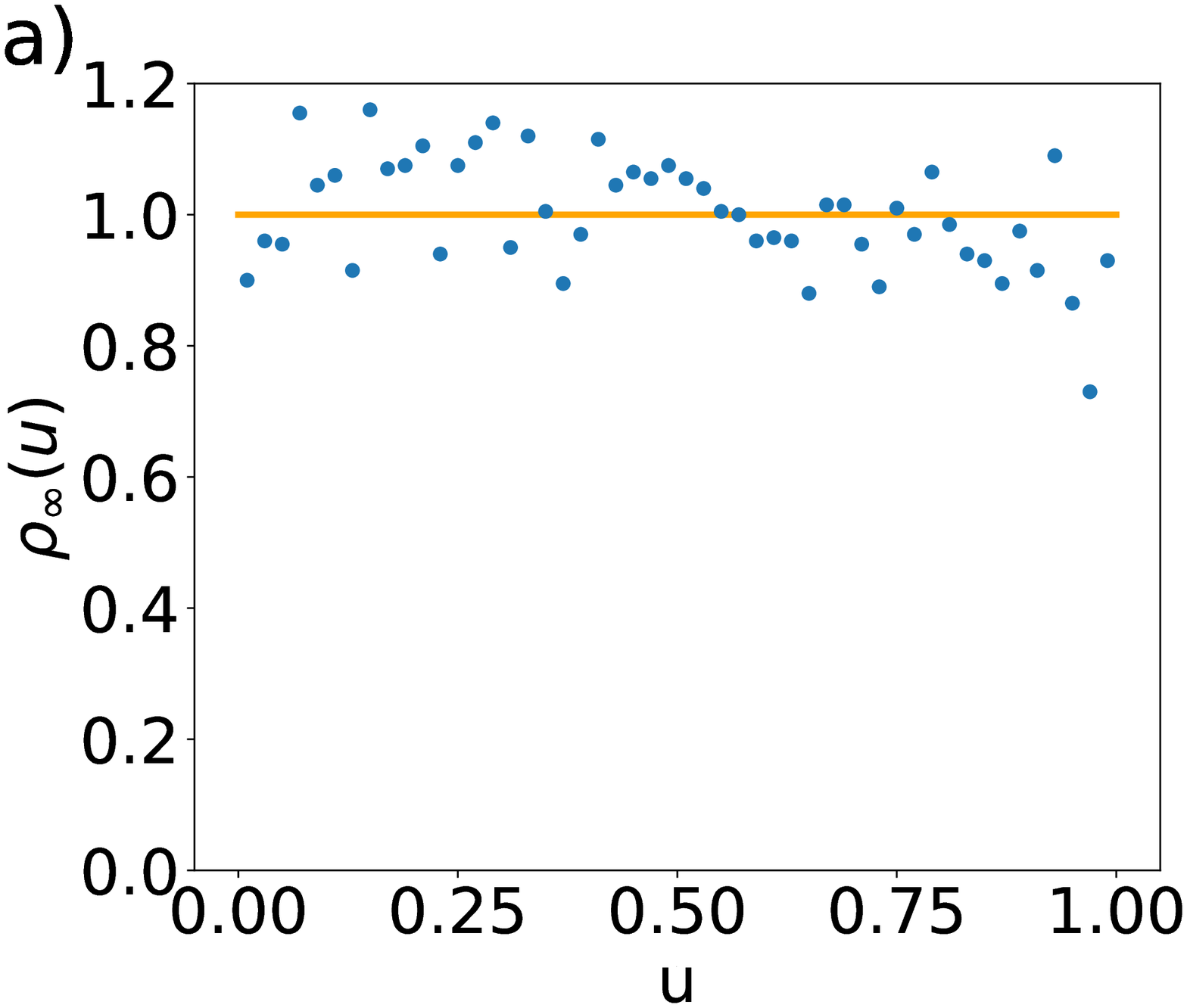}
		\end{subfigure}
		\begin{subfigure}[b]{0.32\textwidth}  
			\centering 
			\includegraphics[width=\textwidth]{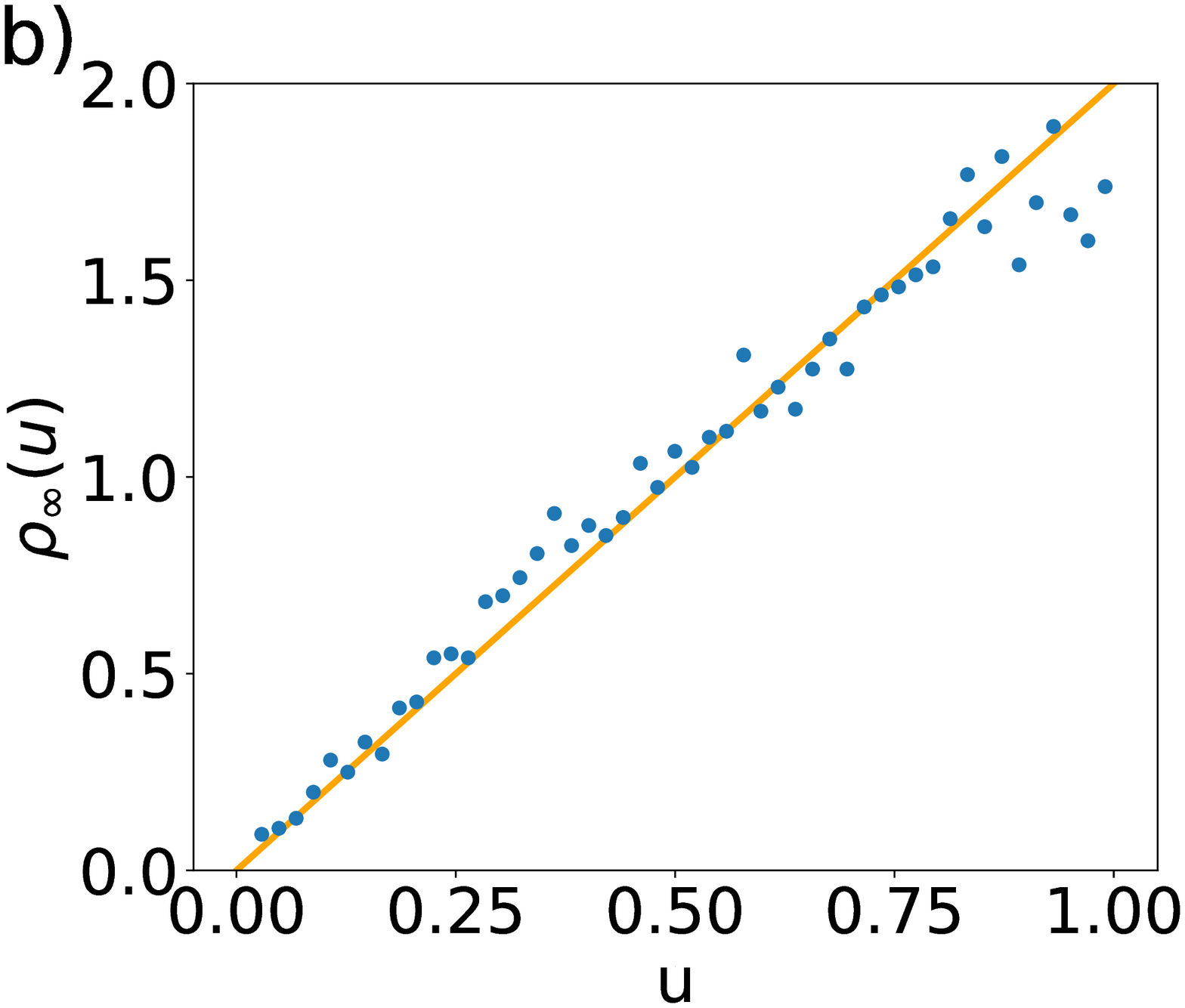}
		\end{subfigure}
		\begin{subfigure}[b]{0.32\textwidth}  
			\centering 
			\includegraphics[width=\textwidth]{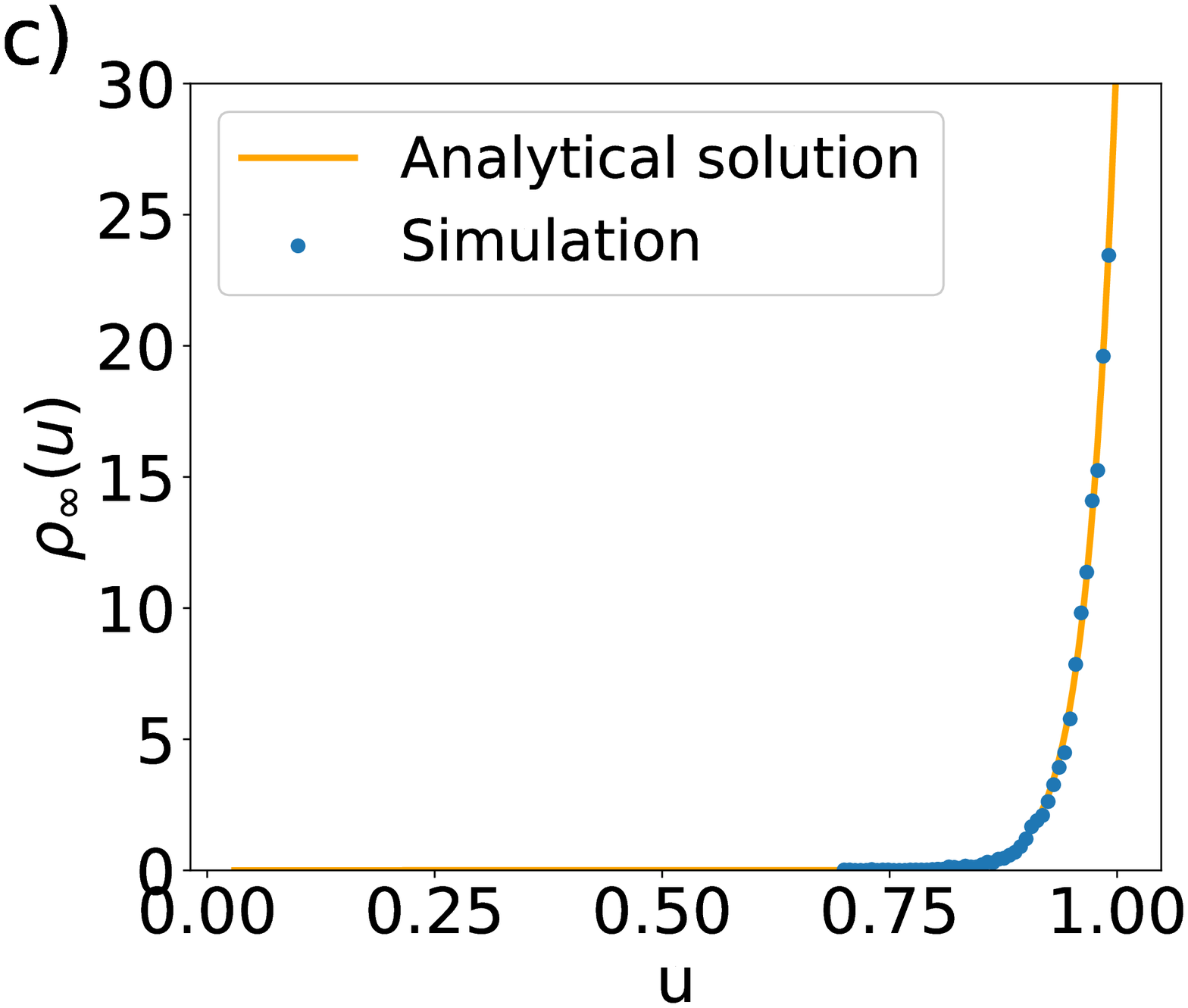}
		\end{subfigure}
		\caption{Comparison between the stationary probability density function $\rho^*(u)$ sampled from 1000 realizations of the simulation's results and the one computed analytically for different values of the dynamical parameters. The parameters set are for Figure~\ref{img:FigureB2}a $\lambda=\mu=0.1, \beta=\gamma=10^{-4}, B=1, \tilde{u}=0, \alpha=NB\beta/\lambda-1=0$; for Figure~\ref{img:FigureB2}b: $\lambda=\mu=0.1, \beta=\gamma=10^{-4}, B=1, \tilde{u}=1/N=0.001$; for Figure~\ref{img:FigureB2}c $\lambda=\mu=0.1, \beta=\gamma=10^{-4}, B=1, \tilde{u}=\frac{30}{N}=0.03$.
		}
		\label{img:FigureB2}
	\end{figure}

\begin{figure}[!ht]
		\centering
		\begin{subfigure}[b]{0.32\textwidth}
			\centering
			\includegraphics[width=\textwidth]{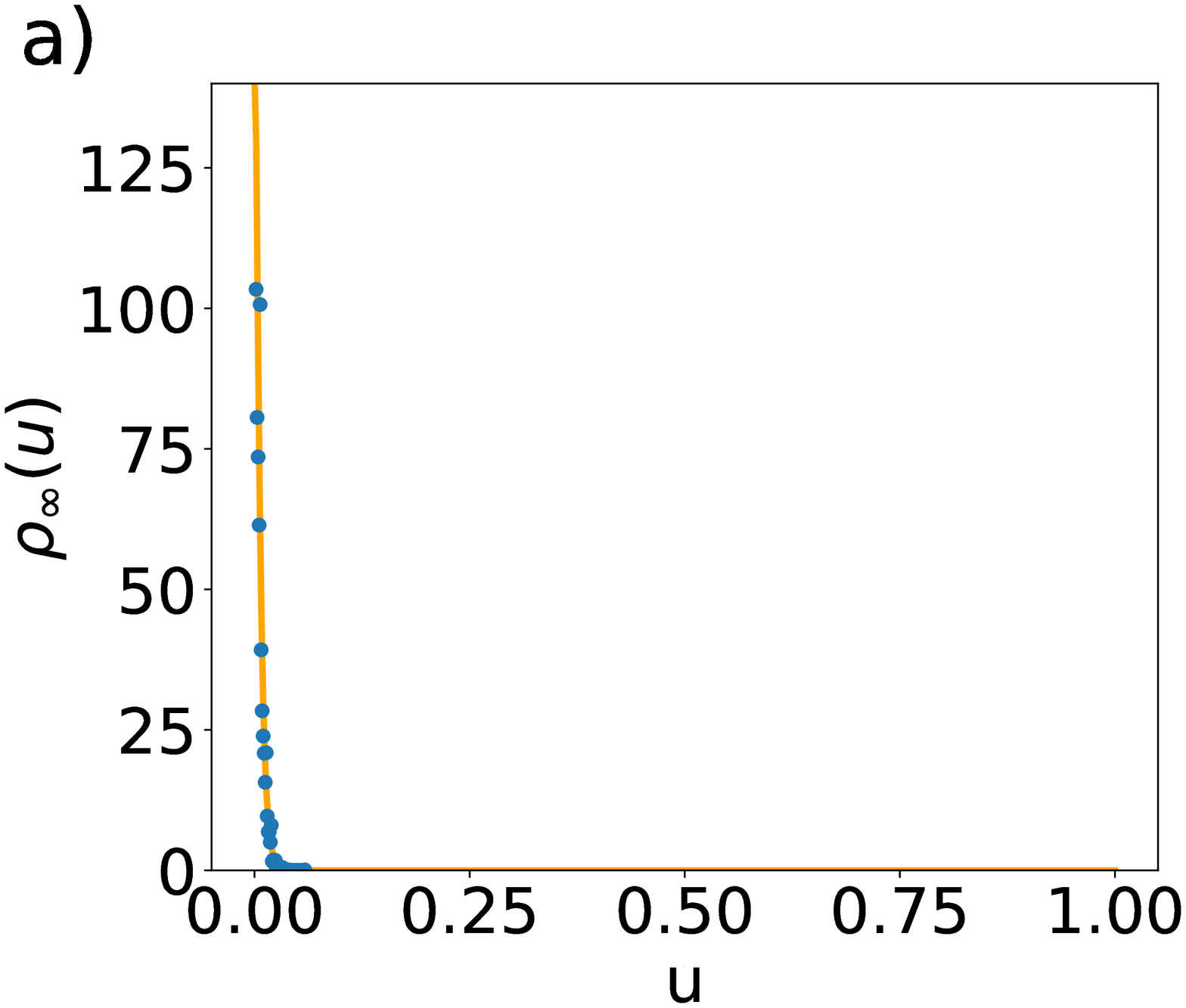}
		\end{subfigure}
		\begin{subfigure}[b]{0.32\textwidth}  
			\centering 
			\includegraphics[width=\textwidth]{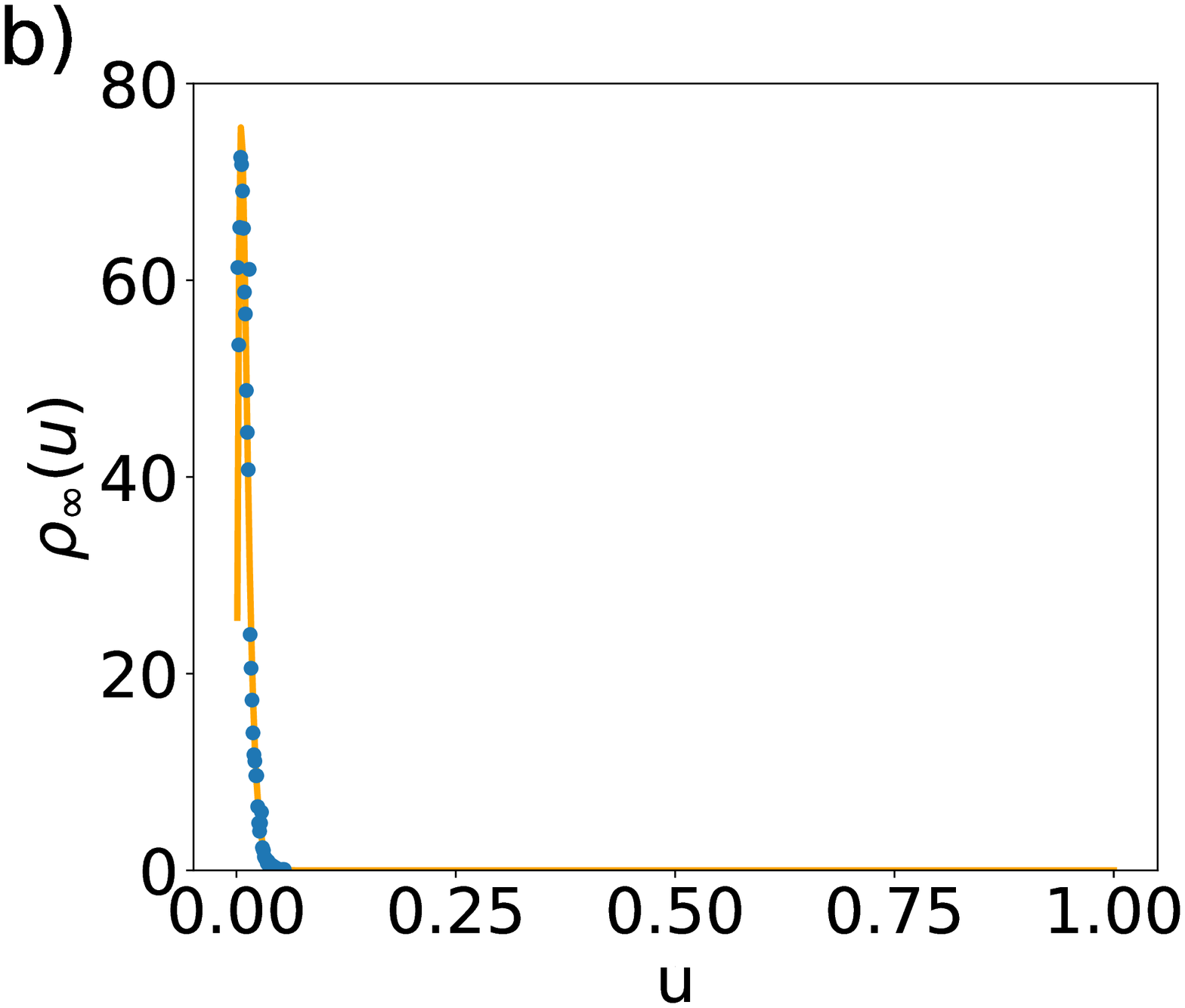}
		\end{subfigure}
		\begin{subfigure}[b]{0.32\textwidth}  
			\centering 
			\includegraphics[width=\textwidth]{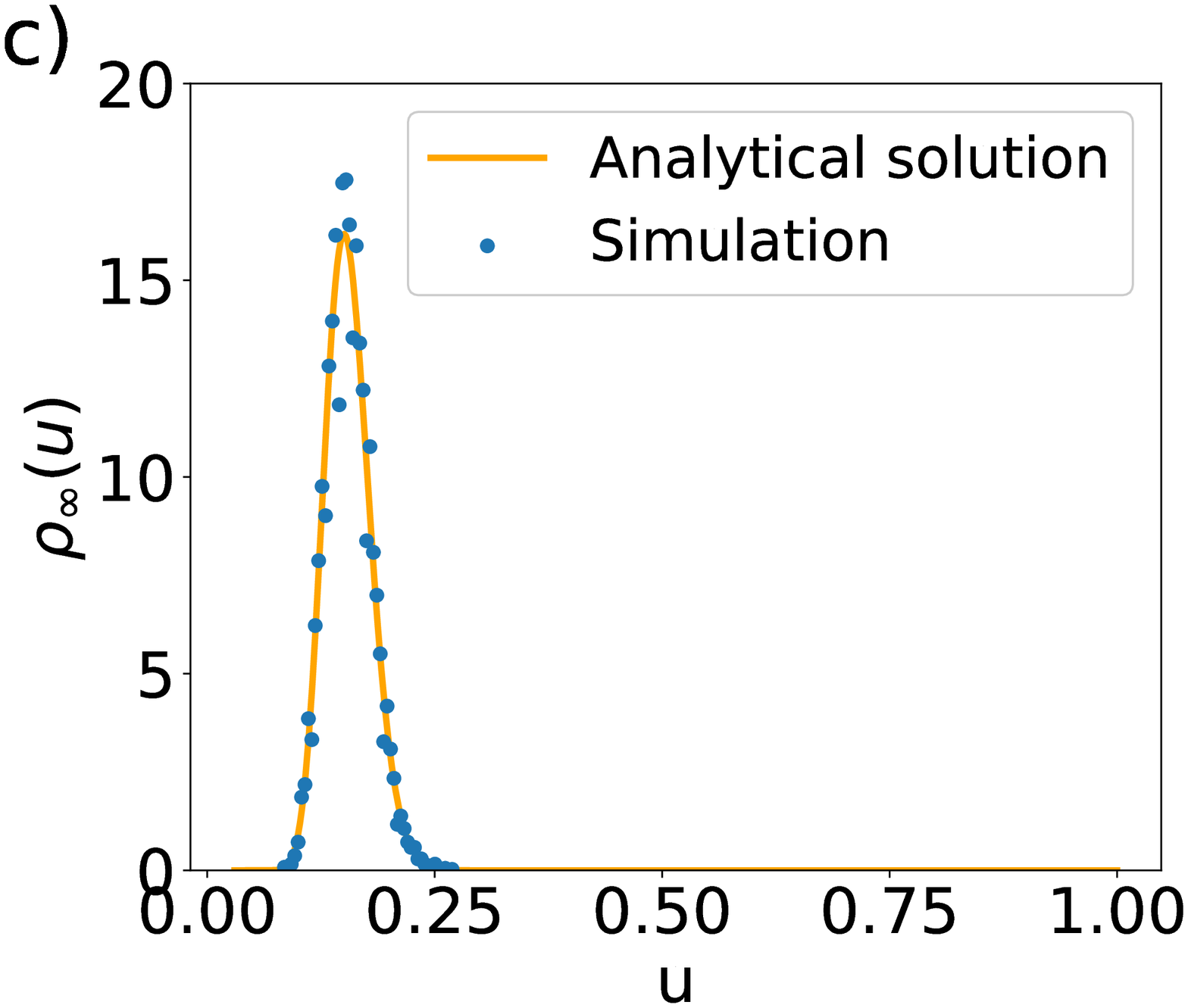}
		\end{subfigure}
		\caption{Comparison between the stationary probability density function $\rho^*(u)$ sampled from 1000 realizations of the simulation's results and the one computed analytically for different values of the dynamical parameters. The parameters set are for Figure~\ref{img:FigureB3}a $\lambda=0.1, \mu=0.08, \beta=\gamma=10^{-4}, B=1, \tilde{u}=0$; for Figure~\ref{img:FigureB3}b $\lambda=0.1, \mu=0.08, \beta=\gamma=10^{-4}, B=1, \tilde{u}=1/N=0.001$; for Figure~\ref{img:FigureB3}c: $\lambda=0.1, \mu=0.08, \beta=\gamma=10^{-4}, B=1, \tilde{u}=1/N=0.03$.
		}
		\label{img:FigureB3}
	\end{figure}

	Figures~\ref{img:FigureB1},~\ref{img:FigureB2}, and \ref{img:FigureB3} compare the stationary distribution built via simulations with the analytical solution of Equation~\eqref{eq:stationaryPrevalence} for different combinations of the parameters. The stationary distribution for a single set of parameters via simulation was obtained running the Monte Carlo simulations multiple times with different initialization and sampling the end of each evolution, checking that the stationary behavior was, in fact, reached. The different outcomes where then put together to build the distributions in blue.
	
	\subsection{Comparing rates in the homogeneous mixing and the network approaches}
	When transitioning from a model that incorporates network structure to a homogeneous mixing approach it is necessary to reconsider the role of the rates. As an example, let us consider the  transition in Equation~\eqref{eq:transitionBehavUA}, whose rate for a single node when considering network structure is given by the first term in Equation~\eqref{eq:evolutionProbBehavU}. Firstly, let us approximate the term in Equation~\eqref{eq:evolutionProbBehavU} as
	\begin{equation}
		1 - \prod_{j=1}^{N}(1 - \lambda A_{ij}p_j^A) \simeq \lambda \sum_{j=1}^N A_{ij}p_j^A
		\label{eq:transitionBehavUA_rate_approx}
	\end{equation}
	which generally holds for small values of $\lambda A_{ij}p_j^A$. It is easy to interpret this quantity in the following way: node $i$ establishes $k_i$ contacts at each time step (with $k_i$ being the degree of node $i$) and the probability of one of these to be at risk is, on average, $\frac{A(t)}{N}$. Thus:	
	\begin{equation}
		\lambda \sum_{j=1}^N A_{ij}p_j^A \simeq \lambda k_i \frac{A(t)}{N}
	\end{equation}
	To avoid confusion, let us refer to $\lambda$ in equation~\eqref{eq:stationaryPrevalenceOmegas} as $\lambda_{\TextInMath{complete}}$. It is easy to see that the relationship that holds between $\lambda_{\TextInMath{complete}}$ and $\lambda$ is $\lambda = \lambda_{\TextInMath{complete}}k_i^{-1}$. In particular, for a complete graph, it becomes $\lambda = \lambda_{\TextInMath{complete}}{(N-1)}^{-1} \simeq \lambda_{\TextInMath{complete}}{N}^{-1}$. In practice, this gimmick is necessary when comparing the results of the simulation with the analytical form of $\rho^*(u)$.
	
	\subsection{Behavioral dynamics without broadcasting}\label{section:appendix:noBroadc}
\begin{figure}[!h]
		\centering
		\begin{subfigure}[!h]{0.48\linewidth}
			\centering
			%\hspace{-50pt}
		\includegraphics[width=\linewidth]{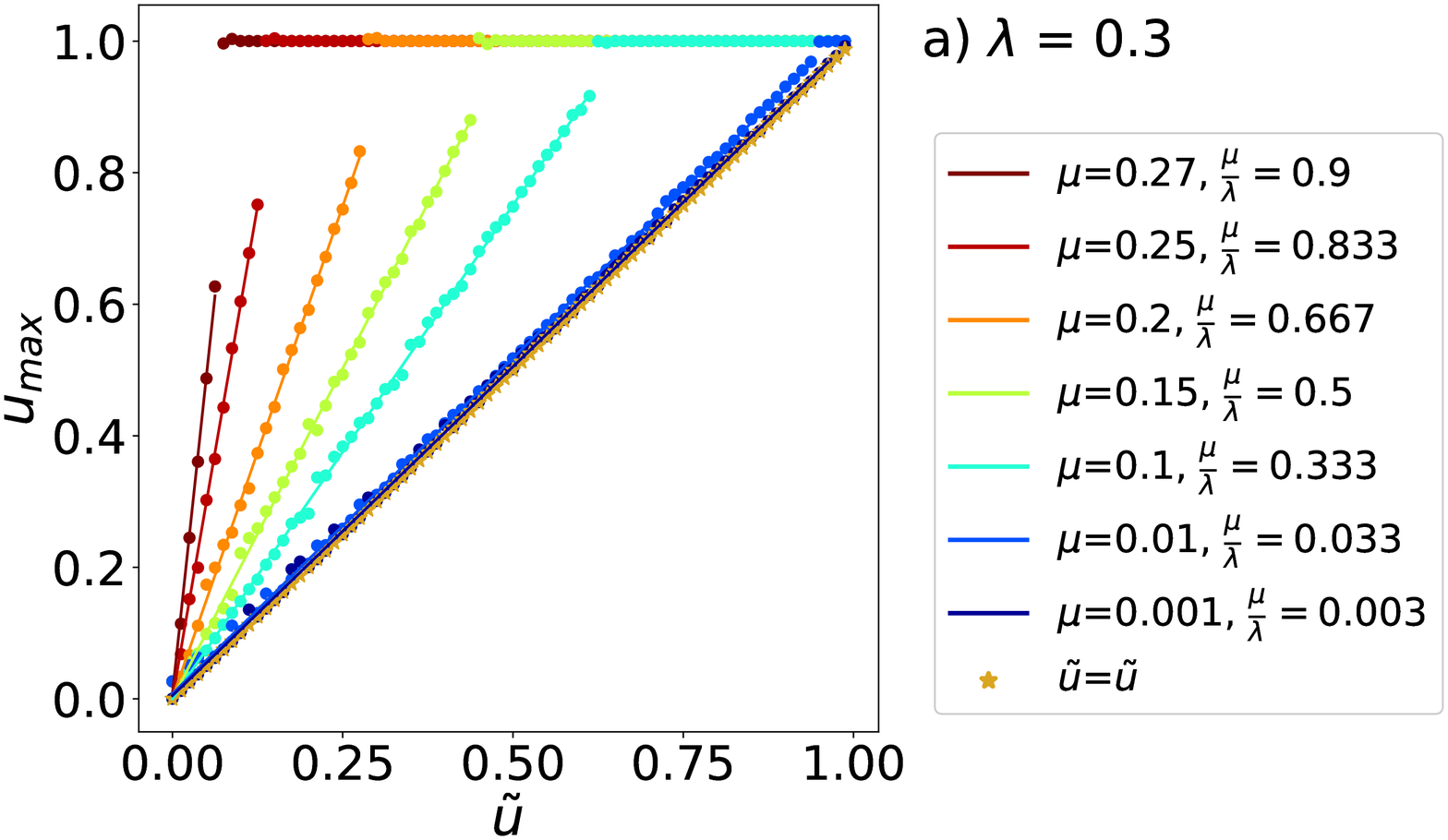}
		\end{subfigure}
		\begin{subfigure}[!h]{0.48\linewidth}
			\centering
			\includegraphics[width=1\textwidth]{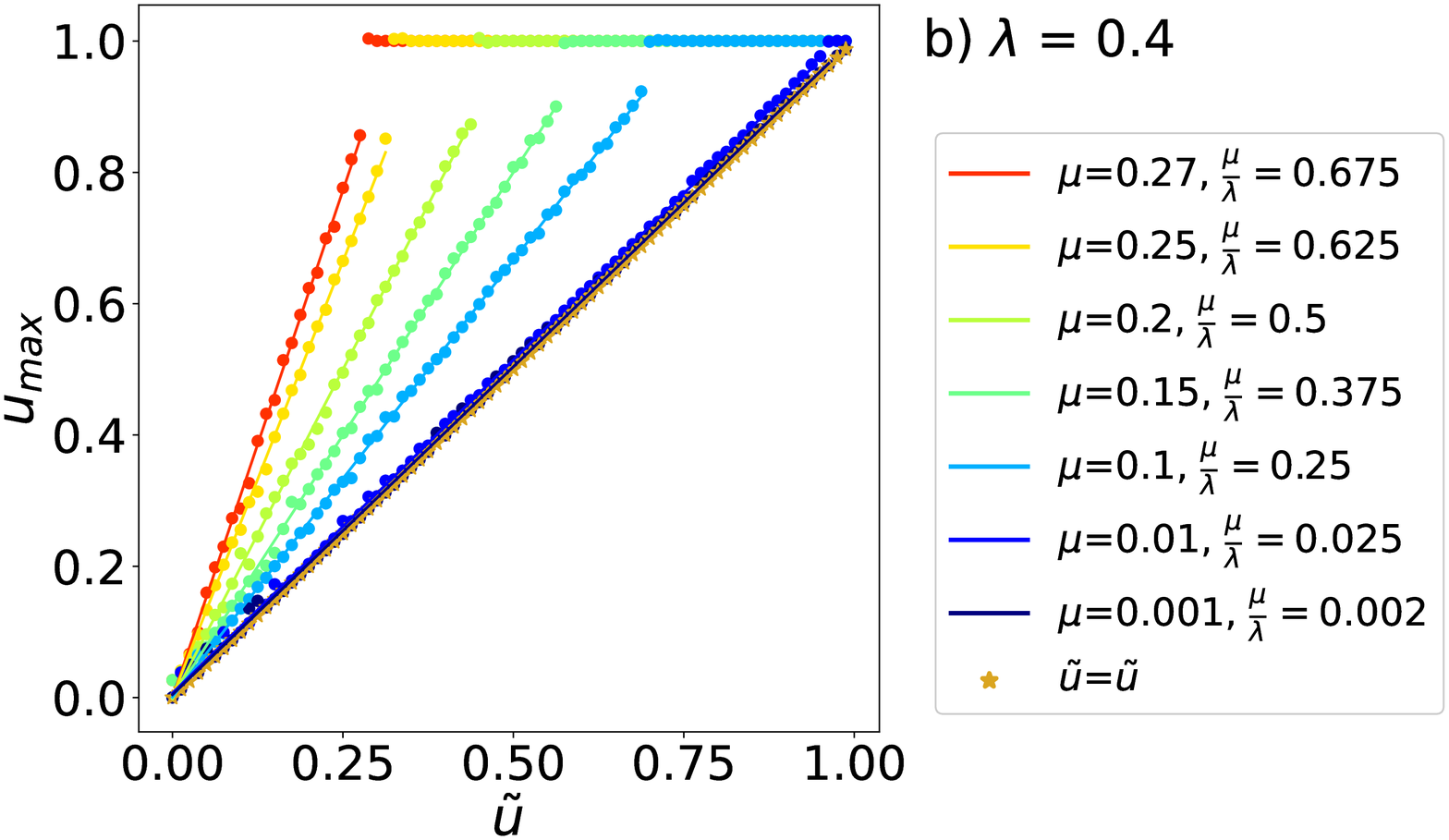}
		\end{subfigure}
		\caption{Trend of the value of $u_{\TextInMath{max}}$, the fraction of unaware nodes $u$ corresponding to the maximum value of $\rho^*(u)$, when varying the fraction of zealot nodes, $\tilde{u}$, and for different combinations of the behavioral parameters $\mu$ and $\lambda$.}
		\label{img:FigureB4}
	\end{figure}

	Before getting into the actual solution of Equation~\eqref{eq:stationaryPrevalenceAppendix} let us consider for a moment the behavioral dynamics by itself. If we do not include broadcasting the rates in Equation~\eqref{eq:stationaryPrevalenceOmegasAppendix} become
	\begin{equation}
		\left\{
		\begin{array}{lr}
			\Omega^- (u) = (u - \tilde{u}) \left(\lambda (1 - u) \right) \\
			\\
			\Omega^+ (u) =  (1 - u) \left( \mu u\right),
		\end{array}
		\right.
	\end{equation}
	that is,
	\begin{equation}
		\left\{
		\begin{array}{lr}
			D_1(u) = (1 - u) \left(\mu u - (u - \tilde{u}) \lambda \right)\\
			\\
			D_2(u) =  (1 - u) \left( \mu u + (u - \tilde{u}) \lambda \right).
		\end{array}
		\right.  
	\end{equation}
	We notice that for $\tilde{u} = 0$ both $D_1$ and $D_2$ are zero at the ends of integration. As a matter of fact, without broadcasting and for $\tilde{u} = 0$ both $u=0$ and $u=1$ are absorbing states as there is no spontaneous transition and the dynamics stops when all nodes are of the same kind. The presence of $\tilde{u}$ nodes breaks the symmetry by eliminating the absorbing boundary in $u=0$, but the presence of absorbing boundaries prevents us from solving Equation~\eqref{eq:stationaryPrevalenceAppendix}. However, we can still identify the following scenarios and make some intuitive considerations:
	
	\begin{itemize}
	    \item $\mu > \lambda, \quad \forall \, \tilde{u}$: The dynamics is unbalanced in favour of the unaware nodes and eventually will be absorbed in the state $u=1$.
	
	    \item $\mu = \lambda$: If $\tilde{u} = 0$ the dynamics is balanced and both $u=1$ and $u=0$ are absorbing boundaries. Being absorbed at the boundaries is the only possible outcome and they will have equal probability if the system is initially balanced. If $\tilde{u} \neq 0$ the only absorbing boundary is $u=1$ and, statistically, the dynamics will be absorbed there.
	
    	\item $\mu < \lambda$: If $\tilde{u} = 0$ then aware nodes are favoured and the dynamics will stop, statistically, in $u=0$. However, when $\tilde{u} \neq 0$ the outcome is not as trivial, as the presence of zealots could counterbalance the advantage given by $\lambda > \mu$. To get some insight into this specific case we run simulations for different combinations of $\lambda > \mu$ and $\tilde{u}$ and sampled from the stationary process in order to build the stationary distribution $\rho^*(u)$. Figure~\ref{img:FigureB4} shows the trend of the value of $u$ corresponding to the maximum value of $\rho^*(u)$, $u_{\TextInMath{max}}$, with $\tilde{u}$ for different combinations of the behavioral parameters. Colors from red to blue correspond to lowering values of the parameter $\mu\lambda^{-1}$. We notice a linear dependency between $u_{\TextInMath{max}}$ and $\tilde{u}$ up until a certain cutoff value of $\tilde{u}$, whose value seems to be dependent on the value of the parameter $\mu\lambda^{-1}$. After the cutoff $u_{\TextInMath{max}}=1$.	
	\end{itemize}
	
	Table~\ref{table:chap2:baseline_fit} shows the results obtained by fitting the curve in Figure~\ref{img:FigureB4} with a degree one polynomial $u_{\TextInMath{max}}(\tilde{u}) = x + y\tilde{u}$. The parameter $\tilde{u}_{\TextInMath{cutoff}}$ corresponds to the first value of $\tilde{u}$ for which $u_{\TextInMath{max}} = 1$. We find that in general the value of $x$ is compatible with $0$ (which is expected as for $\tilde{u} = 0$ we expect the dynamics to be absorbed in $u=0$. The value of $y$ and $\tilde{u}_{\TextInMath{cutoff}}$ seem to be closely related to the parameter $\mu\lambda^{-1}$, as for different couples $\lambda$ and $\mu$ but same value of $\mu\lambda^{-1}$ we find equal value of $y$ and $\tilde{u}_{\TextInMath{cutoff}}$.	
	\begin{table}[!h]
		\centering
		\begin{tabular}{|c|c|c|c|c|c|}
			\hline
			$\lambda$ & $\mu$ & $\mu\lambda^{-1}$ &$x$      & $y$      & $\tilde{u}_{\TextInMath{cutoff}}$ \\ \hline
			0.4                & 0.001          & 0.002                          & 0.004 $\pm$ 0.001  & 1.00 $\pm$ 0.01  & 0.9625                        \\ \hline
			0.3                & 0.001          & 0.003                          & 0.005 $\pm$ 0.001  & 1.00 $\pm$ 0.01  & 0.9625                        \\ \hline
			0.4                & 0.01           & 0.025                          & 0.007 $\pm$ 0.001  & 1.01 $\pm$ 0.01  & 0.9375                        \\ \hline
			0.3                & 0.01           & 0.033                          & 0.008 $\pm$ 0.001  & 1.02  $\pm$ 0.01 & 0.9375                        \\ \hline
			0.4                & 0.1            & 0.25                           & 0.001 $\pm$ 0.001  & 1.33 $\pm$ 0.01  & 0.6875                        \\ \hline
			0.3                & 0.1            & 0.333                          & 0.001 $\pm$ 0.001  & 1.50 $\pm$ 0.01  & 0.6125                        \\ \hline
			0.4                & 0.15           & 0.375                          & -0.001 $\pm$ 0.001 & 1.60 $\pm$ 0.01  & 0.5625                        \\ \hline
			0.3                & 0.15           & 0.5                            & 0.0001 $\pm$ 0.003 & 2.01 $\pm$ 0.01  & 0.4375                        \\ \hline
			0.4                & 0.2            & 0.5                            & -0.002 $\pm$ 0.003 & 2.01$\pm$ 0.01   & 0.4375                        \\ \hline
			0.4                & 0.25           & 0.625                          & 0.001 $\pm$ 0.001  & 2.65 $\pm$ 0.02  & 0.3125                        \\ \hline
			0.3                & 0.2            & 0.667                          & 0.001 $\pm$ 0.001  & 3.00 $\pm$ 0.02  & 0.275                         \\ \hline
			0.4                & 0.27           & 0.675                          & 0.001 $\pm$ 0.001  & 3.09 $\pm$ 0.03  & 0.275                         \\ \hline
			0.3                & 0.25           & 0.833                          & 0.001 $\pm$ 0.001  & 6.0 $\pm$ 0.1    & 0.125                         \\ \hline
			0.3                & 0.27           & 0.9                            & 0.007 $\pm$ 0.001  & 9.7 $\pm$ 0.3    & 0.0625                        \\ \hline
		\end{tabular}
          \caption{Parameters and fit results of the curves in Figure~\ref{img:FigureB4} for different combinations of the behavioral parameters.}
		\label{table:chap2:baseline_fit}
  
	\end{table}
	
	\subsection{Symmetric rates and absence of zealots $(\tilde{u}=0)$}\label{section:appendix:StatSolSymATAUtilde0}
	
	In the case of symmetric rates
	$\lambda = \mu \longrightarrow \lambda$, $\beta = \gamma \longrightarrow \beta$ and $\tilde{u}=0$, Equations~\eqref{eq:stationaryPrevalenceABAppendix} and \eqref{eq:stationaryPrevalenceOmegasAppendix} become, respectively,
	\begin{equation}
		\left\{
		\begin{array}{lr}
			D_1(u) = B \beta (1 - 2u)\\
			\\
			D_2(u) =   - 2 \lambda u^2 + 2\lambda u + B \beta.
		\end{array}
		\right.
		\label{eq:ABSymUtilde0}
	\end{equation}
	\begin{equation}
		\left\{
		\begin{array}{lr}
			\Omega^- (u) = (u) \left(\lambda (1 - u) + B \beta \right) \\
			\\
			\Omega^+ (u) =  (1 - u) \left( \lambda u + B \beta \right)
		\end{array}
		\right. 
		\label{eq:OmegasSymUtilde0}
	\end{equation}
	We notice that, since we introduced broadcasting, even in the case of $\tilde{u}=0$ none of the boundaries is absorbing anymore. 
	
	We can write
	\[ \diffu{D_2(u)} = - 4 \lambda u + 2 \lambda = \frac{2 \lambda }{B \beta} D_1(u)\]	
	making it possible to solve Equation~\eqref{eq:stationaryPrevalenceAppendix}
	\begin{equation}
		\left\{
		\begin{array}{lr}
			\rho^*(u) = D_2(u)^{-1}k D_2(u)^{\frac{N B \beta }{\lambda}} = k \left(  - 2 \lambda u^2 + 2\lambda u + B \beta \right)^{\frac{N B \beta }{\lambda} - 1}\\
			\\
			k = \left[ \int_0^1 \left( - 2 \lambda u^2 + 2\lambda u + B \beta \right)^{\frac{N B \beta }{\lambda} - 1}\right]^{-1}.
		\end{array}
		\right.
	\end{equation}
	
	\subsection{Symmetric rates and presence of zealots $(\tilde{u} \neq 0)$}\label{section:appendix:StatSolSymATA}	
	Now that we consider the presence of zealots, $\tilde{u} \neq 0$, Equations~\eqref{eq:stationaryPrevalenceABAppendix} and \eqref{eq:stationaryPrevalenceOmegasAppendix} become, respectively,
		\begin{equation}
		\left\{
		\begin{array}{lr}
		    D_1(u) = (1 - u) \left[ \lambda u + B \beta - \lambda (u - \tilde{u})\right] - B \beta (u - \tilde{u})\\
			\\
			D_2(u) =  (1 - u) \left[ \lambda u + B \beta + \lambda (u - \tilde{u})\right] + B \beta (u - \tilde{u}).
		\end{array}
		\right.
		\label{eq:ABSym}
	\end{equation}	
	\begin{equation}
		\left\{
		\begin{array}{lr}
			\Omega^- (u) = (u - \tilde{u}) \left[\lambda (1 - u) + B \beta \right] \\
			\\
			\Omega^+ (u) =  (1 - u) \left( \lambda u + B \beta \right)
		\end{array}
		\right. 
		\label{eq:OmegasSym}
	\end{equation}
	The integral at the exponential of \eqref{eq:stationaryPrevalenceAppendix} was computed using \textit{Mathematica}, obtaining,
	\begin{align}
		\int_{\tilde{u}}^u \diffint u' \frac{D_1(u')}{D_2(u')} = & \frac{1}{4 \lambda} \frac{2 \lambda^{\frac{1}{2}} (-2 \beta  B +\lambda (-2+\tilde{u} )) \tilde{u}  \arctan \left[\frac{\lambda^{\frac{1}{2}} (2-3 \tilde{u} )}{(-\lambda (-2+\tilde{u} )^2+8 \beta  B  (-1+\tilde{u} ))^{\frac{1}{2}}}\right]}{(-\lambda (-2+\tilde{u} )^2+8 \beta  B  (-1+\tilde{u} ))^{\frac{1}{2}}} + \nonumber \\
		& - \frac{1}{4 \lambda} \frac{2 \lambda^{\frac{1}{2}} (-2 \beta  B +\lambda (-2+\tilde{u} )) \tilde{u}  \arctan\left[\frac{\lambda^{\frac{1}{2}} (2-4 u+\tilde{u} )}{(-\lambda (-2+\tilde{u} )^2+8 \beta  B  (-1+\tilde{u} ))^{\frac{1}{2}}}\right]}{(-\lambda (-2+\tilde{u} )^2+8 \beta  B  (-1+\tilde{u} ))^{\frac{1}{2}}} + \nonumber \\
		& + \frac{1}{4 \lambda} (2 \beta  B +\lambda \tilde{u} )
		\log \left[ (-1+u) \lambda (2 u-\tilde{u} )+\beta  B  (-1+\tilde{u} )\right] + \nonumber \\
		& - \frac{1}{4 \lambda} (2 \beta  B +\lambda \tilde{u} ) \log \left[{(-1+\tilde{u} ) (\beta  B +\lambda \tilde{u} )} \right].
		\label{eq:stationaryPrevalenceSymExponent}
	\end{align}
	We can leave the first and the last term aside as they do not depend on $u$ and can thus be incorporated in the normalization constant. We first notice that for all $ \lambda, \beta, \tilde{u}$ and $B \, \in [0,1]$ it holds that
	\[[-\lambda (-2+\tilde{u} )^2+8 \beta  B  (-1+\tilde{u} )]^{\frac{1}{2}} \leq 0,\]
	that can thus be rewritten as $d = i[\lambda (-2+\tilde{u} )^2+8 \beta  B  (1-\tilde{u} )]^{\frac{1}{2}}$. We now use the identity
	\[\arctan(z) = \frac{i}{2} \log \left[ \frac{1 - iz}{1 + iz}\right],\] 
	so the second term in Equation~\eqref{eq:stationaryPrevalenceSymExponent} becomes
	\begin{align}
	= & \frac{i}{2} \frac{1}{4 \lambda} \frac{2 \lambda^{\frac{1}{2}} (2 \beta  B +\lambda (2-\tilde{u} )) \tilde{u}}{i[\lambda (-2+\tilde{u} )^2+8 \beta  B  (1-\tilde{u} )]^{\frac{1}{2}}} \ \nonumber \\
		 & \times  \left[  \log \left(1 - \frac{\lambda^{\frac{1}{2}} (2-4 u+\tilde{u} )}{[\lambda (-2+\tilde{u} )^2+8 \beta  B  (1-\tilde{u} )]^{\frac{1}{2}}}\right) +  \right. \nonumber \\
    &  \left. - \log \left(1 + \frac{\lambda^{\frac{1}{2}} (2-4 u+\tilde{u} )}{[\lambda (-2+\tilde{u} )^2+8 \beta  B  (1-\tilde{u} )]^{\frac{1}{2}}}\right )  \right] .
	\end{align}
	We also notice that the argument of the logarithm at the third row is actually equal to $-D_2(u)$.
	We include in a new normalization constant, $K_1$, both the previous normalization term and the two terms we set aside. By substituting back \eqref{eq:stationaryPrevalenceSymExponent} into the exponential of Equation~\eqref{eq:stationaryPrevalenceAppendix}, the exponential and logarithm functions simplify, leading to
	\begin{equation}
		\left\{
		\begin{array}{lr}
			\rho^*(u) = K_1 \left[ \frac{1 - x(u)}{1 + x(u)}\right] ^{E_1} D_2(u)^{E_2} \\
			\\
			x(u) = \frac{\lambda^{\frac{1}{2}} (2-4 u+\tilde{u} )}{[\lambda (-2+\tilde{u} )^2+8 \beta  B  (1-\tilde{u} )]^{\frac{1}{2}}}\\
			\\
			D_2(u) = -2 \lambda u^2 + u (\tilde{u} + 2) \lambda + B \beta - \lambda\tilde{u} - B \beta \tilde{u}\\
			\\
			E_1 =  {\frac{N}{2 \lambda^{\frac{1}{2}}} \frac{ (2 \beta  B +\lambda (2 - \tilde{u} )) \tilde{u}}{[\lambda (2 - \tilde{u} )^2+8 \beta  B  (1-\tilde{u} )]^{\frac{1}{2}}}} \\
			\\
			E_2 = \frac{N}{2 \lambda} (2 \beta  B +\lambda \tilde{u} )- 1.
		\end{array}
		\right.  
	\end{equation}
	By substituting $\phi=B \beta/\lambda$, we finally obtain
	\begin{equation}
		\left\{
		\begin{array}{lr}
			\rho^*(u) = K_1 \left[ \frac{1 - x(u)}{1 + x(u)}\right] ^{E_1} D_2(u)^{E_2} \\
			\\
			x(u) = \frac{(2-4 u+\tilde{u} )}{[(-2+\tilde{u} )^2+8 \phi  (1-\tilde{u} )]^{\frac{1}{2}}}\\
			\\
			D_2(u) = -2 u^2 + u (\tilde{u} + 2) + \phi - \tilde{u} - \phi \tilde{u}\\
			\\
			E_1 =  \frac{N}{2} \frac{ (2 \phi +(2 - \tilde{u})) \tilde{u}}{[(2 - \tilde{u} )^2+8 \phi  (1-\tilde{u})]^{\frac{1}{2}}} \\
			\\
			E_2 = \frac{N}{2 } (2 \phi + \tilde{u} )- 1.
		\end{array}
		\right.  
	\label{eq:stationaryPrevalenceSym}
	\end{equation}

        \setcounter{figure}{0}
        \setcounter{table}{0}
	\section{Dataset}\label{section:appendix:dataset} 
	In this section we describe the procedure that led to the selection of the data employed for the analysis in Section~\ref{section:data} in the main text.
	
	The data against which we compare our model come from \textit{Twitter}. Using data from \textit{Twitter} has a double advantage. The first one is the amount of data it provides: in 2019 \textit{Twitter}'s userbase counted 330 million monthly active users~\cite{Twitter_Statista}, making it one of the most popular micro-blogging platform nowadays. Even though numerous biases and  representativeness issues have been highlighted in its userbase~\cite{blank2017}, \textit{Twitter} remains an incredible resource when it comes to collecting social data. The second advantage is the fairly easy access that the platform offers to its users' activity (deprived of all personal information): through its application programming interface (API) it is possible to efficiently filter content according to the desired topic and the access to its data is completely free.
	
	The raw data for this analysis was obtained from the infrastructure of the \textit{COVID-19 Infodemic Observatory}, an interactive dashboard whose ultimate goal is to provide information on the country-wise risk of infodemics~\cite{infodemy}, derived from the online discussions about the COVID-19 pandemic. Even though we are not directly interested in the risks connected to the spreading of fake news about the pandemic, the observatory collected with minimum sampling bias the activity of users on \textit{Twitter} from January 22 2020, regardless of the their language. All the data collected is open.
	
	Data was collected through the \textit{Twitter} \textit{COVID-19 streaming endpoint}, thus targeting tweets concerning the current epidemic situation~\cite{covid19stream}. The stream provides an incredible amount of content, which was filtered in order to only keep track of ``geotagged'' activities, meaning activities whose geographical localization was derived directly from the sending device and not algorithmically. In order to have a specific dataset on the topic of vaccines, the dataset was furtherly filtered to specifically target terms concerning immunization (the equivalents in different languages of the words ``vaccination'', ``vaccine'', ``vax'', ecc) and the names of the most popular vaccines: \textit{Pfizer-Biontech}, \textit{Moderna}, \textit{AstraZeneca}, \textit{Sputnik}, \textit{Johnson \& Johnson}. The full list of words and considered declinations can be found in the Appendix section of Gallotti \textit{et al.}~\cite{gallotti2021comment} (Section: Overview of social media data).
	
	The initial dataset at our disposal contained information on $750698$ tweets by $321180$ different users from September the 1st 2020 to July 15 2021. The available information on each tweet includes:
	\begin{itemize}
		\item TweetID: single numeric identifier for the activity
		\item User: unique identifier for the user involved (\textit{tweeter})
		\item Timestamp: date and time of the activity
		\item isBot: a flag whose value is equal 1 if the User is marked as bot and 0 otherwise. See Section~\ref{section:appendix:dataset:bot} for further details.
		\item Text: text of the tweet
		\item Location: country code associated to the position of the user at the time of the activity
		\item a flag that is equal to $1$ if the text contains explicit mention to one of each of the following vaccine names: \textit{Pfizer}, \textit{Moderna}, \textit{AstraZeneca}, \textit{Sputnik}, \textit{Johnson \& Johnson}
	\end{itemize}
	Further details on the data can be found in~\cite{infodemy}.
	
\begin{figure}[!ht]
		\centering
		\begin{subfigure}[h!]{0.9\textwidth}
		   \begin{subfigure}[b]{0.5\textwidth}
    			\centering
    	    	\includegraphics[width=\textwidth]{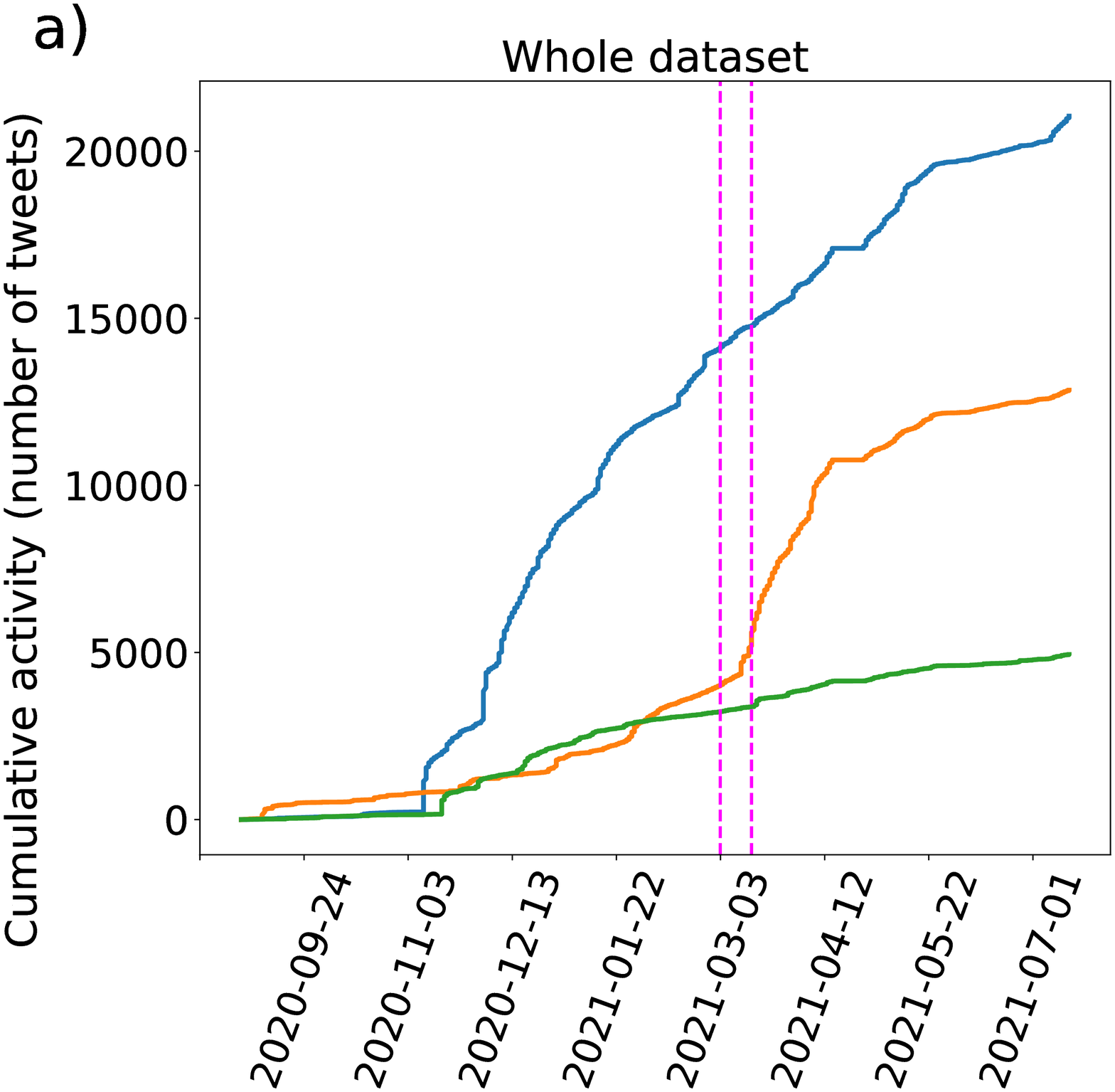}
    		\end{subfigure}
    		\begin{subfigure}[b]{0.5\textwidth}  
    			\centering 
    			\includegraphics[width=\textwidth]{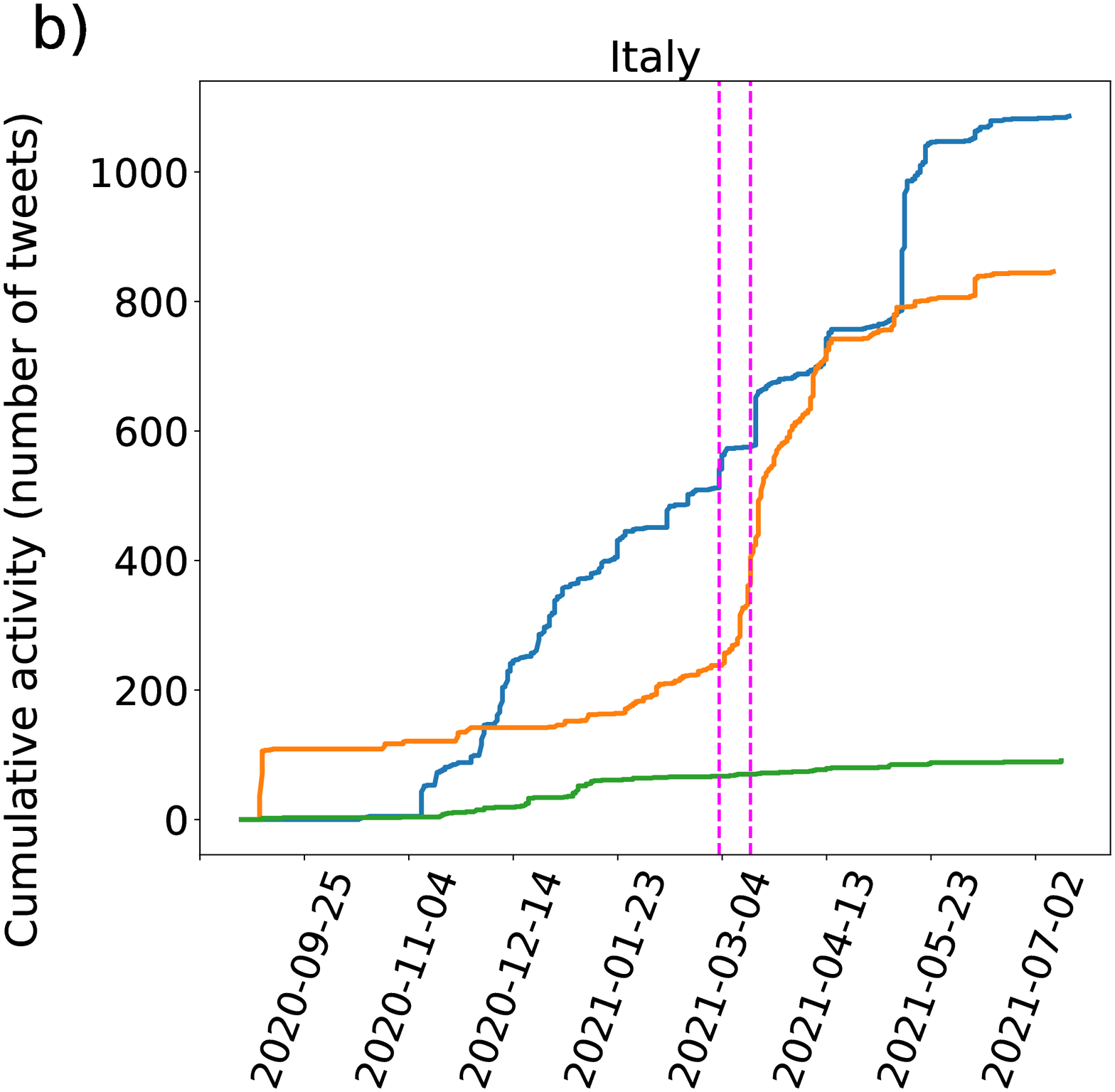}
    		\end{subfigure}
    		\begin{subfigure}[b]{0.5\textwidth}   
    			\centering 
    			\includegraphics[width=\textwidth]{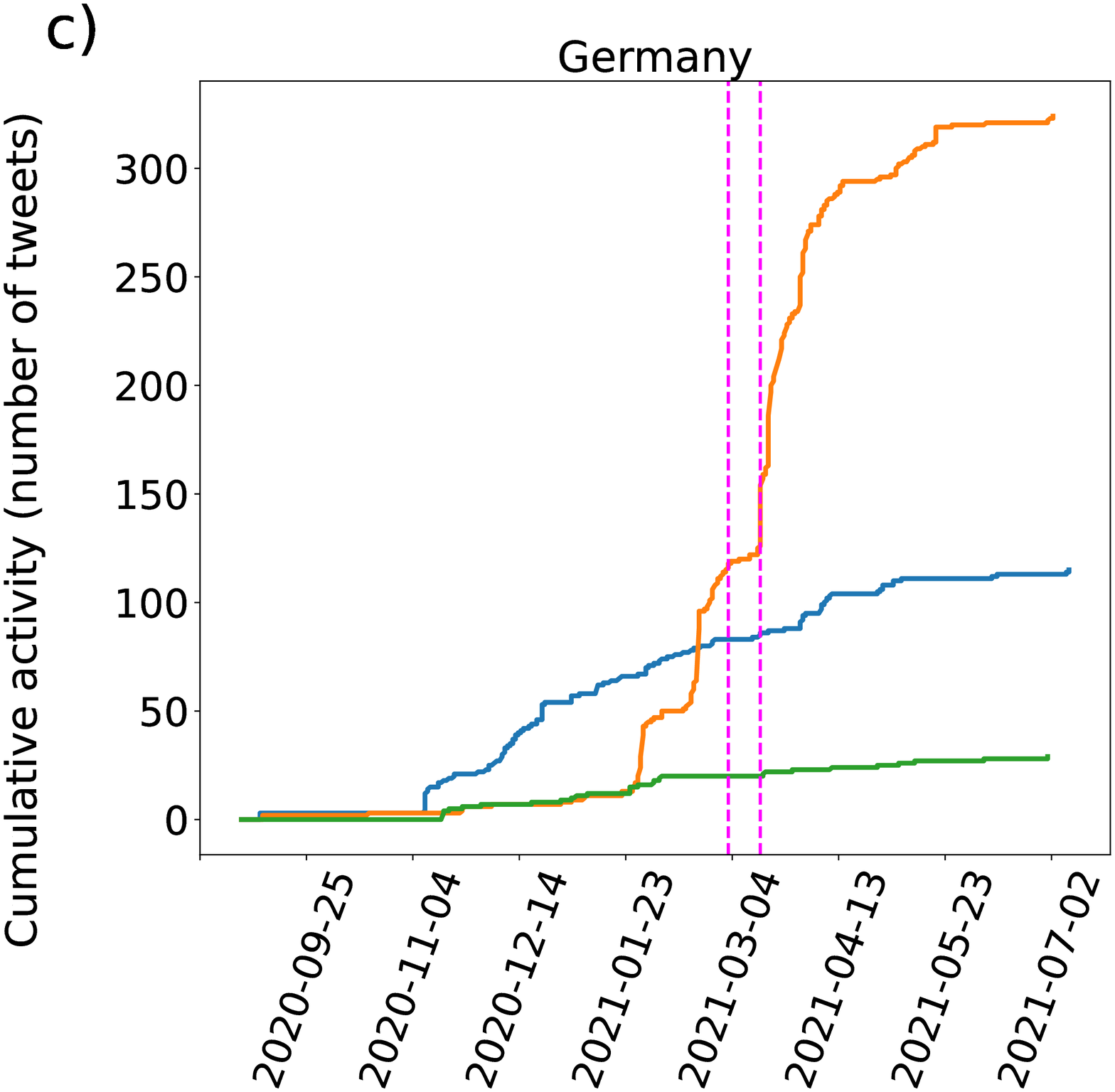}
    		\end{subfigure}
    		\begin{subfigure}[b]{0.5\textwidth}   
    			\centering 
    			\includegraphics[width=\textwidth]{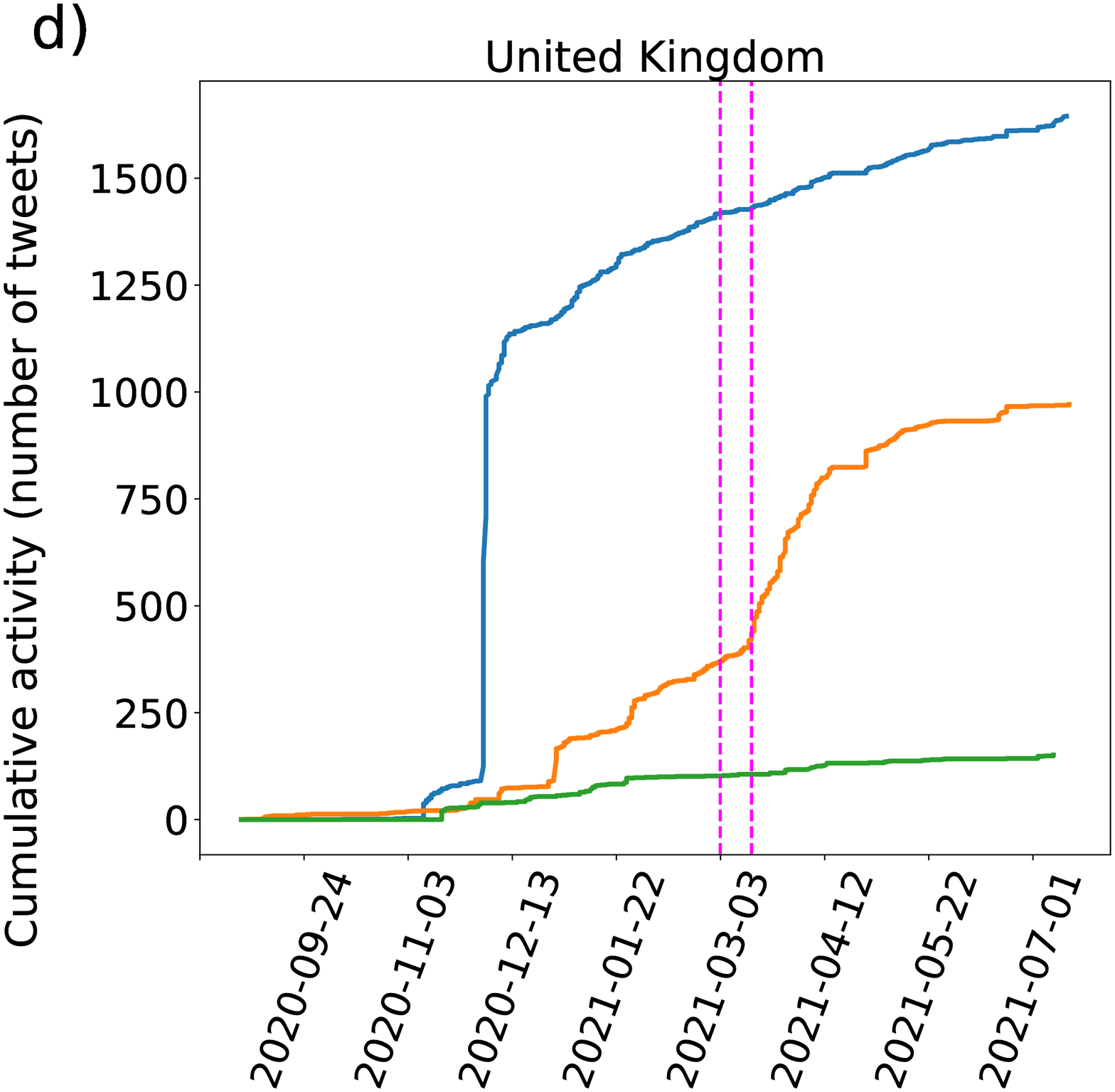}
    		\end{subfigure} 
		\end{subfigure}
		\begin{subfigure}[h!]{0.3\textwidth}
		\centering
			\includegraphics[width=0.95\linewidth]{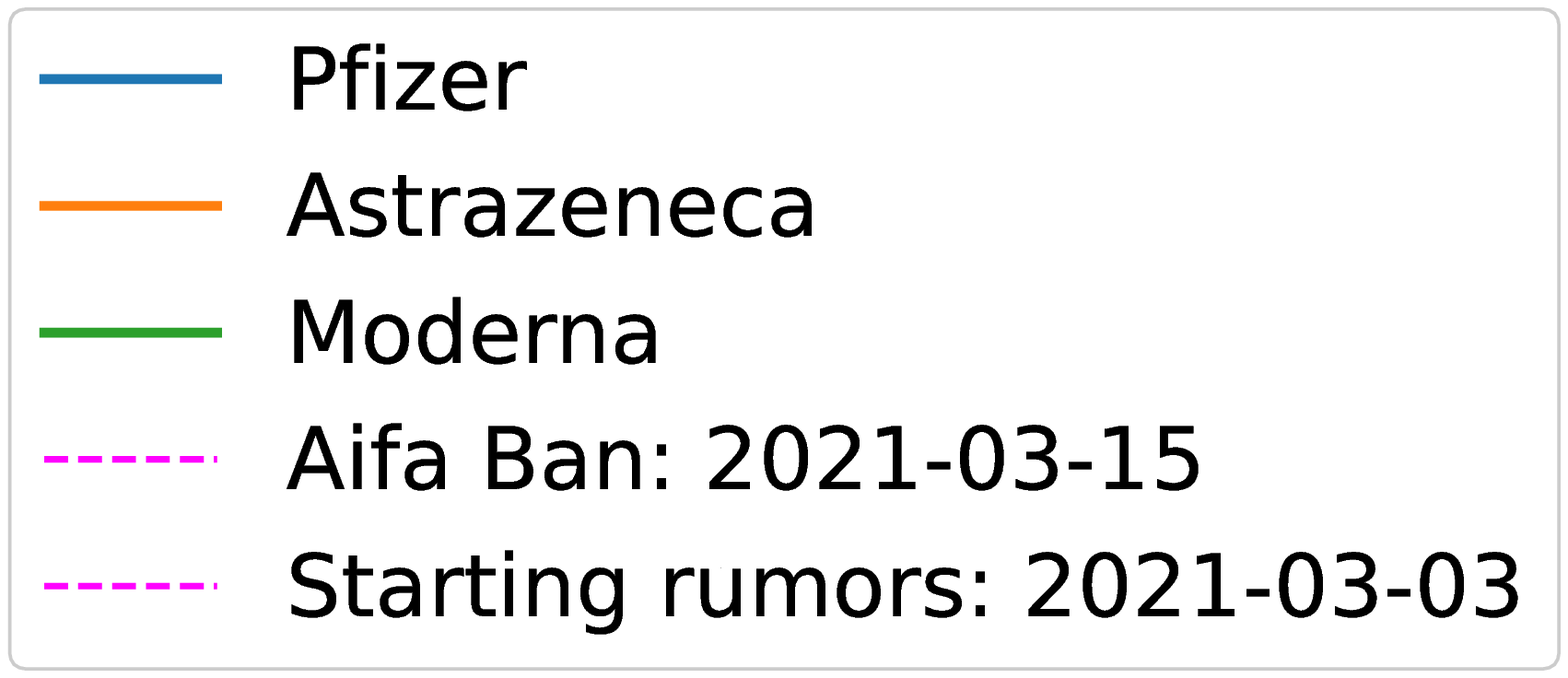}
			
		\end{subfigure}
		\caption{Cumulative number of tweets mentioning a given vaccine for different vaccines and countries. Notice the steep increases in the \textit{AstraZeneca} curves in Italy and Germany around the beginning of March 2021.}
		\label{img:FigureC1}
	\end{figure}

	The cumulative activity related to the vaccination discussion on Twitter can be found in Figure~\ref{img:FigureC1}, which shows the comparison among the cumulative tweet volumes for different countries. The colors refer to different vaccines. It is possible to notice that, compared to other countries and vaccines, the activity related to the keyword \textit{AstraZeneca} in Italy shows a great increase in the first days of March 2021, in correspondence of the ban imposed by \textit{AIFA}. The vaccine underwent a similar ban in Germany and it is indeed possible to notice that a similar trend is shown in this case, although the volumes remain more contained. Volumes are higher in the case of the United Kingdom, although we do not notice a change in the steepness of the curve as the one seen for Italy and Germany. The case of the United Kingdom is particularly interesting since, differently from other European countries, the UK government never put a ban on the \textit{AstraZeneca} vaccine.
	
	In the following paragraphs we illustrate the filtering procedure to obtain the datasets used in Section~\ref{section:data} in the main text. The use we make of the dataset is twofold: it provides the data against which we test our model and it also provides the information to extend the model in a data-driven direction. For this reason we will be interested in producing two datasets: the number of tweets per timeslot during a certain period of time and the distribution associated to the activity of users with the same number of followers in a timeslot. Firstly we remove the activity of users marked as \textit{bots} (Section~\ref{section:appendix:dataset:bot}). We then reduce our dataset to a subset of activities located in \textit{Italy} (Section~\ref{section:appendix:italy_dataset}) and finally extrapolate the information needed for our analysis. 
	
	\subsection{Removing \textit{bot} accounts }\label{section:appendix:dataset:bot}
	The word \textit{bot} (short for \textit{robot}) generally refers to an artificial agent that performs some continuous activity, usually in online environments. There are many different definitions of what an artificial agent is, how it distinguishes itself from a program and what it does ~\cite{what_is_bot1,what_is_bot2}, but in general the definition does not include a specific positive or negative connotation of its doing. However, in the context of online social media platforms the activity of bots is often associated to the promotion of fake news, spam and other disruptive ``behaviors''. Automated and non-automated accounts exhibit some differences in their behavior, making it possible to distinguish between the two. We do not get into the details of how the users were flagged in our dataset but for more information the authors of the dataset explain the procedure in the Supplementary Materials of~\cite{bot_ref} and to \cite{stella2018bots,stella2019influence,gonzalez2021bots,gonzalez2022advantage}.
	
	Since \textit{bots} are generally associated to some form of anomalous behavior, either it being the amount of content they produce or the frequency at which it is created, we decided to exclude from our analysis the activities associated to automated accounts. We are interested in users' online activity as an expression of their interests and possible behaviors, and we considered therefore natural to remove bot accounts' contributions.
	
	After the removal we count $421186$ tweets produced by $222690$ users, meaning that the removed users accounted for roughly $30\%$ of the total userbase and $44\%$ of the total activity.
	
	\subsection{Italy-based dataset}\label{section:appendix:italy_dataset}
	
	\begin{figure}[!ht]
		\centering
		\begin{subfigure}[h!]{0.9\textwidth}
		   \begin{subfigure}[b]{0.5\textwidth}
    			\centering
    	    	\includegraphics[width=\textwidth]{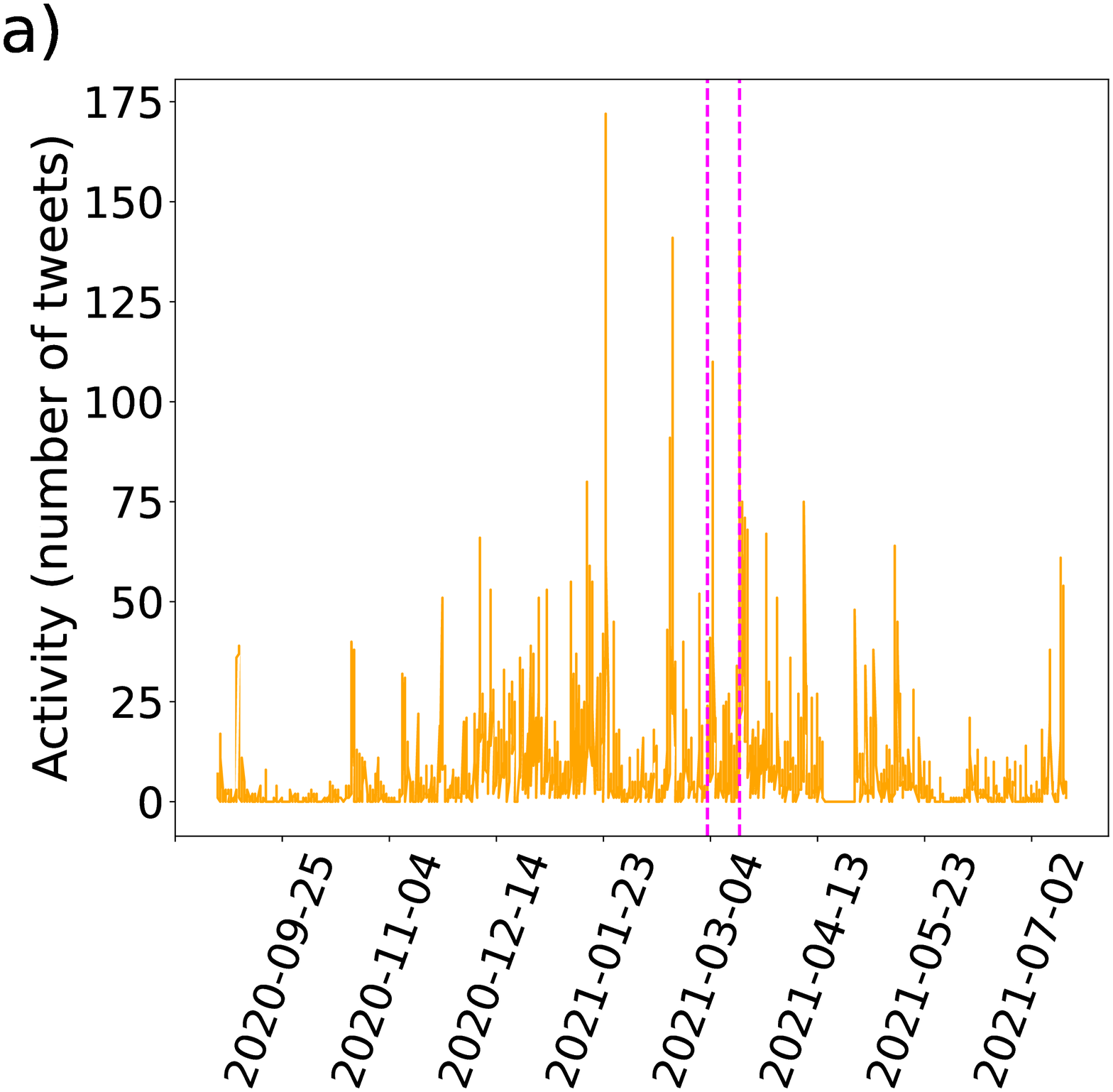}
    		\end{subfigure}%
    		\begin{subfigure}[b]{0.5\textwidth}  
    			\centering 
    			\includegraphics[width=\textwidth]{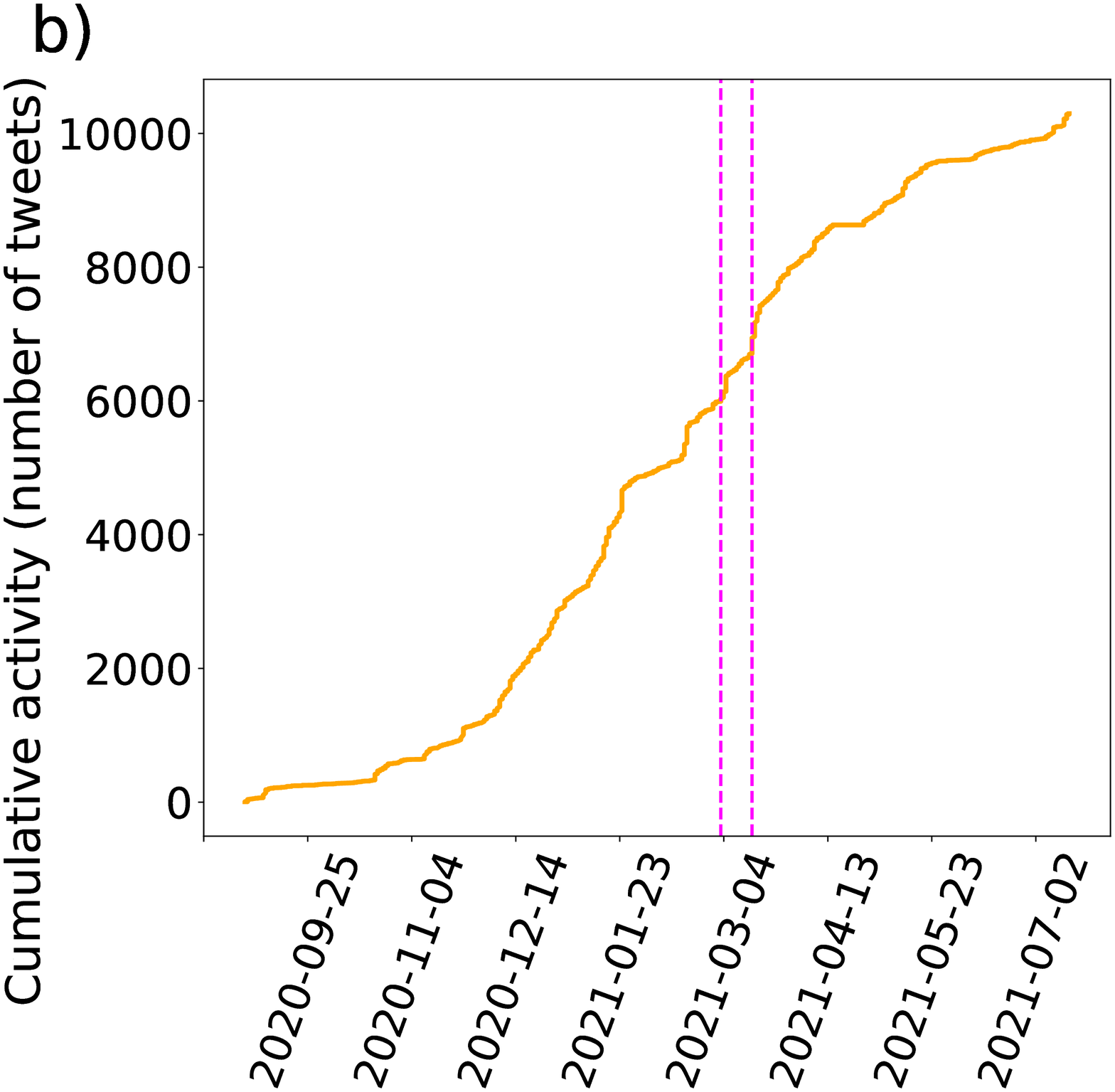}
    		\end{subfigure}\\
    		\begin{subfigure}[b]{0.5\textwidth}   
    			\centering 
    			\includegraphics[width=\textwidth]{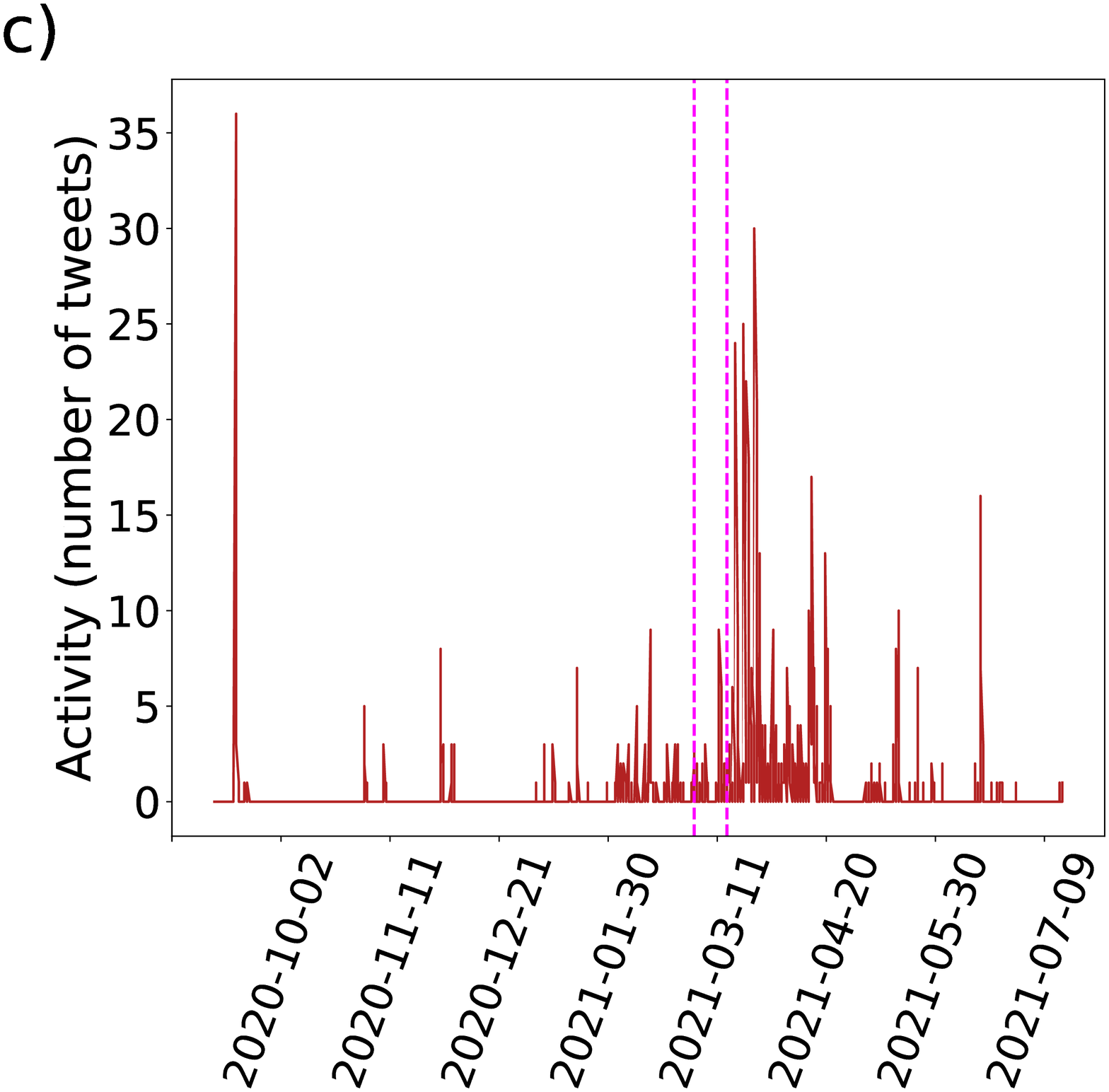}
    		\end{subfigure}%
    		\begin{subfigure}[b]{0.5\textwidth}   
    			\centering 
    			\includegraphics[width=\textwidth]{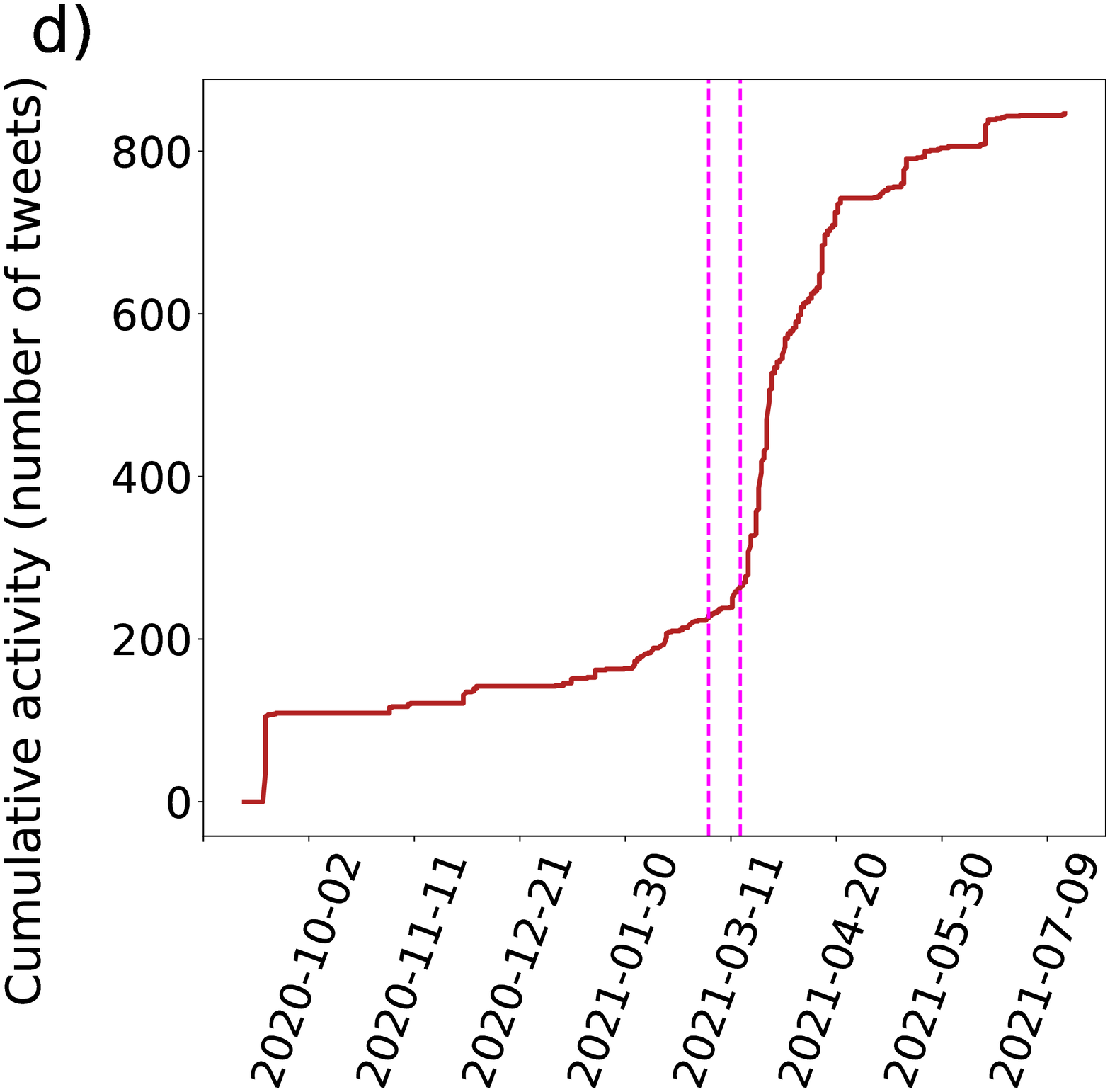}
    		\end{subfigure} 
		\end{subfigure}
		\begin{subfigure}[h!]{0.3\textwidth}
		\centering
			\includegraphics[width=0.95\linewidth]{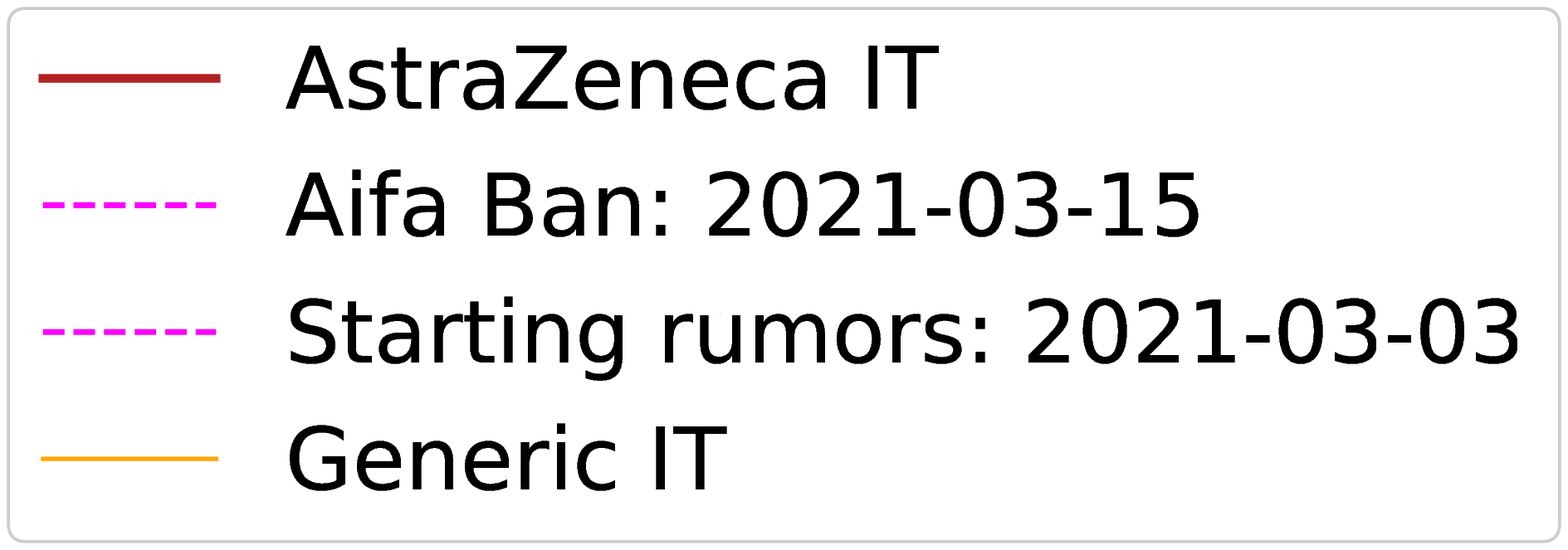}			
		\end{subfigure}
		\caption{Activity (number of tweets) and cumulative activity of tweeters in Italy: the first row includes all the topics surrounding vaccination whereas the second row only includes tweets mentioning \textit{AstraZeneca} }
		\label{img:FigureC2}
	\end{figure}
	
    By selecting only activity corresponding to the Italian \textit{TweetCountryCode} ``IT'' the dataset amounts to $10295$ Tweets by $4078$ users. By selecting a custom timeframe of activity of $6$ hours we can analyze the activity of users over time. It was checked that different choices for the length of the timeslot do not change the overall trend exhibited by the data.
	
	Figure~\ref{img:FigureC2} shows both the amounts of tweets (first column) produced in each timeslot of $6$ hours and the cumulative value (second column). The first row refers to the whole Italian dataset, whereas the second row only includes activities in Italy specifically mentioning \textit{AstraZeneca}. The comparison between model and data carried out in Section \ref{section:results} in the main text refers to posts produced by Italian users specifically mentioning the \textit{AstraZeneca} vaccine, hence to the dataset shown in the lower part of Figure \ref{img:FigureC2}. Such dataset, in the considered timeframe (hence between February 9, 2021 and April 12, 2021), amounts to 501 tweets. Note that this dataset is independent of the choice of keywords made to filter the output of the \textit{Covid-19} endpoint. As explained at the beginning of \ref{section:appendix:dataset} the output of the \textit{Twitter Covid-19 streaming endpoint} has been furtherly filtered to mirror the discussion around vaccines. The keywords chosen for this procedure include vaccine-related words and the name of the most popular vaccines at the time of analysis. Since the term \textit{AstraZeneca} is explicitly included in this list, our dataset is not affected by the choice of words, nor by their possible declination, an issue that could be relevant in some languages.
	
	\subsubsection{Cumulative activity dataset}
	
	After choosing an appropriate time window $\overline{t}$ (in our case $\overline{t} = 6$ hours) in order to be able to bin the users' activity over time, we are interested in building a dataset reporting for each timeslot the total number of tweets mentioning \textit{AstraZeneca} produced up to that time. This dataset is the one against which our model will be tested. In principle we are interested in the data displayed in Figure~\ref{img:FigureC2}d. However, we notice a peak in activity around September 2020 followed by very low-to no activity until the end of January 2021. We decided to consider the first peak in activity as an outlier, as it happened way outside our period of interest. We also decided to remove all those spurious activities between September 2020 and the beginning of February 2021. The starting date of our dataset is February 9 2021 and the ending date is April 12 2021. We decided to not include the final part of the activity as it corresponds to a ``physiological'' decrease in interest that is not accounted for in our model. Our model only aims at modeling an abrupt shift in interest due to broadcasting and it is thus not able to account for a subsequent loss of popularity. 
	
	\subsection{Distribution of the activity in a timeslot given the user's followers count}\label{section:appendix:dataset:pkt}
	As mentioned in Section~\ref{section:results} in the main text we extend our model so that it not only accounts for the fraction of users interested in a certain topic but also for the resulting number of posts they produce. Our focus is thus shifted from a user-centered perspective to an activity-centered one.
	
	We decided to adopt a data-driven approach to this extension and extrapolate information on users' activity behavior directly from the dataset at our disposal~\cite{artime2017}, collecting the amount of content a user produces in a timeslot directly from the dataset. We connect this dataset and our model through each user's degree, sampling users' activity according the number of their followers.
	
	With another dataset at our disposal containing the followers count of the users involved in the activity dataset, we built the data-driven distribution $p_{k,\overline{t}}$. Let $k$ be the followers count of a user and $\overline{t}$ the chosen timeframe (in our case, $6$ hours): for each $k$, $p_{k,\overline{t}}$ is the collection of the number of posts produced by every user whose followers count amounts to $k$ in a timeslot of duration $\overline{t}$. Empty timeslots, i.e. timeslots in which a user has not produced any content, count as $0$. Note that the number of followers of a user is, from a network perspective, the in-degree of the corresponding node. However we are working with an undirected network and thus it just corresponds to the degree.
	
	In order to build $p_{k,\overline{t}}$ we decided to include the activity related to vaccines in Italy in general (and thus the activity reported in the first row of Figure~\ref{img:FigureC2}). Considering mentions to all the vaccines allows us to account for a more general activity pattern and to build a wider and more reliable distribution for the activity. 
	
	\section{Particle Swarm Optimization}\label{section:PSO}
	In this section we provide a general introduction to the \textit{Particle Swarm Optimization} method. In the last paragraph of this section further details on how this technique was employed for this work are given.
	
	Firstly introduced by J. Kennedy and R. Eberhart in 1995, 
	\textbf{Particle Swarm Optimization} (\textit{PSO})\cite{PSO} is a \textit{meta-heuristic}, \textit{population-based} optimization algorithm. Compared to exact optimization methods, meta-heuristic algorithms provide model-independent frameworks or procedures for optimization. They are employed to sample the optimization space efficiently in those cases in which very little is known about it: in fact, what makes them so appealing and widely, successfully used is that the function to be optimized does not necessarily have to be convex or differentiable, a condition that has to be necessarily met when using other non meta-heuristic methods, like, for example, gradient descent. This flexibility comes at a price, as meta-heuristic methods \textit{are not guaranteed} to find an optimal solution to the optimization problem. 
	
	Among meta-heuristic algorithms, \textit{PSO} falls under the category of the population-based ones, as the solution for the optimization problem is found considering many possibilities and selecting the best one. Conversely, single-solution methods, such as simulated annealing for example, focus on iteratively improving a single candidate solution.
	
	The idea underlying the \textit{PSO} algorithm is that of \textit{swarm intelligence}, a biology-inspired concept according to which collective behaviors may arise from random and local interactions among individuals in a group. Biological examples of such phenomenon are birds flocking, ant colonies and fish schooling. In the framework of optimization, a swarm is a set of candidate solutions that are, in turn, points in the optimization space. These points are moved according to some rules until an ending condition is met. The update rules depend on the specific method and will take into account both local and global information.
	
	There are many different \textit{PSO} methods. We will refer to \textit{Global Best Optimization}~\cite{PSO_review}, as this is the one we employed for the following analysis.
	
	\subsection{The algorithm}
	Let us consider a minimization process. This is the type of optimization we will be interested in, as we will ultimately minimize a cost function. We can define this process as:
	\begin{equation}
		\TextInMath{find} \quad \mathbf{b}\in \mathcal{S} \subseteq {\rm I\!R^d} \quad \TextInMath{s.t.} \quad \forall \mathbf{x} \in \mathcal{S}, \quad f(\mathbf{b}) \leq f(\mathbf{x})
	\end{equation}
	where $\mathcal{S}$ is the search space of the minimization procedure, it has dimension $d$ and for every dimension is defined by the boundaries ${\mathbf{x} : l_i \leq x_i \leq u_i}$. $f(\cdot)$ is the objective function to be minimized. 
	
	A swarm is a set of $n > 1$ particles moving in the $\mathcal{S}$ space. Each particle  $p$ at time (iteration) $t$ is characterized by three $d$-dimensional vectors:
	\begin{itemize}
		\item Position ($\mathbf{x}_t^p$): The position of each particle is a candidate solution for the minimization problem at that iteration.
		\item Velocity ($\mathbf{v}_t^p$): It defines the direction and intensity of the movement of each particle at every iteration towards a new position
		\item Personal best ($\mathbf{b}_t^p$): It is the best known position of particle $p$ up to iteration $t$, meaning the point in $\mathcal{S}$ visited by particle $p$ such that $f(\mathbf{b}_t^p)<f(\mathbf{x}_{h}^p) \quad \forall \, h < t$. It serves as memory to the particle.
	\end{itemize}
	
	These vectors are updated at each iteration according to specific update rules for each of them. The details of the update rule depend on the specific \textit{PSO} algorithm. In the case of \textit{global best optimization}, we have
	\begin{align}
			\mathbf{v}_{t+1}^p & = \omega \mathbf{v}_{t}^p + \varphi_{1t} R_{1t}^p(\mathbf{b}_t^p - \mathbf{x}_{t}^p) + \varphi_{2t} R_{2t}^p(\mathbf{g}_t - \mathbf{x}_{t}^p) \label{eq:PSOVelocity}\\
			\mathbf{x}_{t+1}^p &  = \mathbf{x}_{t}^p + \mathbf{v}_{t+1}^p\label{eq:PSOPosition}\\
			\mathbf{b}_{t+1}^p & = 
			\left\{
			\begin{array}{lr}
				\mathbf{x}_{t+1}^p \quad \TextInMath{if} \quad f(\mathbf{x}_{t+1}^p) < f(\mathbf{b}_{t}^p )\\
				\mathbf{b}_{t}^p \quad \TextInMath{otherwise}.
			\end{array}
			\right.  
			\label{eq:PSOBest}			
	\end{align}
	Equations \eqref{eq:PSOPosition} and \eqref{eq:PSOBest} read very easily: position is updated according to the velocity whereas the personal best of each particle changes only if a position that corresponds to a new, lower, local minimum is found.
	
	Equation~\eqref{eq:PSOVelocity} needs more explanation, as the new velocity is the sum of three contributions. Its first term accounts for the previous velocity weighted with an \textit{inertia term}, $\omega$. 
	The second term is called \textit{cognitive influence} and accounts for the memory of the particle, adjusting its current velocity towards its best known position $\mathbf{b}_{t}^p$. This term is weighted with a parameter $\varphi_{1t}$, called \textit{cognitive parameter}, whose value may or may not vary at each iteration. The term $R_{1t}^p$ is a random $d$-dimensional diagonal matrix with values in $[0,1]$ extracted at each iteration. Its role can be better understood once the third term in \eqref{eq:PSOVelocity}, related to the so-called \textit{social influence}. We refer to $\mathbf{g}_{t}$ as the \textbf{global best position} of the whole swarm, meaning that value of $\mathbf{x}_t^p$ corresponding to the lowest possible value of $f(\cdot)$ among all particles. This term accounts for the collective intelligence of the swarm, as each particle adjusts its velocity towards the common, global best position. The weight $\varphi_{2t}$ is called \textit{social parameter}. $R_{2t}^p$ is analogous to $R_{1t}^p$.
	This part of the update rule is the one that makes this implementation of \textit{PSO} a ``global best optimization'' algorithm, as the interaction at the collective level among the particles involves \textit{all} the particles of the swarm. Conversely, one may also choose to involve in the update of each particle's velocity just those particles that are close enough to it, which would be the case of \textit{local best optimization}.
	
	The relative strength between the social and cognitive parameters determines whether the swarm is moving in an exploratory regime (the cognitive parameter prevails and each particle moves fairly independently) or exploitative regime (social parameter prevails and the swarm's movement strongly leans toward the global best position). $\varphi_1$ and $\varphi_2$ are common to all particles and, depending on the details of the algorithm, can change value over time. For example, one might choose to adopt an exploratory regime at the beginning of the search and then slowly turn on the social parameter as iterations go on in order to favour convergence towards the same position.
	
	The terms $R_{1t}^p$ and $R_{2t}^p$ are two $d \times d$ diagonal matrices that are updated at each iteration for each particle. If the values on the diagonal are the same for each dimension, the cognitive and social parameters are scaled along the directions $\mathbf{b}_t^p - \mathbf{x}_t^p$ and $\mathbf{g}_t - \mathbf{x}_t^p$ respectively (\textit{linear particle swarm optimization}), whereas using different values for each dimension slightly changes the direction of update, improving the explorability of the space~\cite{clerc2010}.
	
	Normally, at the beginning of each search, the best position of each particle corresponds to its first position, which is initialized at random in the available space. Velocity is usually initialized to $0$. The updates of each particle are synchronous and happen for a fixed number of iterations or until a certain ending condition is met. The final position returned is the global best position, and the final cost function value is the one corresponding to the global best position. For our analysis, we will use \textit{mean square error} (MSE) between the data and the output of our model as the cost function $f(\cdot)$.
	
	\subsection{Python Implementation and hyperparameters search}
	For the purpose of this work, we used the \textit{PySwarms} python library. \textit{PySwarms} provides an optimization routine that performs the updates in Equations~\eqref{eq:PSOVelocity}---\eqref{eq:PSOBest} for a given number of iterations as well as the option to early stop the search if the function to be optimized does not improve beyond a certain threshold for a given number of iterations. The library also offers easy access to the optimization function's value history, as well as the particles' positions at each iteration.
	
	The $PySwarms$ library does not offer a method to check the convergence of the swarm: it was noticed that in many situations the returned set of parameters $\mathbf{g}$ does indeed correspond to a low value of the cost function but that the overall behavior of the swarm is not convergent, meaning that at the end of the search each particle is in a different position. This usually means that in a different run with the same set hyperparameters the swarm will not reliably find a good solution again. As an approximate measure of convergence we compute the mean distance between the particles over the last steps of the swarm's search and look for sets of hyperparameters that guarantee a convergence measure as low as possible.
	
	The library does include the possibility to change the value of the hyperparameters $\omega$, $\varphi_1$ and $\varphi_2$ at each iteration, as well as a set of possible update strategies. However, it is not possible to choose custom starting and ending points and it was necessary to slightly modify the library to be able to include this option, which was observed to be crucial for the algorithm to be stable and convergent. In particular we are interested in reproducing a trade-off between a more exploratory behavior at the beginning of the search, followed by a more exploitative phase that activates gradually as $\varphi_1$ decreases. 
	In order to find a set of hyperparameters that could guarantee good quality of the results we performed a random search over the three hyperparameters $\omega$, $\varphi_1$ and $\varphi_2$, sampling their starting value, their final value and the decay method, and over the number of particles $n$. Given the amount of time a single search requires and since we want to collect some statistics to be able to evaluate the performance of each set of hyperparameters, we adopted a two-step search strategy: in a preliminary phase of the search we explore the behavior of a high number of hyperparameters sets, collecting some statistics on the mean cost associated to the final set of parameters and the actual convergence of the swarm. We then select the three best sets according to convergence and cost and re-iterate the search for a larger number of repetitions in order to consolidate the previous statistics. Again, the final set of $\omega$, $\varphi_1$, $\varphi_2$ and $n$ values is selected according to convergence and cost. 
	
	\setcounter{figure}{0}
        \setcounter{table}{0}
	\section{Prevalence models and further results}\label{section:appendix:results} 
	In this section we provide further details on the results obtained in Section~\ref{section:results} in the main text. Moreover we provide an overview of additional models that we tests (\textit{symmetric rates and heterogeneous mixing}, \textit{asymmetrc rates and complete graph}) and discuss the results obtained.
	
	\subsection{Handling initialization}
	We decided to adopt the following methods for initializing the dynamics, i.e. for setting the initial conditions for the solution of Equations~ \eqref{eq:evolutionProbBehavBroadc}.
	\begin{itemize}
		\item \textbf{uniform initialization}: we consider an initial probability $p_i^A(t=0)=1/N$ for each node
		\item \textbf{hub initialization}: a single node $i$ in the network is initialized with $p_i^A(t=0)=1$. Such node is extracted at random among the $5$ nodes with highest degree.
	\end{itemize}

	Both initialization methods correspond to an initial prevalence $\rho(t=0)=1/N$.
	We chose this type of initialization since, in principle, one could assume that the whole dynamics starts from a single \textit{aware}/\textit{infected} node and then spreads to the other individuals. Since the moment in time at which this occurs is not known, we opted for the following strategy: in the \textit{PSO} parameters search we also include a \textit{padding} parameter, which accounts for the time at which the dynamics should, in principle, have started. The prevalence dynamics and then the cumulative number of tweets are thus computed for this extended interval and then appropriately cut so that the comparison with the data and the computation of the cost only involve the dynamics from the end of the padding onward.
	
	The padding parameter can have values ranging from $0$ (no padding) to $100$ days. This quantity was arbitrarily chosen looking at the activity plot for the discussion around vaccines and choosing approximately the date at which the discussion started. This date correspond to November 1st 2020.
	
	We care to highlight that the results obtained show no substantial influence of the padding parameter when dealing with symmetric rates: this is due to the fact that the sets of parameters that better seem to reproduce the data always resolve in the behavioral dynamics expiring before reaching the date at which we start the comparison with the data (February 9 2021). 
	
	\subsection{Model with symmetric rates and all-to-all approximation}\label{section:appendix:results:sym_ata}
	
	We recall the formula for the evolution of the probability of a single node to be in the \textit{aware} state $A$, as in Equation~\eqref{eq:evolutionProbBehavBroadc}:
	\begin{align}
		\difft{} p_i^A(t) = & - p_i^A(t) \left[ 2 + (\gamma + \beta)\Theta(t - \tau)p_i^b   - \prod_{j=1}^N \left( 1-\lambda A_{ij}p_j^A(t) \right) + \right. \nonumber\\
		& \left. - \prod_{j=1}^N \left ( 1-\mu A_{ij}(1 - p_j^A(t)) \right) \right] + \nonumber \\
		& + ( 1 - p_i^{\tilde{U}}) \left[1 - \prod_{j=1}^N(1-\lambda A_{ij}p_j^A) + \beta \Theta(t - \tau)p_i^b  \right].
		\label{eq:appendix:evolutionProbNodeIAware}
	\end{align}
	Recalling the change we need to perform over the parameters when moving into an all-to-all description and using symmetric parameters for the behavioral and broadcasting rates ($\mu = \lambda \longrightarrow \lambda$, $\gamma = \beta \longrightarrow \beta$) we obtained
	\begin{align}
		\difft{} p_i^A(t) = & - p_i^A(t) \left[ 2\beta\Theta(t - \tau) +\lambda \sum_{j=1}^N \frac{A_{ij}p_j^A}{N} +\lambda \sum_{j=1}^N \frac{A_{ij}(1 -p_j^A)}{N} \right] + \nonumber\\
		& + ( 1 - p_i^{\tilde{U}}) \left[ \lambda \sum_{j=1}^N \frac{A_{ij}p_j^A}{N} + \beta \Theta(t - \tau) \right].
		\label{eq:appendix:evolutionProbNodeIAwareSymATATemp}
	\end{align}
	
	To make computations faster we used uniform initialization and uniform probability for the zealot nodes. The \textit{all-to-all} scenario, where all individuals are connected among them, allows us to completely neglect network information from the start as the prevalence evolves with the same dynamics for every node. Equation~\eqref{eq:appendix:evolutionProbNodeIAwareSymATATemp} then becomes
	\begin{equation}
	\difft{} p_i^A(t) =  - p_i^A(t) \left(2\beta\Theta(t - \tau) +\lambda \right) + ( 1 - p_i^{\tilde{U}}) \left(\lambda p_i^A(t) + \beta \Theta(t - \tau) \right).
		\label{eq:appendix:evolutionProbNodeIAwareSymATA}
	\end{equation}
	Note that we consider the distribution $p_i^b$ to be uniform and we thus incorporated the fraction $p_i^b$ in the broadcasting parameters.

	\subsubsection{PSO hyperparameters search and further results}
	
	We now describe the procedure for the hyperparameters search during the \textit{PSO} procedure and some additional information regarding the results obtained. This section is meant to be complementary to Section~\ref{section:data} in the main text.
	
	Table~\ref{table:appendix:hyperparsSymATA} shows the best set of hyperparameters for the \textit{PSO} algorithm found using the two-step random search procedure explained in~\ref{section:PSO}. We sample the number of particles uniformly in an interval $n \in [5,15]$. For the inertia ($\omega$) and cognitive ($\varphi_1$) parameters we consider three types of decay possibilities (refer to the \textit{Pyswarms} library for further details): \textit{linear}, \textit{non-linear}, \textit{None} (the parameters do not decay). The social parameter $\varphi_2$ is sampled and then does not decay. All initial values of the parameters are sampled in an interval $[0,1]$. Final values are computed sampling a number in $[0,1]$ to be multiplied to the initial value.
	
	\begin{table}[!ht]
		\centering
		\begin{tabular}{|c|ccc|}
			\hline
			& \multicolumn{1}{c|}{initial value} & \multicolumn{1}{c|}{final value} & decay type \\ \hline
			$\omega$                         & \multicolumn{1}{c|}{0.29}          & \multicolumn{1}{c|}{0.038}       & linear \\ \hline
			$\varphi_1$ (cogntive parameter) & \multicolumn{1}{c|}{0.11}          & \multicolumn{1}{c|}{0.07}        & non linear     \\ \hline
			$\varphi_2$ (social parameter)   & \multicolumn{3}{c|}{0.43}                                                          \\ \hline
			particles                        & \multicolumn{3}{c|}{13}                                                            \\ \hline
		\end{tabular}
	\caption{Final hyperparameters for the homogeneous mean field model and symmetric rates.}
		\label{table:appendix:hyperparsSymATA}
	\end{table}
	
\begin{figure}[!ht]
		\centering
		\includegraphics[width=0.9\textwidth]{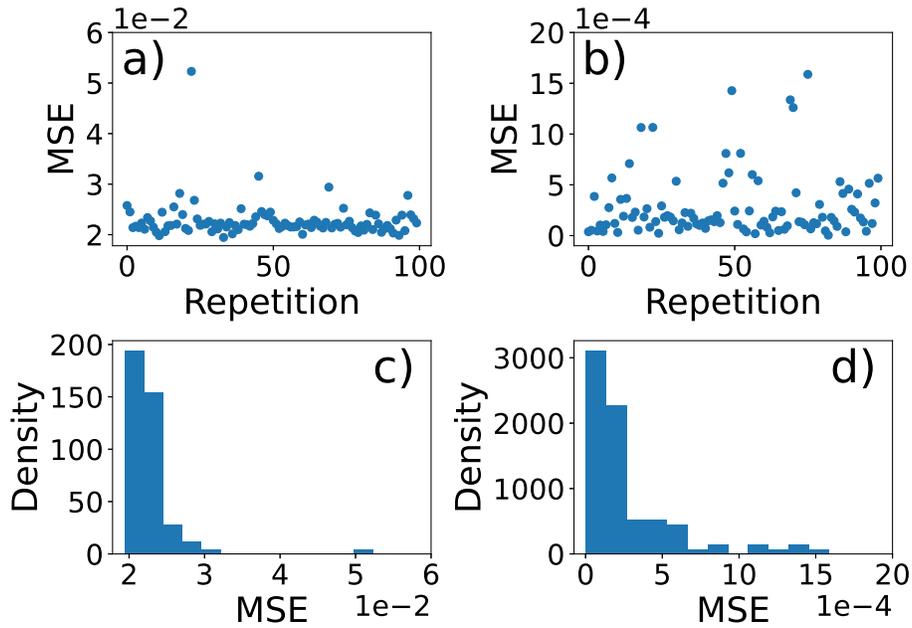}
		\caption{Box plot and histograms for the cost and final convergence measure over the $100$ repetitions of the \textit{Particle Swarm Optimization} procedure for the homogeneous mean field model and symmetric rates.}
		\label{img:FigureE1}
	\end{figure}

	Figure~\ref{img:FigureE1} shows the distribution of cost and convergence measure over the $100$ repetitions of the search. We believe that both measures show overall good values, hinting that the model manages to qualitatively reproduce the trend in the data.
	
	\subsection{Model in the heterogeneous mean-field approximation with symmetric rates}\label{section:appendix:results:hmf}
	
	As a further step to include topology information into the prevalence model we employed a heterogeneous mean-field approach. Being $k_i$ the degree of node $i$ and $\rho(t)$ the prevalence at time $t$, Equation~\eqref{eq:appendix:evolutionProbNodeIAwareSym} becomes
	\begin{equation}
		\difft{} p_i^A(t) = - p_i^A(t) \left(2\beta\Theta(t - \tau) +\lambda \right) + ( 1 - p_i^{\tilde{U}}) \left(\lambda k_i \rho(t) + \beta \Theta(t - \tau) \right).
		\label{eq:appendix:evolutionProbNodeIAwareSym}
	\end{equation}
	
	We build a synthetic scale-free network with $N=1000$ nodes and with exponential parameter for the degree distribution $\nu = 2.2$. Due to computational limitations, we decided to employ a smaller network compared to the case treated in Section~\ref{section:results} in the main text. We find that the results are stable under different systems sizes. In this case, hub initialization was used.
	
	\subsubsection{PSO hyperparameters and results}
	
	Table~\ref{table:appendix:hyperpars_HMF} reports the best set of hyperparameters for the \textit{PSO} algorithm.
	
	\begin{table}[!ht]
		\centering
		\begin{tabular}{|c|ccc|}
			\hline
			& \multicolumn{1}{c|}{initial value} & \multicolumn{1}{c|}{final value} & decay type \\ \hline
			$\omega$                         & \multicolumn{1}{c|}{0.028}         & \multicolumn{1}{c|}{0.017}       & non linear \\ \hline
			$\varphi_1$ (cognitive parameter) & \multicolumn{1}{c|}{0.11}         & \multicolumn{1}{c|}{0.018}       & linear     \\ \hline
			$\varphi_2$ (social parameter)   & \multicolumn{3}{c|}{0.75}                                                          \\ \hline
			particles                        & \multicolumn{3}{c|}{10}                                                            \\ \hline
		\end{tabular}
            \caption{Final hyperparameters for the heterogeneous mean-field approximation with symmetric rates.}
		\label{table:appendix:hyperpars_HMF}
	\end{table}
	
\begin{figure}[!ht]
		\centering
		\includegraphics[width=\textwidth]{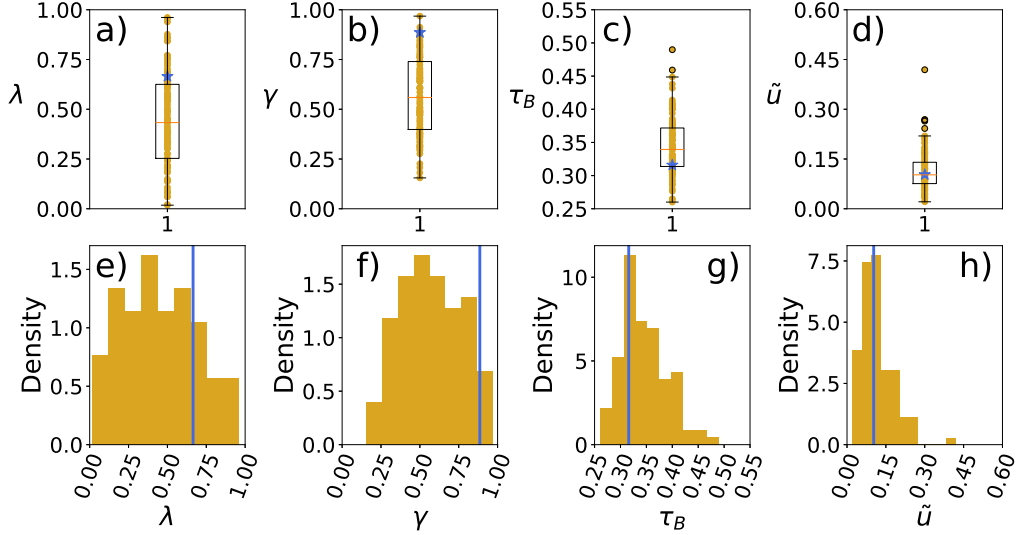}
		\caption{Distribution of occurrences of the parameters for the heterogeneous mean-field approximation with symmetric rates. Top row: box plots showing the distribution of the occurrences of the four parameters throughout the repetitions of the \textit{Particle Swarm Optimization} procedure. The four parameters correspond to the behavioral transition rate $\lambda$, the broadcasting transition rate $\gamma$, the broadcasting onset $\tau_B$ and the fraction of unaware nodes $\tilde{u}$. Bottom row: the same distributions in the form of histograms. The blue marks represent the values found in the set with lowest cost.}
		\label{img:FigureE2}
	\end{figure}

\begin{figure}[!ht]
		\centering
		\includegraphics[width=0.9\textwidth]{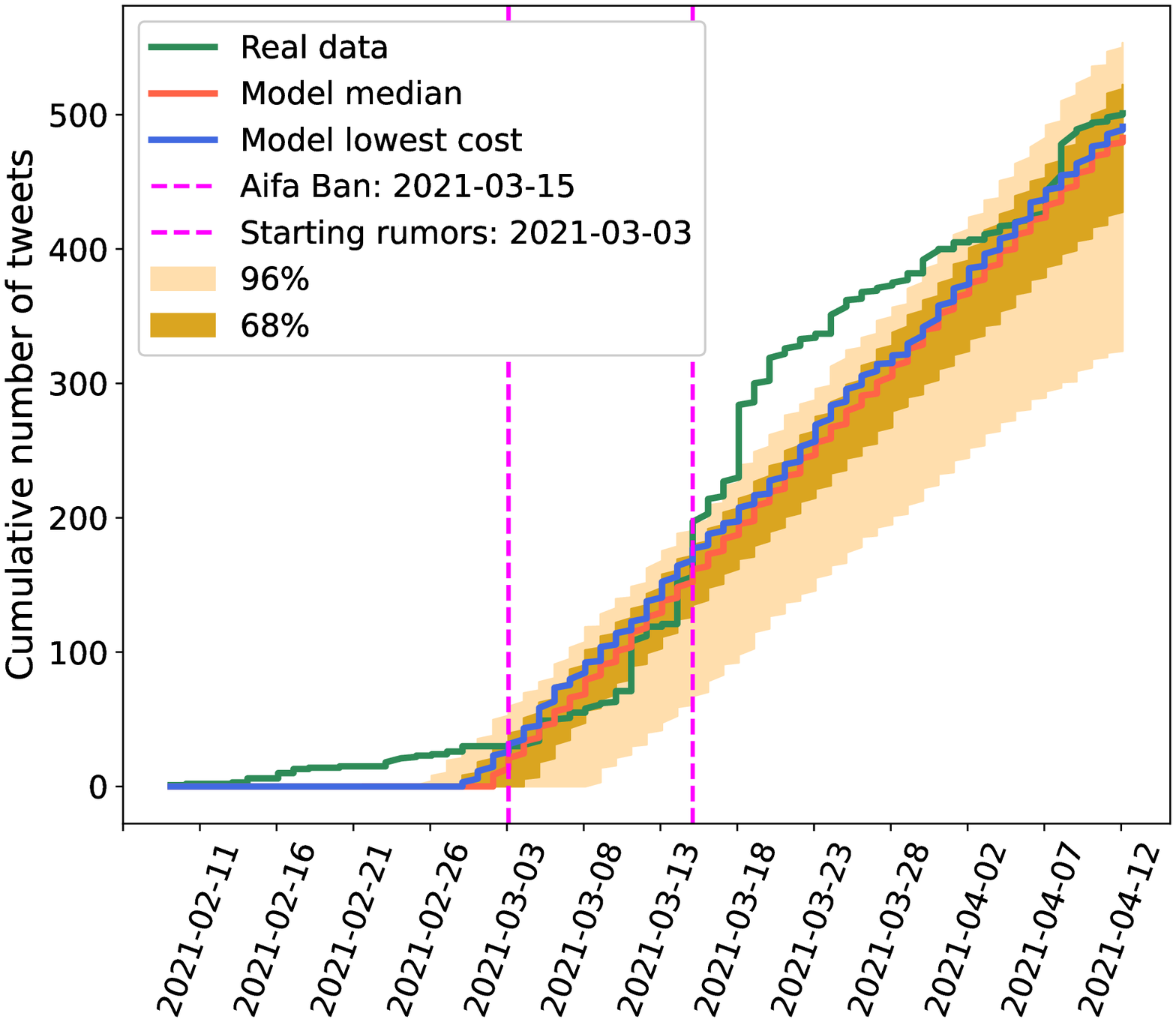}
		\caption{Comparison between the fitted model and the dataset (in green) for the heterogeneous mean-field approximation with symmetric rates. The blue line corresponds to the model associated to the set of parameters with lowest cost. The red line corresponds to the model obtained using the median parameters. The different shades of yellow aim at reproducing the same information as in the box plot: the cumulative curve was computed for all the $100$ parameters sets, thus the red line correspond to the median values of the curves at each time and the shaded yellow areas to the $68\%$ quantile (dark yellow) and $98\%$ (light yellow) quantile respectively. Mind that the red and yellow lines do not correspond to the model obtained using the median set of parameters and might not correspond to any particular set of parameters in general.}
		\label{img:FigureE3}
	\end{figure}

\begin{figure}[!ht]
		\centering
		\includegraphics[width=0.9\textwidth]{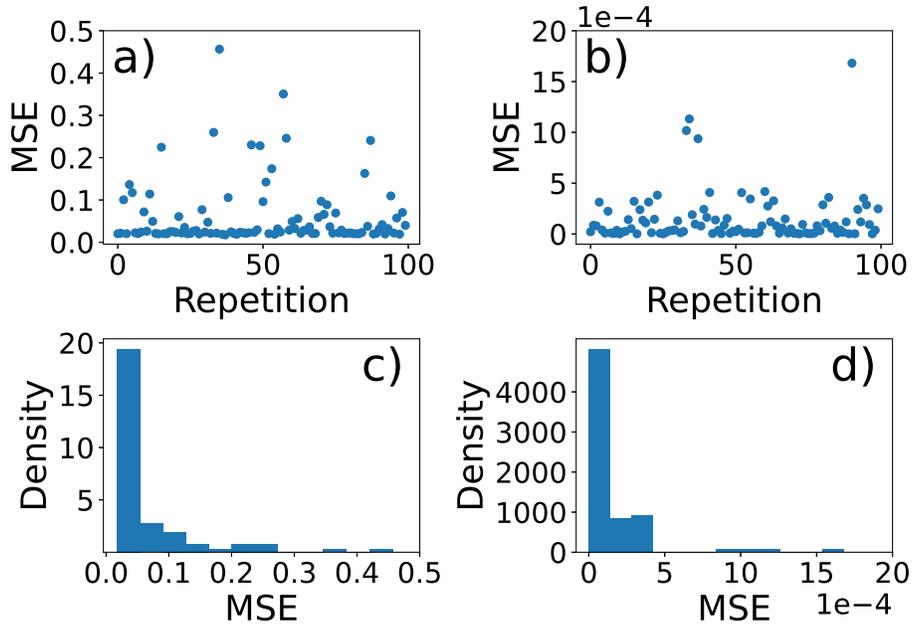}
		\caption{Box plot and histograms for the cost and final convergence measure over the $100$ repetitions of the \textit{Particle Swarm Optimization} procedure for the heterogeneous mean-field approximation with symmetric rates.}
		\label{img:FigureE4}
	\end{figure}

	Figure~\ref{img:FigureE2} shows the distribution of the behavioral parameter $\lambda$, the broadcasting parameter $\gamma$, the broadcasting starting time $\tau$ and the fraction of zealot nodes $\tilde{u}$. The blue marks correspond to the lowest cost set, whose values are reported in Table~\ref{table:appendix:HMF_lowest_median} along with the median. In the heterogeneous case we find that not all the values of the parameters in the set with lowest cost fall between the $Q_1$ and $Q_3$ quantiles. In particular we notice a poorer performance of the $\lambda$ and $\gamma$ parameter.
	Figure~\ref{img:FigureE3} shows the comparison between the dataset and the model. Compared to the results in Section~\ref{section:results} in the main text we notice more variability. The overall lower precision of these results, also in terms of MSE (Figure~\ref{img:FigureE4}) and the great variability that $\lambda$ and $\gamma$ show could be due to the fact that we did not place any bounds on those dynamical parameters. We find that with \textit{PSO} it can be very important to be able to make some guesses on the values of the parameters of the model and to place the corresponding bounds during the search. Better results could be also obtained using asymmetric rates (see~\ref{section:appendix:results:asym_ata}).
	
	\begin{table}[!ht]
		\centering
		
  \begin{tabular}{|l|c|c|c|c|c|}
			\hline
			& $\lambda$ & $\gamma$ & $\tau$             & $\tilde{u}$ & MSE   \\ \hline
			lowest cost & 0.6      & 0.88    & February 28 2021 & 0.1        & 0.018 \\ \hline
			median      & 0.43      & 0.56    & March 2nd 2021 & 0.10        &       \\ \hline
		\end{tabular}
          \caption{Lowest cost and median parameters for the heterogeneous mean-field approximation with symmetric rates. Mind that there is no MSE associated to the median set of parameters as the median is computed for every single parameter and not as a set.}
		\label{table:appendix:HMF_lowest_median}
	\end{table}
	
	\subsection{Model in the all-to-all approximation with asymmetric rates}\label{section:appendix:results:asym_ata}
	
	If we lift the symmetric rates assumption from Equation~\eqref{eq:appendix:evolutionProbNodeIAwareSymATATemp}, we obtain
	
	\begin{align}
		\difft{} p_i^A(t) = & - p_i^A(t) \left[ (\beta + \gamma)\Theta(t - \tau) +(\lambda - \mu)p_i^A +\mu \right] + \nonumber \\
		&+ ( 1 - p_i^{\tilde{U}}) \left[ \lambda p_i^A + \beta \Theta(t - \tau) \right]
		\label{eq:appendix:evolutionProbNodeIAware_asym_ata_temp}
	\end{align}
	For the initialization, the degree assignment and the sampling of $p_{k,\overline{t}}$ distribution, we use the same method described in Section~\ref{section:results} in the main text.
	
	\subsubsection{PSO hyperparameters search and results}
	
\begin{figure}[!ht]
		\centering
		\includegraphics[width=\textwidth]{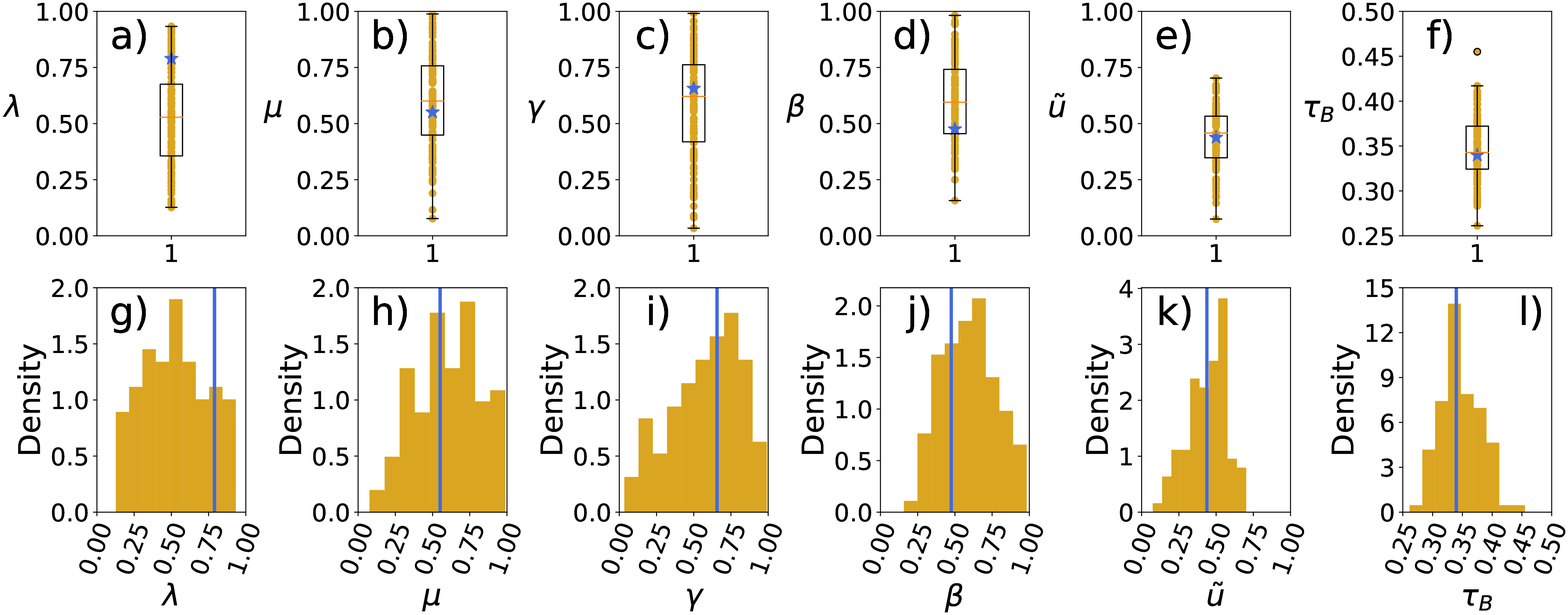}
		\caption{Distribution of occurrences of the parameters for the all-to-all approximation with asymmetric rates. Top row: box plots showing the distribution of the occurrences of the four parameters throughout the repetitions of the \textit{Particle Swarm Optimization} procedure. The four parameters correspond to the behavioral transition rate $\lambda$, the broadcasting transition rate $\gamma$, the broadcasting onset $\tau_B$ and the fraction of unaware nodes $\tilde{u}$. Bottom row: the same distributions in the form of histograms. The blue marks represent the values found in the set with lowest cost.}
		\label{img:appendix:results:ASYM_ATA_boxplot}
	\end{figure}

	Table~\ref{table:appendix:hyperpars_ASYM_ATA} reports the best set of hyperparameters for the \textit{PSO} search.
	Figure~\ref{img:appendix:results:ASYM_ATA_boxplot} shows the distribution of the parameters: as in the case with symmetric rates we find good adherence between the point corresponding of the maximum of the distribution and the lowest cost values, hinting that the swarm tends to converge towards a good set of parameters in terms of cost. Figure~\ref{img:FigureE7} shows that the MSE values and the convergence measures are similar to the symmetric case. Again, we believe this shows that there is good accordance between the output of the model and the data.
	
\begin{figure}[!ht]
		\centering
		\includegraphics[width=0.9\textwidth]{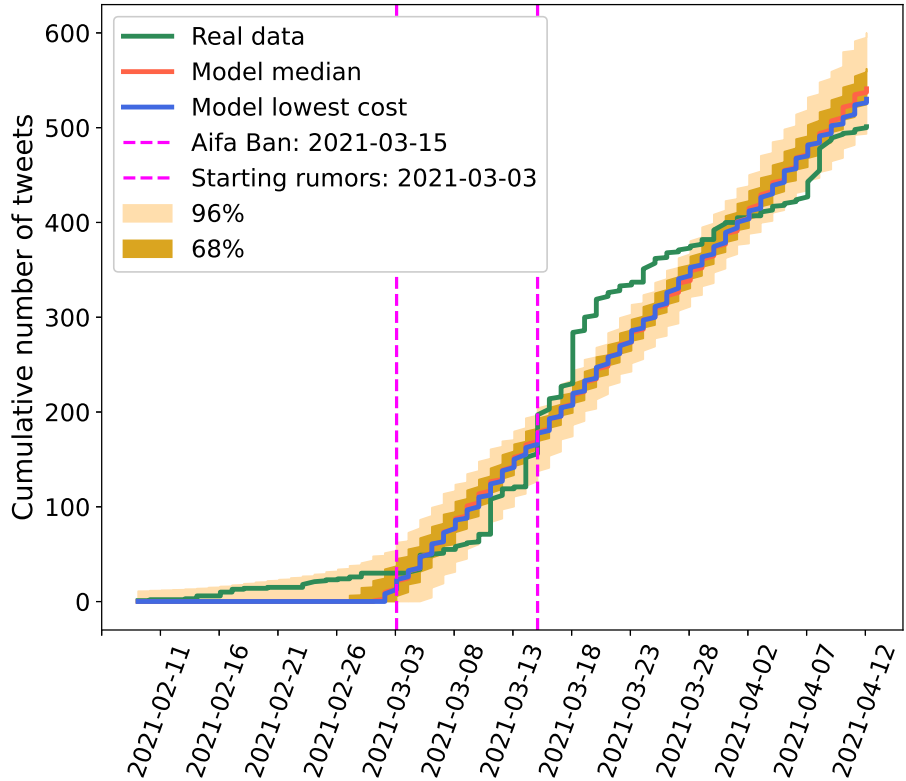}
		\caption{Comparison between the fitted model and the dataset (in green) for the all-to-all approximation with asymmetric rates. The blue line corresponds to the model associated to the set of parameters with lowest cost. The red line corresponds to the model obtained using the median parameters. The different shades of yellow aim at reproducing the same information as in the box plot (see previous figures for further explanation).Note that differently from previous cases the introduction of an asymmetry in the process parameters seems to be able to better reproduce the first part of the dynamics, before the activation of broadcasting.}
		\label{img:FigureE6}
	\end{figure}

\begin{figure}[!ht]
		\centering
		\includegraphics[width=0.9\textwidth]{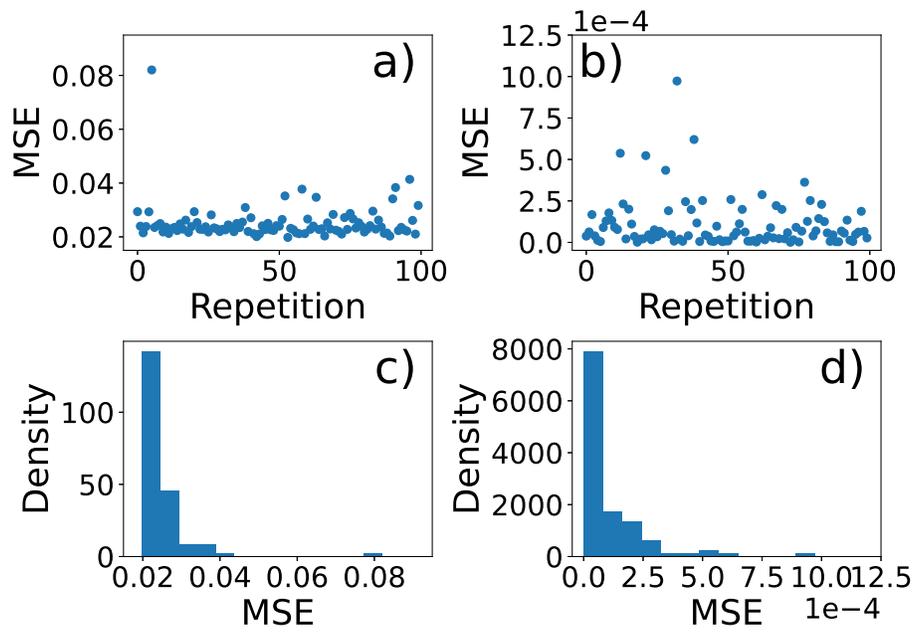}
		\caption{Box plot and histograms for the cost and final convergence measure over the $100$ repetitions of the \textit{Particle Swarm Optimization} procedure for the all-to-all approximation with asymmetric rates.}
		\label{img:FigureE7}
	\end{figure}

	The substantial difference between this model and the previous ones is that it is possible to notice from Figure~\ref{img:FigureE6} that in this case there are some configurations of the model that also capture the initial behavioral dynamics alone. In these cases the behavioral dynamics does not die out before our starting date (February 9 2021) and this is due to the asymmetry in the parameters.

	\begin{table}[!ht]
		\centering
		\begin{tabular}{|c|ccc|}
			\hline
			& \multicolumn{1}{c|}{initial value} & \multicolumn{1}{c|}{final value} & decay type \\ \hline
			$\omega$                         & \multicolumn{1}{c|}{0.13}         & \multicolumn{1}{c|}{0.06}       & linear       \\ \hline
			$\varphi_1$ (cogntive parameter) & \multicolumn{3}{c|}{0.13}                                                          \\ \hline
			$\varphi_2$ (social parameter)   & \multicolumn{3}{c|}{0.87}                                                          \\ \hline
			particles                        & \multicolumn{3}{c|}{11}                                                            \\ \hline
		\end{tabular}
	\caption{Final hyperparameters for the asymmetrical all-to-all model. Mind that in this case the best decay method found for the cognitive parameter was to keep it constant.}
		\label{table:appendix:hyperpars_ASYM_ATA}
	\end{table}
	
	\begin{table}[!ht]
		\centering
		\begin{tabular}{|c|c|c|c|c|c|c|c|}
			\hline
			& $\lambda$ & $\mu$ & $\gamma$ & $\beta$ & $\tau$             & $\tilde{u}$ & MSE  \\ \hline
			lowest cost & 0.79      & 0.55  & 0.66    & 0.48     & March 2nd 2021     & 0.44        & 0.02 \\ \hline
			median      & 0.53      & 0.6   & 0.62    & 0.59     & March 2nd 2021 & 0.46        &      \\ \hline
		\end{tabular}
		\caption{Lowest cost and median parameters for the all-to-all approximation with asymmetric rates. Mind that there is no MSE associated to the median set of parameters as the median is computed for every single parameter and not as a set.}
		\label{table:appendix:ASYM_ATA_lowest_median}
	\end{table}

\end{document}